\documentclass{jfm}
\usepackage{graphicx}
\usepackage{epstopdf, epsfig}
\usepackage{hyperref}
\usepackage{upgreek}

\usepackage{amsmath}
\usepackage{longtable,tabularx}
\usepackage{esvect}
\usepackage{color}
\DeclareMathAlphabet{\mathpzc}{OT1}{pzc}{m}{it}

\shorttitle{Lagrangian modal analysis}
\shortauthor{V. J. Shinde and D. V. Gaitonde}

\title{Lagrangian approach for modal analysis of fluid flows}
\author{Vilas J. Shinde
  \corresp{\email{shinde.33@osu.edu}}
  \and Datta V. Gaitonde}

\affiliation{Department of Mechanical and Aerospace Engineering,
The Ohio State University,
Columbus, OH 43210, USA
}
\begin{document}

\maketitle

\begin{abstract}

Common modal decomposition techniques for ﬂowﬁeld analysis, data-driven modeling and ﬂow control, such as proper orthogonal decomposition (POD) and dynamic mode decomposition (DMD) are usually performed in an Eulerian (ﬁxed) frame of reference with  snapshots from measurements or evolution equations. The Eulerian description poses some diﬃculties, however, when the domain or the mesh deforms with time as, for example, in ﬂuid-structure interactions.  For such cases, we ﬁrst formulate a Lagrangian modal analysis (LMA) ansatz by a posteriori transforming the Eulerian ﬂow ﬁelds into Lagrangian ﬂow maps through an orientation and measure-preserving domain diﬀeomorphism. The development is then verified for Lagrangian variants of POD and DMD using direct numerical simulations (DNS) of two canonical ﬂow conﬁgurations at Mach 0.5, the lid-driven cavity and ﬂow past a cylinder, representing internal and external ﬂows, respectively, at pre- and post-bifurcation Reynolds numbers. The LMA is demonstrated for several situations encompassing unsteady ﬂow without and with boundary and mesh deformation as well as non-uniform base ﬂows that are steady in Eulerian but not in Lagrangian frames. We show that LMA application to steady nonuniform base ﬂow yields insights into ﬂow stability and post-bifurcation dynamics. LMA naturally leads to Lagrangian coherent ﬂow structures and connections with finite-time Lyapunov exponents (FTLE). We examine the mathematical link between FTLE and LMA by considering a double-gyre ﬂow pattern. Dynamically important flow features in the Lagrangian sense are recovered by performing LMA with forward and backward (adjoint) time procedures.
\end{abstract}


\section{Introduction}

Flow fields of interest contain a broad range of coherent flow structures undergoing non-linear interactions.
Recent advances in computational and experimental techniques have generated high-fidelity representations of these dynamics in the form of enormous databases.
The extraction of knowledge from such high-dimensional fields is greatly facilitated by modal decomposition, which is playing an increasingly crucial role in discerning the relevant kinematic and dynamic flow features.
Modal decomposition describes the spatio-temporally varying flow field in terms of spatially correlated, or coherent, flow features ordered by some property such as energy content or growth rate, together with their temporal variation.

Some of the commonly used modal decomposition techniques have been recently reviewed by \cite{rowley2017model}, and \cite{taira2017modal,taira2020modal}.
Among these, the most popular are proper orthogonal decomposition (POD) and dynamic mode decomposition (DMD).
In these, data from high-fidelity numerical or experimental efforts are \textit{a posteriori} processed to extract  a set of energetically ranked 
orthogonal modes (POD)~\citep{kosambi1943statistics,karhunen1946spektraltheorie,loeve1948functions} or dynamically significant 
modes (DMD)~\citep{schmid2009dynamic,rowley2009spectral,schmid2010dynamic} each with {associated temporal dynamics.} 
Other ways of assessing these modes include their energies and growth/decay rates.
POD is optimal among all decompositions in terms of the maximization of energy for a given number of modes~\citep{berkooz1993proper}.
The corresponding temporal coefficients are closely linked to the spatial modes, resulting in a bi-orthogonal decomposition~\citep{aubry1991spatiotemporal,shinde2020proper}.
DMD, on the other hand, provides modes that are associated with unique frequencies and growth or decay rates. It is based on Koopman theory~\citep{koopman1931hamiltonian}, whose connection to aspects of the Koopman operator/modes has been developed in ~\cite{mezic2005spectral,mezic2013analysis} and to DMD modes is illustrated in~\cite{rowley2009spectral}. 
Although both POD and DMD are linear procedures, {the underlying dynamics of system that generated the flow fields may be nonlinear.}

Most decomposition techniques are formulated and applied in the Eulerian (fixed) frame of reference, which complicates their application in deforming or moving domains.
An example is the problem of fluid-structure interactions where structural response causes boundary shape changes with corresponding mesh deformation.
This difficulty has been recognized in the literature as, for instance, by \cite{menon2020dynamic} and \cite{mohan2016model}, who performed DMD of pitching/plunging airfoils.
One solution is to simplify the application by restricting attention to a part of the domain as in  \cite{goza2018modal,schmid2010dynamic,shinde2019transitional}, who used POD and DMD for fluid-structure interactions.
A method that factors mesh deformation is presented by~\cite{shinde2019galerkin}, who obtained POD modes on deforming mesh solutions in the context of reduced-order modeling of the vortex induced vibration and supersonic flutter.
Nevertheless, a formal mathematical framework for modal decomposition on deforming and moving domains applicable to fluid flow analysis remains a pressing need.
To address this gap, we develop a Lagrangian modal analysis (LMA) approach that eases application of modal decomposition techniques to flows involving domain deformation, by recasting the analysis in a suitably selected moving reference frame.
The formulation is couched in general terms, though for concreteness, we consider both Lagrangian POD (LPOD) as well as Lagrangian DMD (LDMD)

An important practical difference between the Eulerian and Lagrangian descriptions lies in the number of variables required to represent the flow \citep{price2006lagrangian}. 
For example, a steady non-uniform flow in the Eulerian (fixed point) representation is  time-dependent in the Lagrangian formulation.
The difference between the reference frames manifests in the definition of acceleration and is related to spatial velocity gradients in the flow.
Thus, the Eulerian (local) acceleration for a steady non-uniform flow is zero, whereas the Lagrangian (convective) acceleration remains non-zero.
In the present work, we exploit this Lagrangian time dependence of steady non-uniform flows, to pose modal decompositions in the Lagrangian frame of reference.
The significance of Lagrangian modal analysis of an otherwise steady (Eulerian) base flow is also discussed from the standpoint of flow stability.

A natural question arises on the connection between LMA and Lagrangian techniques employed in chaos and mixing studies. 
In particular, Lyapunov exponents are commonly used to quantify the divergence or stretching of a filament in time, and are related to specific stretching rates and mixing efficiencies~\citep{ottino1989kinematics}.
The relatively popular FTLE (finite-time Lyapunov exponent) technique invokes a Lagrangian frames of reference and has been employed for a variety of assessments~\citep{haller2000lagrangian,shadden2005definition,peacock2010introduction,mancho2013lagrangian,samelson2013lagrangian,haller2015lagrangian,nelson2015dg,gonzalez2016finite}.
The largest such exponent identifies high strain regions exhibiting stable/unstable manifolds or hyperbolic trajectories~\citep{balasuriya2016hyperbolic}.
The present LMA aims to decompose the stretching of the flow fabric into coherent modes pertaining to the specific modal decomposition technique (in our case, POD or DMD).
Thus, the largest FTLE, which represents the largest eigenvalue of the right Cauchy-Green strain tensor, is analogous to the first Lagrangian POD mode.
We establish this correspondence in general terms through a mathematical relation between the FTLE and Lagrangian modal analysis ansatz.

To demonstrate LPOD and LDMD, we consider two canonical flow configurations, namely, lid-driven cavity and flow past a cylinder, representing, respectively, an internal and an external flow.
Direct numerical simulations (DNS) are performed for each in the Eulerian frame of reference at $M_\infty=0.5$ by solving the compressible Navier-Stokes equations.
A range of Reynolds numbers is considered for each to encompass steady (pre-critical) and unsteady (post-critical) regimes.
For the lid-driven cavity, the first Hopf bifurcation occurs at $Re_L\approx 10{,}500$, where $Re_L$ is the Reynolds numbers based on the cavity length.  
Thus the range chosen is $5{,}000 \leq Re_L \leq 15{,}000$.
A suitable surrogate representing key properties of fluid-structure interactions is constructed by subjecting the lid-driven cavity to a forced-domain deformation.
The flow past a cylinder considers the Reynolds number (based on the cylinder diameter, $D$ ) range between $20 \leq Re_D \leq 100$, encompassing the first Hopf bifurcation at $Re_c\approx 50$.
A simple analytical model of the double-gyre flow pattern is used to examine the mathematical link derived between the FTLE and the LPOD and LDMD modes.

The article is organized as follows.
Section~\ref{sec:theory} presents the theoretical framework for the LMA, which includes i) transformation of the Eulerian flow fields to the Lagrangian flow fields, ii) formulation of the LPOD and LDMD and iii) the derivation of the mathematical link between the FTLE and LMA.
The details of the numerical methodology and case studies are provided in Sec.~\ref{sec:numerical}.
The results and discussion section (Sec.~\ref{sec:res_disc}) presents the application of LMA to the different flow types, namely, unsteady flow, flow with mesh deformation, Eulerian steady  but Lagrangian unsteady flow, and the double-gyre case study elucidating the relation with FTLE.
Lastly, we provide some concluding remarks in Sec.~\ref{sec:concl}.

\section{Theory} \label{sec:theory}
The Lagrangian (moving) and Eulerian (fixed) descriptions of fluid flow are necessarily equivalent in terms of the dynamics.
{
Although the Lagrangian perspective offers some mathematical and conceptual advantages, the lack of direct access to spatial velocity gradients poses difficulties for the solution of Navier-Stokes equations~\citep{batchelor2000introduction}.}
The Eulerian perspective is more convenient such as for example in the comparison with fixed point measurements obtained from experiment, and is thus commonly employed for the flow simulations.
In the same vein, modal decomposition techniques such as POD and DMD have been developed for Eulerian description.
In the present work, we retain the Eulerian approach to simulate the flow, but the data is then recast into a suitable Lagrangian frame that is more convenient for the application of modal analysis techniques, even when the domain is deforming.

\subsection{Eulerian to Lagrangian transformation} \label{sec:Eul_Lag}

Consider a real Euclidean vector space $\mathsf{E}$ of dimension $d$, 
with the inner product $\langle \pmb{x}, \pmb{x} \rangle > 0$ for non-zero $\pmb{x}$, and real norm $\| \pmb{x} \|=\sqrt{\langle \pmb{x}, \pmb{x} \rangle}$.
Here, for convenience we consider $\textsf{E}$ as a $d$-dimensional point space with a coordinate system and a frame of reference, on which the Euclidean space with $\mathbb{R}^{d=3}$ can be realized by considering an orthonormal basis.
A flow in a suitable closed domain $\mathsf{D} \subseteq \mathsf{E} = \mathbb{R}^3$ may be represented in terms of a vector field $\pmb{u}$ through the mapping~\footnote{The mathematical terminology on the Eulerian and Lagrangian flow descriptions, to some degree, follows~\cite{Talpaert2002}.}
\[ \pmb{u}:\mathsf{D}\times [0,\mathsf{T}] \rightarrow \mathbb{R}^3:(\pmb{x},t)\mapsto \pmb{u}(\pmb{x},t)\text{ with } \pmb{x}\in \mathsf{D},\]
where $t$ is an instant from the total time $\mathsf{T}\subset \mathbb{R}$.
In the Eulerian description, all physical quantities (scalar, vector or tensor) are expressed at each instant and at every fixed spatial location with respect to the frame of reference.
Thus, the fixed spatial coordinates $x_i$ of vector $\pmb{x}$ and time $t$ constitute the Eulerian coordinates with respect to a fixed (Eulerian) frame of reference of $\mathsf{E}$.
The Eulerian description refers to  flow fields at an instant $t$ mapping on another time $t+dt$, where $dt$ is the time differential.

The Lagrangian description of the flow, on the other hand, identifies a flow state at an instant with respect to a time dependent frame of reference.
The flow field, $\pmb{\mathcal{U}}$, over a closed domain $\mathcal{D} \subseteq \mathsf{E} = \mathbb{R}^3$ and time interval $[0,\mathcal{T}] \subset \mathbb{R}$ may be mapped as:
\[ \pmb{\mathcal{U}}:\mathcal{D}\times [0,\mathcal{T}] \rightarrow \mathbb{R}^3:(\pmb{\chi},\tau)\mapsto \pmb{\mathcal{U}}(\pmb{\chi},\tau)\text{ with } \pmb{\chi}\in \mathcal{D}, \tau\in [0,\mathcal{T}].\]
The flow evolves from a reference state and maps on a deformed geometrical configuration.
Thus, an initial reference configuration, $\Omega_0\in \mathcal{D}$ at $\tau=\tau_0$ and a current configuration $\Omega \in \textit{SDiff}(\mathcal{D})$ at $\tau$, may be defined, where $\textit{SDiff}(\mathcal{D})$ is an \textit{orientation and measure-preserving diffeomorphism} of $\mathcal{D}$.
Mathematically, the flow map can be expressed as
\begin{eqnarray} \label{eq:mapping}
\mathcal{M}:\mathcal{D}\times [0,\mathcal{T}] \rightarrow \textit{SDiff}(\mathcal{D})\subseteq \mathsf{E}=\mathbb{R}^3: (\pmb{\chi},\tau) \mapsto \mathcal{M}(\pmb{\chi},\tau)=(\pmb{x},t), \text{ and }  \nonumber \\
\mathcal{M}(\pmb{\chi}_0,\tau_0)=\text{identity map}.
\end{eqnarray}
The triple components $\chi_i$ of vector $\pmb{\chi}$ and time $\tau$ comprise the Lagrangian coordinates, which can be explicitly expressed using the Eulerian frame of reference as
\begin{equation} \label{eq:x=chi}
(x_i,t) = \mathcal{M}^i(\pmb{\chi},\tau) = \mathcal{M}^i(\chi_1,\chi_2,\chi_3,\tau).
\end{equation}

The Lagrangian flow mapping from an initial configuration $\Omega_0$ to a current configuration $\Omega$ must meet the regularity conditions of the transformation, mainly that it be injective and $\mathcal{M}$ be a bijection.
The inverse $\mathcal{M}^{-1}$ exists due to the regularity conditions, and by considering the existence of the inverse at any instant $\tau$, we can state
\begin{equation} \label{eq:M=M-1}
(\pmb{\chi},\tau) = \mathcal{M}^{-1}(\pmb{x},t) \Leftrightarrow (\pmb{x},t)=\mathcal{M}(\pmb{\chi},\tau).
\end{equation}
Consequently, the Jacobian matrix $\mathcal{J}=\partial (\pmb{\chi},\tau)/\partial (\pmb{x},t)$ is invertible, which plays an important role in the domain deformations.
The vector fields of the Lagrangian and Eulerian frame of references are related as
\begin{equation}\label{eq:ul-ue}
\pmb{\mathcal{U}}(\pmb{\chi},\tau)=\pmb{u}(\mathcal{M}(\pmb{\chi},\tau)) \Leftrightarrow \pmb{u}(\pmb{x},t)=\pmb{\mathcal{U}}(\mathcal{M}^{-1}(\pmb{x},t)),
\end{equation}
which also applies to each physical quantity of the flow.

The total or material derivative of a quantity, \textit{e.g.,} flow velocity vector, in the Lagrangian frame of reference is simply its partial derivative with respect to time $\tau$, written as
\begin{equation}\label{eq:tot_Lag}
\frac{D \pmb{\mathcal{U}}}{D\tau}=\frac{\partial \pmb{\mathcal{U}}}{\partial \tau}\Big|_{\pmb{\chi}}.
\end{equation}
On the other hand, the Eulerian frame of reference accounts for the local and convective rates of change of a quantity.
The total derivative from Eq.~\ref{eq:ul-ue} is then:
\begin{eqnarray}
\frac{D\pmb{u}}{Dt}&=&\frac{\partial \pmb{u}}{\partial t}\Big|_{\pmb{x}} + \frac{\partial \pmb{u}}{\partial \pmb{x}}\cdot \frac{\partial \pmb{x}}{\partial t}\Big|_{\pmb{\chi}} \\
&=&\underbrace{\frac{\partial \pmb{u}}{\partial t}}_{\text{local rate of range}}+\underbrace{ (\pmb{u}_{\pmb{\chi}}\cdot\nabla) \pmb{u}}_{\text{convective rate of change}}. \label{eq:tot_Eul}
\end{eqnarray}
The Eulerian convective flow velocity $\pmb{u}_{\pmb{\chi}}$ is the mapping of the vector field from the fixed spatial coordinates $\pmb{x}$ to $\pmb{x}+d\pmb{x}$, where $d\pmb{x}$ is the differential of space.
The flow velocity vector fields can be expressed in terms of total derivative of the space vector fields in the Eulerian and Lagrangian approaches, respectively, as 
\begin{equation}\label{eq:Eul_Lag_vel}
\underbrace{\pmb{u}(\pmb{x},t)=\frac{D \pmb{x}}{Dt}}_{Eulerian} \text{ and } \underbrace{\pmb{\mathcal{U}}(\pmb{\chi},\tau)=\frac{\partial \pmb{\chi}}{\partial \tau}\Big|_{\pmb{\chi}}}_{Lagrangian}.
\end{equation}
Note that the velocity field in the Lagrangian frame of reference is always a function of time for a non-uniform flow.

\subsection{Lagrangian proper orthogonal decomposition (LPOD)} \label{sec:lPOD}

The POD method is based on forming a two-point correlation tensor leading to an eigenvalue problem~\citep{lumley67}.
The procedure yields an expansion in terms of orthogonal real basis functions or modes, which are coherent flow structures with associated modal energies.
The technique may be applied in the space or spectral domains~\citep{lumley67,moin1989characteristic,citriniti2000reconstruction,towne2018spectral}, each with its own advantages.
Various mathematical properties of POD, such as the optimal modal energy representation and spatio-temporal modal dynamics~\citep{lumley1970stochastic,aubry1991hidden,aubry1991spatiotemporal}, are instrumental in the popularity of the technique.
The most popular method is that of \cite{sirovich1987turbulence}, which uses snapshots gathered from successive flow instants to form an equivalent two-point correlation tensor, adhering to the conventional Eulerian frame of reference.
Here we will consider the equivalent Lagrangian approach; for concreteness, we develop the spatial form of POD, with the understanding that the correspondence to the spectral form is straightforward.
{
Let us consider a real matrix $\pmb{X}\in \mathbb{R}^{m\times n}$ that comprises Lagrangian flow fields in discrete form, where $m$, $n$ are the space and time dimensions, respectively. 
The Lagrangian flow fields matrix $\pmb{X}$ can be redefined accounting for the weight tensor as: $\pmb{Y}=\pmb{\textsc{w}}^T \pmb{X} \in \mathbb{R}^{m\times n}$.
For a given Lagrangian flow field, the objective of POD is to distill out functions $\pmb{\Phi}_l \in \mathbb{R}^m$, such that
\begin{equation}\label{eq:pod_objective}
\lambda_l = \arg \max \left\lbrace \frac{\pmb{\Phi}_l^T \pmb{Y} \pmb{Y}^T \pmb{\Phi}_l}{\pmb{\Phi}_l^T\pmb{\Phi}_l}\right\rbrace.
\end{equation}
The LPOD spatial modes $\pmb{\Phi}_l$ are the eigenfunctions of the eigenvalue problem,
\begin{equation}\label{eq:Fredholm}
\pmb{Y}\pmb{Y}^T \pmb{\Phi}_l = \lambda_l \pmb{\Phi}_l,
\end{equation}
where the matrix $\pmb{Y}\pmb{Y}^T$ is symmetric positive semi-definite, ensuring a set of orthonormal eigenvectors and corresponding eigenvalues: $\lbrace \pmb{\Phi}_l, \lambda_l \rbrace_{l\in\{1,...,m\}}$ ordered as $\lambda_l \geq \lambda_{l+1} \geq 0$.
The matrix of these eigenvectors $\pmb{\Phi} \in \mathbb{R}^{m\times m}$ with $\pmb{\Phi}^T\pmb{\Phi}=\pmb{I}$ forms a complete orthonormal basis of $\mathbb{R}^m$.
Here $\pmb{I}\in \mathbb{R}^{m\times m}$ is an identity matrix.
The Lagrangian flow fields can be expressed as,
\begin{equation}\label{eq:vel_pod}
\pmb{X} = \pmb{\Phi}\pmb{\Lambda}^{\frac{1}{2}}\pmb{\Psi}^T
\end{equation}
where $\pmb{\Lambda}=\text{diag}\{\lambda_1,...,\lambda_m\}$  and $\pmb{\Psi}\in \mathbb{R}^{n\times m}$ are the LPOD temporal coefficients  that are associated with the LPOD spatial modes $\pmb{\Phi}$.
The LPOD time coefficient can be obtained as,
\begin{equation}\label{eq:pod_Psi}
\pmb{\Psi} = \pmb{Y}^T\pmb{\textsc{w}}^{-1}\pmb{\Phi} \pmb{\Lambda}^{-\frac{1}{2}}.
\end{equation}
The LPOD temporal coefficients matrix are also orthonormal, \textit{i.e.}, $\pmb{\Psi}^T\pmb{\Psi}=\pmb{I}$, forming another set of basis functions.
Thus, the eigenvalue problem of Eq.~\ref{eq:Fredholm} can be alternatively stated as
\begin{equation}\label{eq:Fredholm_tau}
\pmb{Y}^T\pmb{Y}\pmb{\Psi} = \pmb{\Psi}\pmb{\tilde{\Lambda}},\hspace{2mm}\text{with}\hspace{2mm} \pmb{\Phi}=\pmb{\textsc{w}}^{-T}\pmb{Y}\pmb{\Psi}\pmb{\tilde{\Lambda}}^{-\frac{1}{2}} = \pmb{X}\pmb{\Psi}\pmb{\tilde{\Lambda}}^{-\frac{1}{2}},
\end{equation}
where $\pmb{\tilde{\Lambda}}=\text{diag}\{\lambda_1,...,\lambda_n\}$.
Typically, the LPOD procedure via Eq.~\ref{eq:Fredholm_tau} is much more efficient compared to Eq.~\ref{eq:Fredholm} due to the fewer degrees of freedom that arise in the time discretization as opposed to the spatial discretization, \textit{i.e.}, $n \ll m$, which is the key aspect underlying the method of snapshots~\citep{sirovich1987turbulence}.
}
\subsection{Lagrangian dynamic mode decomposition (LDMD)} \label{sec:lDMD}

The popularity of DMD has grown recently as a complementary approach to POD.
DMD extracts coherent features based on the Koopman operator and may be applied to snapshots, which usually represent progress in time, though spatially evolving features can also be extracted if desired~\citep{schmid2010dynamic,rowley2009spectral}.
When the snapshots represent time progression, the DMD modes represent spatially coherent structures evolving in time with unique frequencies and growth/decay rates.
For a linearized flow about a steady state (in Eulerian frame of reference), the DMD modes are equivalent to  global stability modes~\citep{schmid2010dynamic}.
As noted earlier, the Lagrangian formulation is inherently unsteady for non-uniform flows; thus the Lagrangian DMD may be performed directly on a (non-uniform Eulerian) steady base flow, leading to modal information pertinent to the stability of the base flow.

Typically, DMD derives a mapping between suitably constructed sequences of flow states.
It then solves for the basis functions (eigenvectors) of a reduced-order representation of the mapping.
The equivalent in a Lagrangian frame of reference may be developed as follows.
We consider $\pmb{X}$ and $\pmb{Y}$ as tensors whose elements are the Lagrangian flow fields, \textit{e.g.}, the velocity vector $\pmb{\mathcal{U}}(\pmb{\chi},\tau)$, where time $\tau \in [0,\mathcal{T}]$, such that $\tau=\{\tau_1, \tau_2,\cdots,\tau_{n} \}$ for $\pmb{X}$ and $\tau=\{\tau_2, \tau_3,\cdots,\tau_{n+1} \}$ for $\pmb{Y}$.
Similarly, the Lagrangian space coordinate vector $\pmb{\chi}$ is considered to be discrete of size $m$, thus $\pmb{X}, \pmb{Y}\in \mathbb{R}^{m\times n}$.
The aim of the DMD procedure is to find $\pmb{A} \in \mathbb{R}^{m\times m}$ such that
\begin{equation}\label{eq:A_operator}
\pmb{A}\pmb{X} = \pmb{Y}\text{ or } \pmb{A}=\pmb{Y}\pmb{X}^+,
\end{equation}
where $\pmb{X}^+$ is the Moore-Penrose pseudoinverse of $\pmb{X}$.
As in the traditional Eulerian approach, in practice $m \gg n$ which complicates the use of Eq.~\ref{eq:A_operator}.
A low-order representation of $\pmb{A}$ is sought  through the compact singular value decomposition of $\pmb{X}$. 
\[ \pmb{X}=\pmb{U}\pmb{\Sigma}\pmb{V}^{T},\]
This leads to an approximate representation of $\pmb{A}$ as,
\begin{equation}\label{eq:A_tilde}
\pmb{\tilde{A}} = \pmb{U}^{T}\pmb{A}\pmb{U} = \pmb{U}^{T}\pmb{Y}\pmb{V}\pmb{\Sigma}^{-1} \in \mathbb{R}^{n \times n},
\end{equation}
where $\pmb{U}\in \mathbb{R}^{m \times n}$ and $\pmb{V}\in \mathbb{R}^{n\times n}$ are orthogonal matrices, while $\pmb{\Sigma}$ is a diagonal matrix of size $n\times n$ with non-zero real singular values.
Lastly, the Lagrangian DMD modes, $\pmb{\phi}_l \in \mathbb{C}^m$, are obtained by
\begin{equation}\label{eq:dmd_modes}
\pmb{\phi}_l = \pmb{U}\pmb{v}_l,
\end{equation}
where the $l$th eigenvector $\pmb{v}_l \in \mathbb{C}^n$ is a solution of the eigenvalue problem: 
\[ \pmb{\tilde{A}}\pmb{v}_l=\kappa_l \pmb{v}_l \] with the corresponding eigenvalue $\kappa_l \in \mathbb{C}$.
The growth rate and angular frequency of the LDMD mode are $\ln |\kappa_l|/\delta \tau$ and $\arg(\kappa_l)/\delta \tau$, respectively, where $\delta \tau$ is the Lagrangian uniform time discretization.

{
The LDMD formulation naturally connects to the Lagrangian flow map, which may comprise fixed points, periodic orbits, stable and unstable manifolds, and chaotic attractors~\citep{ottino1989kinematics,wiggins2005dynamical,lekien2007lagrangian,shadden2005definition,haller2015lagrangian}.
Indeed, the LDMD matrix $\pmb{\tilde{A}}$, which is an approximation for $\pmb{A}$, seeks properties of the Lagrangian flow map $\mathcal{M}(\pmb{\chi},\tau)$ (of Eq.~\ref{eq:mapping}) in terms of the Lagrangian flow fields $\pmb{X}$ (see Fig.~\ref{fig:lma_ftle}).
These properties include eigenvalues, eigenvectors, energy amplification, and resonance behavior~\citep{schmid2010dynamic}, which reveal the dynamic characteristics of the process that is governing the flow map.
The Lagrangian flow fields at any time instant $\tau$ can be expressed as,
\begin{equation}
\pmb{X}_\tau=\pmb{\phi}\exp{\left( \pmb{\kappa}\tau \right)}\pmb{a} \hspace{5mm}\text{with}\hspace{5mm}\pmb{a}=\pmb{\phi}^+\pmb{X}_{\tau_1}.
\end{equation}
Here $\pmb{\phi}=\{\pmb{\phi}_l\}_{l\in\{1,...,n\}}\in \mathbb{C}^{m\times n}$ is a complex set of LDMD modes, while $\pmb{\kappa}=\text{diag}\{\kappa_1,...,\kappa_n\}$ are the eigenvalues (also Ritz values).
The initial conditions $\pmb{a}\in \mathbb{C}^{n}$ are obtained by means of the pseudoinverse $\pmb{\phi}^+$ and the identity map $\pmb{X}_{\tau_1}=\mathcal{M}(\pmb{\chi}_0,\tau_0)$.
}
\subsection{Lyapunov exponents and Lagrangian modal analysis ansatz} \label{sec:LE}

The Lagrangian flow map $\mathcal{M}(\pmb{\chi},\tau)$ of Eq.~\ref{eq:mapping} represents a dynamical system, evolving from an initial state, \textit{i.e.}, from the identity map $\mathcal{M}(\pmb{\chi}_0,\tau_0)$.
Lyapunov exponents characterize the rate of separation between two points on the manifold $\textit{SDiff}(\mathcal{D})$, with divergence between the points being constrained to the linear approximation; in addition, the Lyapunov exponent spectrum is analogous to the eigenvalue spectrum of the linearized stability equations at  steady state~\citep{vastano1991short,goldhirsch1987stability}.
For a $d$-dimensional state space, there are $d$ number of Lyapunov exponents; however, among these, the largest is significant in determining the system behavior.
If $\delta \pmb{\chi}_0$ and $\delta \pmb{\chi}_\tau$ are the separations between any two points at an initial time $\tau_0$ and a later time $\tau$, respectively, then the maximum Lyapunov exponent is given by,
\begin{equation}\label{eq:Ly_max}
\lambda^{LE} = \lim_{\tau \rightarrow \infty} \lim_{|\delta \pmb{\chi}_0| \rightarrow 0} \frac{1}{\tau} \ln \frac{|\delta \pmb{\chi}_\tau|}{|\delta \pmb{\chi}_0|}, 
\end{equation}
where the limits $\tau\rightarrow \infty$ and $|\delta \pmb{\chi}_0| \rightarrow 0$ ensure time asymptotic and linear considerations, respectively.
The Lyapunov exponents provide insights into a vector space that is tangent to the state space. 
The Jacobian matrix $\mathcal{J}$ governs the evolution of the small separation $\delta \pmb{\chi}_0$ as,
\begin{equation}\label{eq:Jac_pert}
\delta \pmb{\chi}_\tau = \exp{\left(\int_0^\tau \mathcal{J}(\tau') d\tau'\right)} \delta \pmb{\chi}_0.
\end{equation}
A matrix $\pmb{\mathcal{X}}$, defined as~\citep{oseledets2008oseledets}
\begin{equation}\label{eq:Lam_LE}
\pmb{\mathcal{X}} = \lim_{\tau \rightarrow \infty} \frac{1}{\tau} \ln \sqrt{\delta \pmb{\chi}_\tau \delta \pmb{\chi}_\tau^T}, 
\end{equation}
provides the Lyapunov exponent spectrum in terms of its eigenvalues, giving the average exponential growth rates of the separation at time $\tau$.
Furthermore, in the time limit $\tau \rightarrow \infty$, the Lyapunov spectrum offers a global measure of the strange attractor of the dynamical system~\citep{yoden1993finite}.

Alternatively, Lyapunov exponents may be estimated locally (in the limit $\tau \to 0$) or for a finite time (for $\tau \in [0,\mathcal{T}]$) in order to investigate the local dynamics of the system~\citep{goldhirsch1987stability,thiffeault2001geometrical,nolan2020finite}.
The finite time Lyapunov exponents for $\tau\in[0,\mathcal{T}]$ are estimated as: 
\begin{equation}\label{eq:ftle}
\lambda^{LE}_d = \frac{1}{\mathcal{T}} \ln{\sqrt{\lambda_d(\pmb{\mathcal{C}})}},
\end{equation}
where $\lambda_d(\pmb{\mathcal{C}})$ denotes the $d$th eigenvalue of the right Cauchy-Green strain tensor,
\begin{equation}\label{eq:cauchy}
\pmb{\mathcal{C}} = \mathcal{J}^T\mathcal{J} = \nabla \mathcal{M}^T \nabla \mathcal{M}.
\end{equation}
The maximum Lyapunov exponent is expressed in terms of the maximum eigenvalue of $\pmb{\mathcal{C}}$ as,
\begin{equation}\label{eq:lamLE}
\lambda^{LE} = \frac{1}{\mathcal{T}} \ln{\sqrt{\lambda_{max}(\pmb{\mathcal{C}})}}.
\end{equation}
$\lambda_{max}(\pmb{\mathcal{C}})$ identifies flow regions with high shear, which are further illuminated by the maximum FTLE field due to the logarithmic definition~\citep{haller2002lagrangian}.

To establish the relation between the Lyapunov exponents and LMA, we express the Lagrangian velocity of Eq.~\ref{eq:Eul_Lag_vel} as
\begin{equation}\label{eq:}
\pmb{\mathcal{U}}(\pmb{\chi},\tau) = \frac{\partial \mathcal{M}(\pmb{\chi},\tau)}{\partial \tau} = \frac{\partial \mathcal{M}(\pmb{\chi},\tau)}{\partial \pmb{\chi}_0} \frac{\partial \pmb{\chi}_0}{\partial \tau} = \nabla \mathcal{M} \pmb{\mathcal{U}}(\pmb{\chi}_0,\tau).
\end{equation}
{Reconsider the real matrix $\pmb{X}\in \mathbb{R}^{m\times n}$ utilized in the formulation of LPOD and LDMD.
Here $\pmb{X}$ comprises the Lagrangian flow fields of the absolute velocity $\|\pmb{\mathcal{U}}(\pmb{\chi},\tau)\|$.
\begin{figure}
\centering
\includegraphics[scale=0.35]{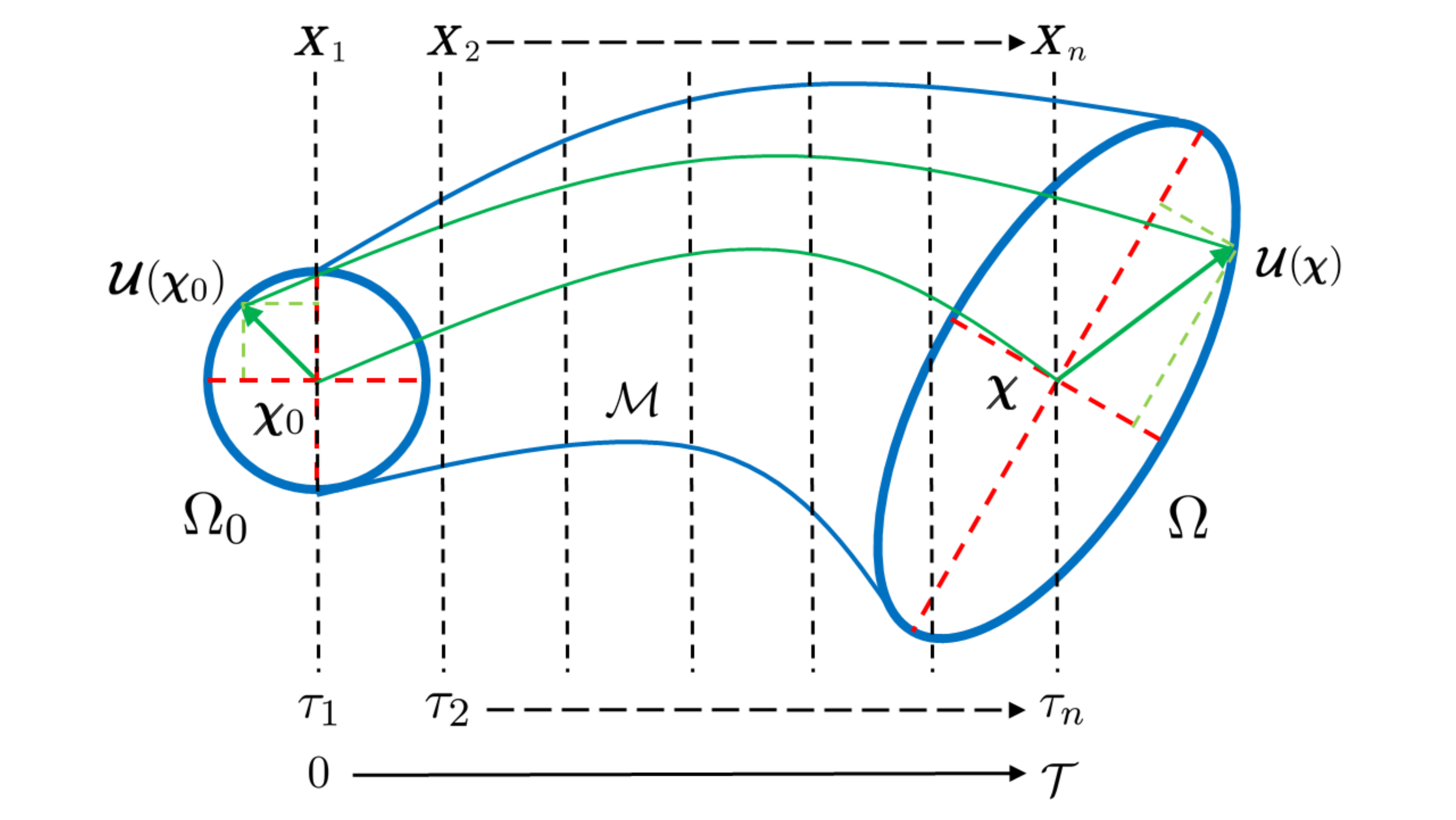}
\caption{Schematic representation of a fluid element in Lagrangian frame of reference, where the deforming flow trajectories lead to Lyapunov exponents and a data matrix for LMA over a finite time.}
\label{fig:lma_ftle}
\end{figure}
For a time instant $\tau$ with $n=1$, we can write
\begin{eqnarray}
\text{diag}\left\lbrace \pmb{X}\pmb{X}^T \right\rbrace &=&  \|\pmb{\mathcal{U}}(\pmb{\chi},\tau)\|^2 = \pmb{\mathcal{U}}^T(\pmb{\chi},\tau)\pmb{\mathcal{U}}(\pmb{\chi},\tau) \\
&=& \pmb{\mathcal{U}}^T(\pmb{\chi}_0,\tau){ \{ \nabla \mathcal{M}}^T\nabla \mathcal{M}\}\pmb{\mathcal{U}}(\pmb{\chi}_0,\tau) \\
&=& \pmb{\mathcal{U}}^T(\pmb{\chi}_0,\tau)\pmb{\mathcal{C}}\pmb{\mathcal{U}}(\pmb{\chi}_0,\tau) \label{eq:uCu}
\end{eqnarray}
where $\pmb{\mathcal{C}} \in \mathbb{R}^{d\times d}$ is the right Cauchy-Green strain tensor of Eq.~\ref{eq:cauchy}.
The alignment between $\pmb{\mathcal{U}}(\pmb{\chi}_0,\tau)$ and the eigenvectors of $\pmb{\mathcal{C}}$ manifests in the value of $\text{diag}\left\lbrace\pmb{X}\pmb{X}^T\right\rbrace$.
For a finite time $\tau\in[0,\mathcal{T}$], we can rewrite Eq.~\ref{eq:uCu} as
\begin{eqnarray}
\text{diag}\left\lbrace (\pmb{X}\pmb{X}^T)_d \right\rbrace &=& \| \pmb{\mathcal{U}}_d(\pmb{\chi}_0,\tau) \|^2 \lambda_d(\pmb{\mathcal{C}}) \\
&=& \| \pmb{\mathcal{U}}_d(\pmb{\chi}_0,\tau) \|^2 \exp\left(2\mathcal{T}\lambda_d^{LE}\right).
\end{eqnarray}
The (maximum) FTLE relates to the maximum of $\text{diag}\left\lbrace (\pmb{X}\pmb{X}^T)_{d} \right\rbrace$ for a specific argument $d$, which corresponds to the alignment of $\pmb{\mathcal{U}}(\pmb{\chi}_0,\tau)$ and the eigenvector of $\pmb{\mathcal{C}}$ with the largest eigenvalue $\lambda_{max}(\pmb{\mathcal{C}})$, as
\begin{equation}
\lambda^{LE} = \frac{1}{\mathcal{T}}\ln\sqrt{\text{diag}\left\lbrace\max_d\left\lbrace(\pmb{X}\pmb{X}^T)_d\right\rbrace\right\rbrace}.
\end{equation}
The FTLE field represents the local maxima of 
\begin{eqnarray}
\text{diag}\left\lbrace\max\limits_d\left\lbrace(\pmb{X}\pmb{X}^T)_d\right\rbrace\right\rbrace &=& \text{diag}\left\lbrace\max\limits_d\left\lbrace\left(\pmb{\Phi}\pmb{\Lambda}\pmb{\Phi}^T\right)_d\right\rbrace\right\rbrace \\ &=& \text{diag}\left\lbrace\max\limits_d\left\lbrace\left(\sum_{l=1}^m\pmb{\Phi}_l{\lambda}_l\pmb{\Phi}_l^T\right)_d\right\rbrace\right\rbrace,
\end{eqnarray}
whereas LPOD provides the global eigenfunctions $\pmb{\Phi}_l$ and associated energies ordered as $\lambda_l\geq \lambda_{l+1} \geq 0$.
In addition to the real symmetry and positive semi-definiteness, the auto-correlation tensor $\pmb{X}\pmb{X}^T$ is also diagonally dominant, which is a consequence of the rearrangement inequality~\citep[Chapter X]{zbMATH03073200}.
Thus, the maximum Lyapunov exponent field closely relates to the first eigenmode (LPOD mode with maximum $\lambda_l$) of the auto-correlation tensor of the velocity magnitude.
}

{
A key feature of FTLE field is that it is objective, \textit{i.e.}, independent of the observer's frame of reference.
This is due to the functional dependence of Lyapunov exponents on the invariants of the right Cauchy-Green strain tensor, which satisfy the principle of material frame-independence~\citep{truesdell2004non}.
In general, the objectivity in terms of Euclidean measures is ensured for an observer transformation from ($\pmb{\chi},\tau$) to ($\pmb{\chi}^\ast,\tau^\ast$) as
\begin{equation}
\pmb{\chi}^\ast=\pmb{Q}\pmb{\chi}+\pmb{c}, \hspace{5mm} \hfill \tau^\ast=\tau+\mathsf{b},
\end{equation}
where $\mathsf{b}$ is an arbitrary constant, $\pmb{c}$ is a time-dependent vector and $\pmb{Q}$ is a time-dependent proper orthogonal tensor.
The scalar, vector, and tensor fields are objective if, respectively,
\begin{equation}
\beta^\ast=\beta, \hspace{5mm} \pmb{b}^\ast=\pmb{Q}\pmb{b}, \hspace{5mm}\text{and}\hspace{5mm} \pmb{B}^\ast=\pmb{Q}\pmb{B}\pmb{Q}^T.
\end{equation}
Let us now consider the orthonormal basis $\pmb{\Phi}=\{\pmb{\Phi}_l\}_{l \in \{ 1,...,m \}}$, and a second orthonormal basis $\pmb{\Phi}^\ast=\{\pmb{Q}\pmb{\Phi}_l\}_{l \in \{1,...,m \}}$.
For a frame-independent vector $\pmb{b}$ in the basis $\pmb{\Phi}$, an equivalent $\pmb{b}^\ast$ in the basis $\pmb{\Phi}^\ast$ is \[ \pmb{b}^*_l=\pmb{b}^{\ast T}\pmb{Q}\pmb{\Phi}_l=\pmb{b}^T\pmb{Q}^T\pmb{Q}\pmb{\Phi}_l=\pmb{b}^T\pmb{\Phi}_l=\pmb{b}_l,\] \textit{i.e.}, the components of $\pmb{b}^\ast$ in basis $\pmb{\Phi}^*$ and $\pmb{b}$ in basis $\pmb{\Phi}$ are identical.
Similarly, an objective tensor $\pmb{B}$ in the basis $\pmb{\Phi}$ can be expressed as $\pmb{B}_{kl}=\pmb{\Phi}_k^T\pmb{B}\pmb{\Phi}_l$, while the components of a second tensor $\pmb{B}^\ast$ in the basis $\pmb{\Phi}^\ast$ are \[ \pmb{B}_{kl}^\ast = (\pmb{Q}\pmb{\Phi}_k)^T\pmb{B}^\ast(\pmb{Q}\pmb{\Phi}_l)= \pmb{\Phi}_k^T\pmb{Q}^T\pmb{Q}\pmb{B}\pmb{Q}^T\pmb{Q}\pmb{\Phi}_l = \pmb{\Phi}_k^T\pmb{B}\pmb{\Phi}_l=\pmb{B}_{kl},\]
leading to the exact same tensor.
Thus, the objectivity of the flow fields, including scalar, vector and tensor fields, is preserved under LMA, ensuring the principle of material frame-independence.
}

The FTLE and Lagrangian DMD relate through the well documented connections between the POD and DMD in the literature~\citep{schmid2009dynamic,schmid2010dynamic}.
As noted before, the POD optimally extracts the most energetic coherent flow structures, whereas the DMD focuses on the coherent structures with unique frequency and growth/decay rate.
In the LMA ansatz, the dominant LPOD modes are the coherent flow structures that comprise maximum stretching of the flow fields, while the LDMD modes are the coherent flow structures that evolve at unique frequencies.
The relation between the FTLE and Lagrangian POD/DMD modes is illustrated in Sec.~\ref{sec:gyre} by considering the simple mathematical model comprised of the double gyre pattern.

\section{Numerical methods and case studies} \label{sec:numerical}
Two canonical configurations, namely: lid-driven cavity and flow past a cylinder are considered for the application of the Lagrangian modal analysis.
For both configurations, direct numerical simulations 
are performed in two-dimensional (2D) compressible but shock-free settings.
The governing flow equations and simulation setups are presented in the following subsections.

\subsection{Flow governing equations and numerical methods} \label{sec:NSE}

The flow fields are governed by the full compressible Navier-Stokes equations, which are solved in non-dimensional form using curvilinear coordinates:
\begin{equation}
\frac{\partial}{\partial\tau}\left(\frac{\pmb{S}}{J}\right) + \frac{\partial \pmb{F}}{\partial\xi_1} + \frac{\partial \pmb{G}}{\partial\xi_2} + \frac{\partial \pmb{H}}{\partial\xi_3} = \frac{1}{Re}\left[ \frac{\partial \hat{\pmb{F}}}{\partial\xi_1} + \frac{\partial \hat{\pmb{G}}}{\partial\xi_2} + \frac{\partial \hat{\pmb{H}}}{\partial\xi_3} \right]
\end{equation}
where $\pmb{S}=[\rho, \rho \pmb{u}, \rho E ]^T$ is the conserved solution vector.
The flow variables are non-dimensionalized by their reference ($\infty$) values, except for pressure, which is normalized by using the reference density and reference velocity.
$\pmb{u}$ is the velocity vector, while $\rho$ and $E$ are the density and internal energy, respectively.
The non-dimensionalized flow variables are defined by:
\begin{equation}\label{eq:non-dim}
\rho = \frac{\rho^\ast}{\rho_\infty^\ast}, \pmb{u} = \frac{\pmb{u}^\ast}{u_\infty^\ast}, p = \frac{p^\ast}{\rho^\ast_\infty u^{\ast 2}_\infty}, T=\frac{T^\ast}{T^\ast_\infty}, \pmb{x}=\frac{\pmb{x}^\ast}{L_{ref}^\ast}, \text{ and } t=\frac{t^\ast u^\ast_\infty}{L_{ref}^\ast},
\end{equation}
where $L_{ref}^\ast$ is a dimensional reference length and the asterisk denotes a dimensional quantity.
The non-dimensional Reynolds and Mach numbers are then,
\begin{equation}\label{eq:Rey_Mac}
Re \equiv \frac{\rho^\ast_\infty u^\ast_\infty L_{ref}^\ast}{\mu_\infty^\ast}\text{ and } M_\infty \equiv \frac{u^\ast_\infty}{\sqrt{\gamma p^\ast_\infty/\rho^\ast_\infty}}.
\end{equation}
The Jacobian, denoted by $J$, of the Cartesian to curvilinear coordinate transformation  ($\pmb{x},t$)$\rightarrow$($\pmb{\xi},\tau$) is given by $J={\partial (\pmb{\xi},\tau)}/{\partial (\pmb{x},t)}$.
The inviscid and viscous fluxes, for instance $\pmb{F}$ and $\hat{\pmb{F}}$ respectively, are given as:
\begin{equation}
\pmb{F} = \frac{1}{J}\begin{bmatrix}
\rho U_1\\[.75ex]
\rho u_1U_1+\frac{\partial \xi_1}{\partial x_1}p\\[.75ex]
\rho u_2U_1+\frac{\partial \xi_1}{\partial x_2}p\\[.75ex]
\rho u_3U_1+\frac{\partial \xi_1}{\partial x_3}p\\[.75ex]
(\rho E+p)U_1-\frac{\partial \xi_1}{\partial t}p
\end{bmatrix},
\hat{\pmb{F}} = \frac{1}{J}\begin{bmatrix}
0\\[.75ex]
\frac{\partial \xi_1}{\partial x_i}\sigma_{1i}\\[.75ex]
\frac{\partial \xi_1}{\partial x_i}\sigma_{2i}\\[.75ex]
\frac{\partial \xi_1}{\partial x_i}\sigma_{3i}\\[.75ex]
\frac{\partial \xi_1}{\partial x_i}\left(u^j\sigma_{ij}-\Theta_i\right) \\
\end{bmatrix}, 
\end{equation}
where $i$ and $j$ are summation indices.
$U_i$ is the contravariant velocity component, which is expressed by using a summation index $j$ as,
\begin{equation}
U_i = \frac{\partial \xi_i}{\partial t}+\frac{\partial \xi_i}{\partial x_j}u_j.
\end{equation}
The internal energy is given by
\begin{equation}
E = \frac{T}{\gamma (\gamma - 1) M^2_\infty} +\frac{1}{2}\| \pmb{u} \|^2,
\end{equation}
where $T$, $\gamma$ and $M_\infty$ are the temperature, the ratio of specific heats and the reference Mach number, respectively.
The fluid is assumed to be a perfect gas, with pressure $p=\rho T/\gamma M^2_\infty$.
The ratio of specific heats for air, $\gamma$ is assumed to be $1.4$.
The components of the stress tensor and the heat flux vector are given by, respectively,
\begin{equation}
\sigma_{ij}= \mu \left( \frac{\partial \xi_k}{\partial x_j} \frac{\partial u_i}{\partial \xi_k} + \frac{\partial \xi_k}{\partial x_i}\frac{\partial u_j}{\partial \xi_k} - \frac{2}{3}\frac{\partial \xi_l}{\partial x_k}\frac{\partial u_k}{\partial \xi_l} \delta_{ij} \right)
\end{equation}
and
\begin{equation}
\Theta_i=- \frac{\mu/Pr}{(\gamma-1)M_\infty^2} \frac{\partial \xi_j}{\partial x_i} \frac{\partial T}{\partial \xi_j}.
\end{equation}
The Prandtl number is set to $Pr=0.72$.
$\mu$ denotes the dynamic viscosity of the fluid, while the bulk viscosity is $-2\mu/3$, assuming the Stokes' hypothesis.
The fluid viscosity change due to the temperature is modeled using Sutherland's law, given as
\[ \mu=T^{3/2}\left(\frac{1+C_1}{T+C_1}\right),\]
where $C_1=0.37$ is the non-dimensionalized Sutherland's constant.

The second-order implicit time marching scheme of \cite{beam1978implicit} is adopted, with two Newton-like subiterations to reduce factorization and explicit boundary condition application errors.
Further details on the time scheme are provided in \cite{visbal2004numerical}.
The spatial derivatives are discretized using a $6^{th}$ order compact finite difference scheme with the central difference, ensuring no dissipation error.
An $8^{th}$ order implicit low-pass Pade-type filtering, with $\alpha_f=0.4$, is used to provide dissipation at high spatial wavenumbers.
Detailed validation studies may be found in \cite{visbal1999high}, \cite{gaitonde2000pade} and \cite{visbal2002use}.

\subsection{Two-dimensional lid-driven cavity} \label{sec:LDC}
The first test case considers a compressible two-dimensional lid-driven cavity flow.
The flow inside a lid-driven cavity exhibits relatively complex vortex dynamics with increasing Reynolds number, including the onset of Hopf bifurcation, making it one of the classical configurations for flow stability and transition~\citep{ghia1982high,shen1991hopf,ramanan1994linear}.
Although three-dimensionality and end-wall effects are significant for the flow physics~\citep{koseff1983three,sheu2002flow,albensoeder2005accurate,lopez2017transition},
high-fidelity two-dimensional numerical simulations continue to be canonical benchmarks~\citep{bruneau20062d}, in situations such as the present.
For generality, the effects of compressibility are retained by considering a Mach number of $M_\infty=0.5$, where the flow stability and dynamics of the lid-driven cavity have been discussed by~\cite{bergamo2015compressible,ohmichi2017compressibility,ranjan2020robust}.
Simulations were performed at Reynolds numbers based on cavity size $L$ ranging from $Re_L=5{,}000$ to $Re_L=15{,}000$ at intervals of $2{,}000$. 
The flow remains steady until $Re_L=9{,}000$ but becomes unsteady at $Re_L=11{,}000$, consistent with the critical Reynolds number $Re_c=10{,}500$ at this Mach number \cite{ohmichi2017compressibility}.
Details on the geometry and grid convergence are provided in Appendix~\ref{sec:grid_ldc}.
For concreteness, the Lagrangian modal analysis is discussed for $Re_L=7{,}000$ (steady) and $Re_L=15{,}000$ (unsteady).

\begin{figure}
\begin{minipage}{0.5\textwidth}
\centering {\includegraphics[scale=0.32]{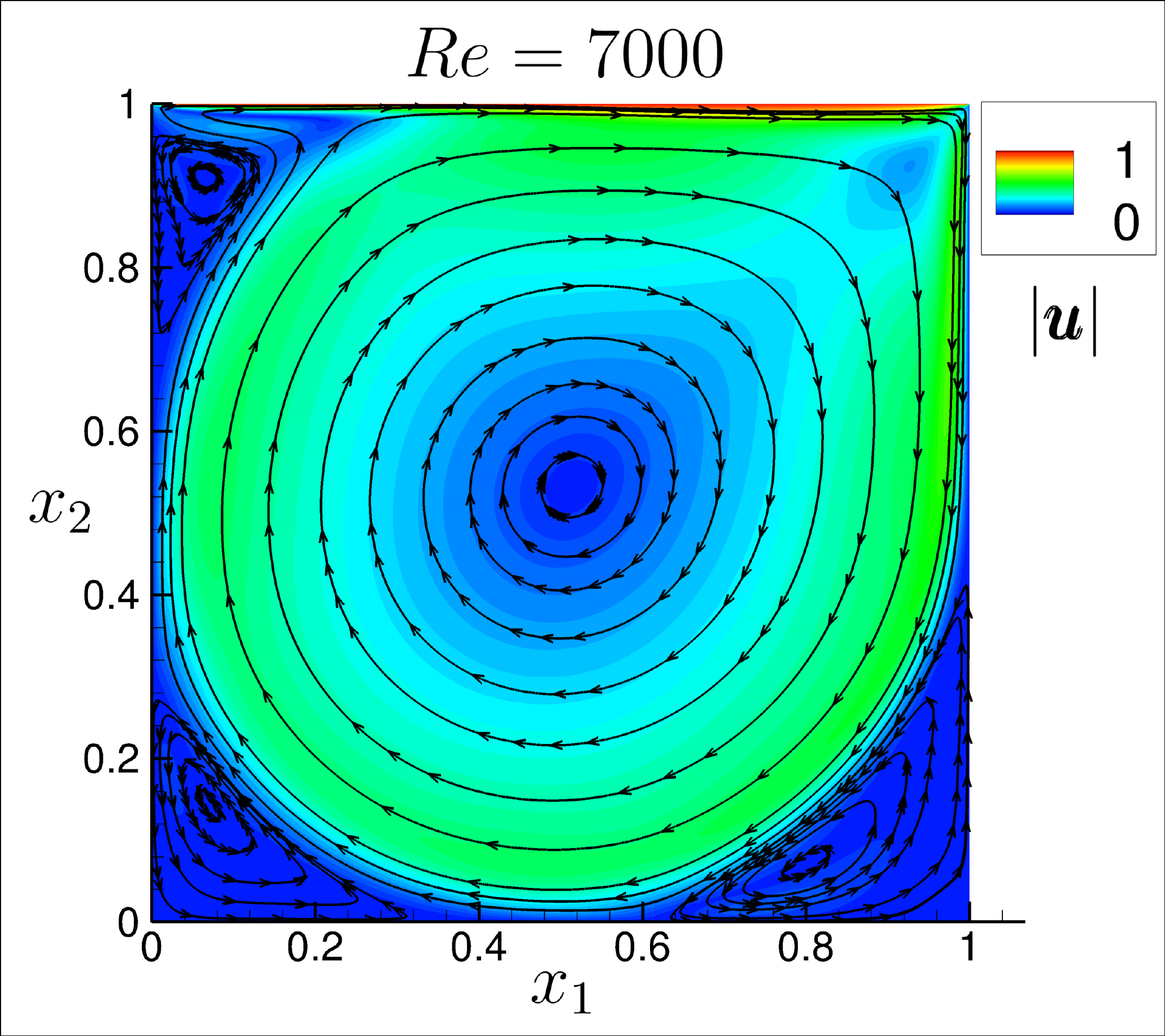}}\\(a)
\end{minipage}
\begin{minipage}{0.5\textwidth}
\centering {\includegraphics[scale=0.32]{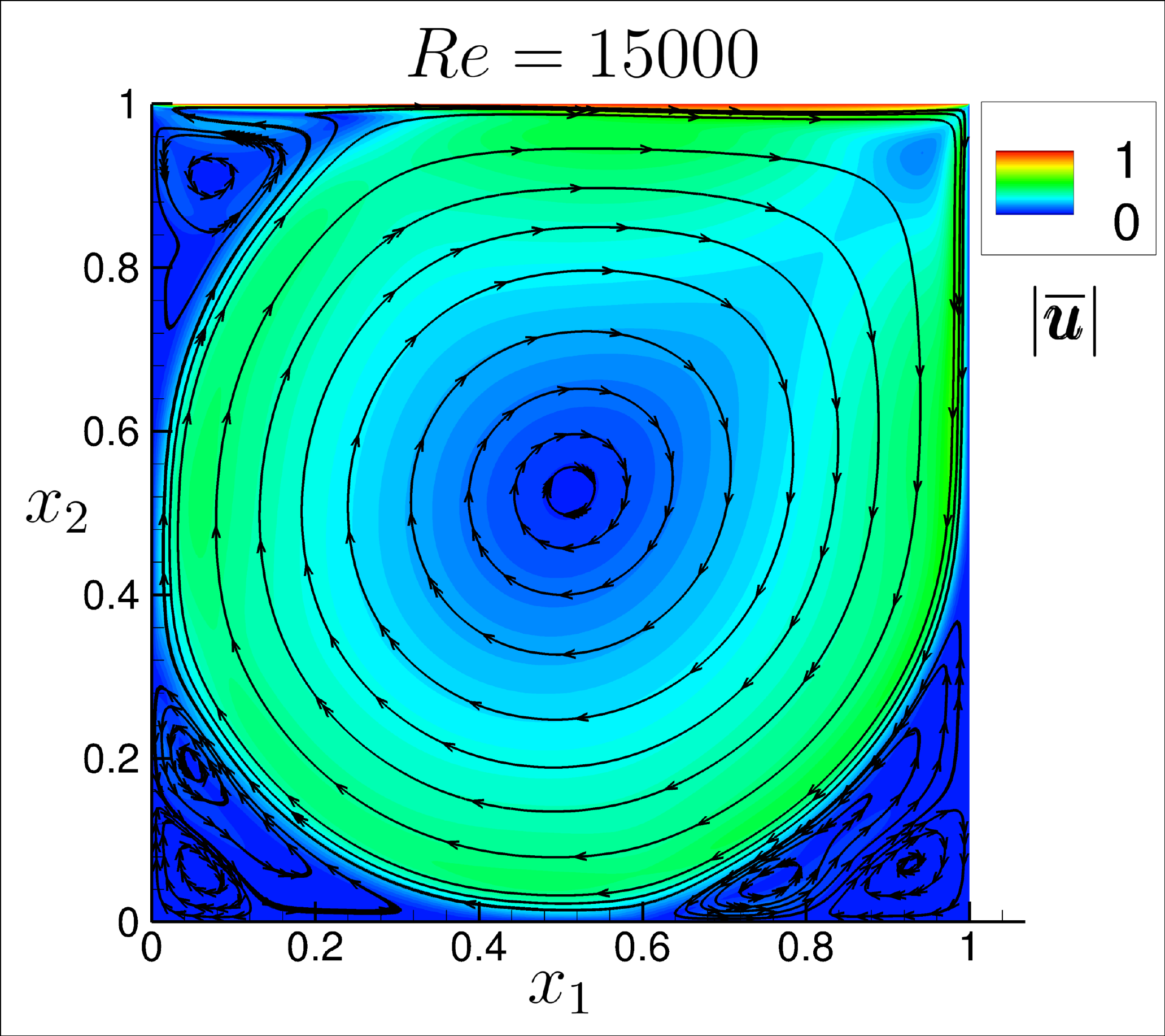}}\\(b)
\end{minipage}
\caption{Flow recirculation patterns inside the lid-driven cavity at Mach number $M_\infty=0.5$ and pre- and post-critical Reynolds numbers. (a) Steady flow velocity $|\pmb{u}|$ at $Re_L=7{,}000$ and (b) time-averaged flow velocity $|\overline{\pmb{u}}|$ at $Re_L=15{,}000$.  Streamlines display the flow recirculation patterns.}
\label{fig:ldc_Uabs}
\end{figure}
Figure~\ref{fig:ldc_Uabs}(a) displays the steady pre-critical Reynolds number flow at $Re_L=7{,}000$ in terms of the absolute flow velocity $\pmb{u}$ and select flow streamlines.
Several recirculation regions are apparent: in addition to the large central feature, three smaller regions are evident near the top-left, bottom-left and bottom-right corners of the cavity.
On the other hand, at the post-critical Reynolds number of $Re_L=15{,}000$, the flow is unsteady. 
The pattern of Fig.~\ref{fig:ldc_Uabs}(b) shows the time-averaged absolute flow velocity $\overline{\pmb{u}}$ at $Re_L=15{,}000$; the mean streamlines in this case indicate additional flow re-circulation patterns near the lower corners.
Furthermore, the skin-friction coefficient (Eq.~\ref{eq:skin-friction}) estimated along the bottom wall (Fig.~\ref{fig:ldc_uvmesh} b) also indicates the regions of re-circulation, corresponding to the flow pattern of Fig.~\ref{fig:ldc_Uabs}(b).

\subsection{Two-dimensional lid-driven cavity with mesh deformation} \label{sec:mesh_def_2D-LDC}



\begin{figure}
\begin{minipage}{1.0\textwidth}
\centering{\includegraphics[scale=0.184]{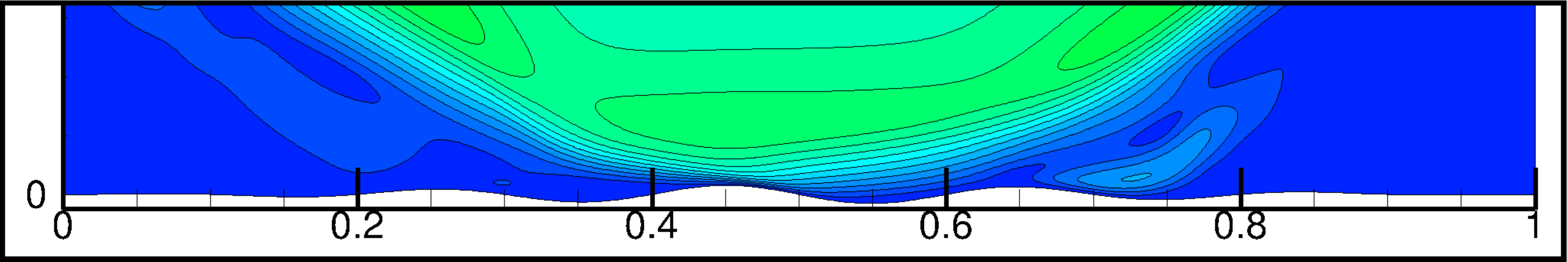}}
\end{minipage}
\begin{minipage}{0.5\textwidth}
\centering {\includegraphics[scale=0.09]{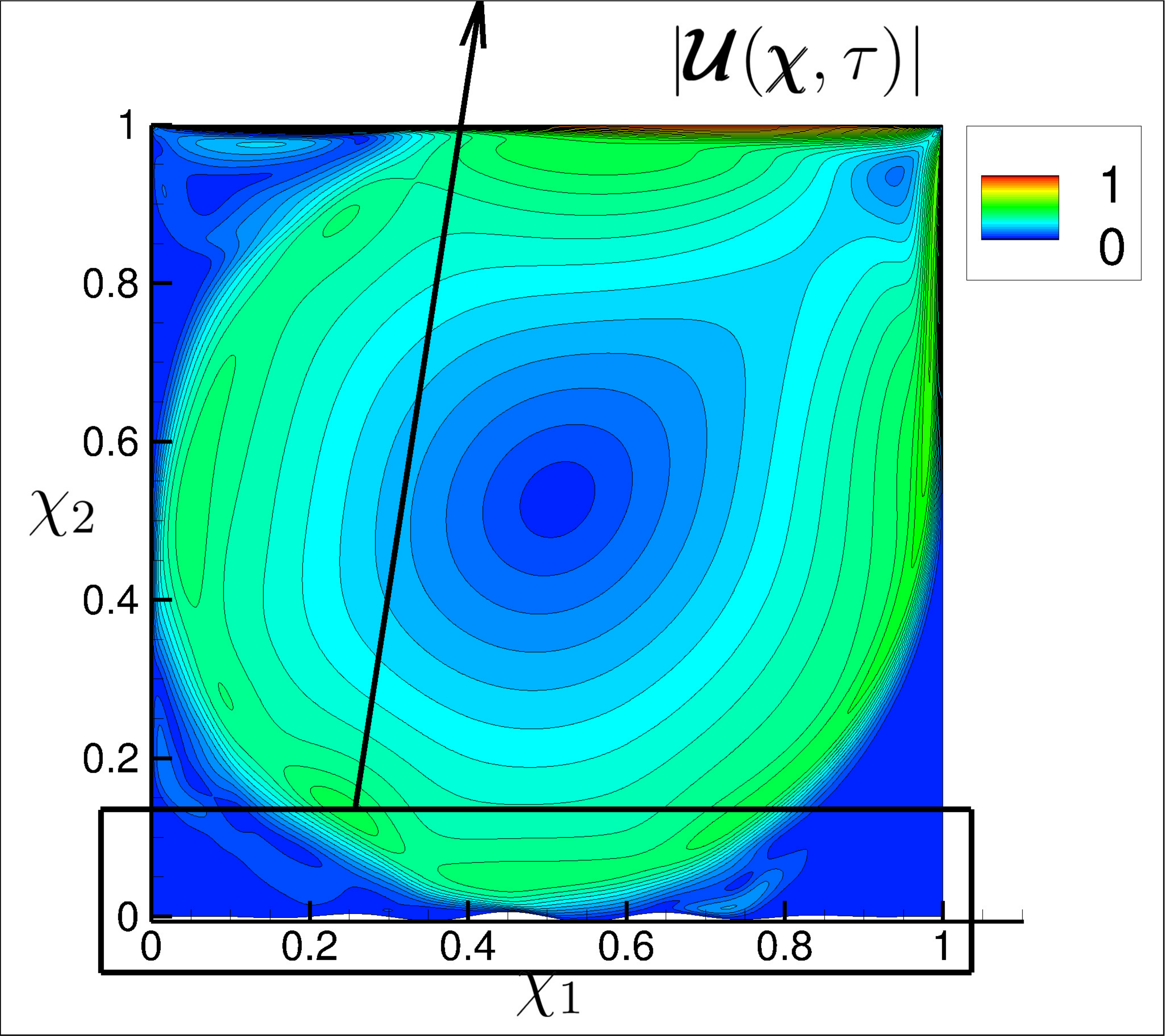}}\\(a)
\end{minipage}
\begin{minipage}{0.5\textwidth}
\centering {\includegraphics[scale=0.8]{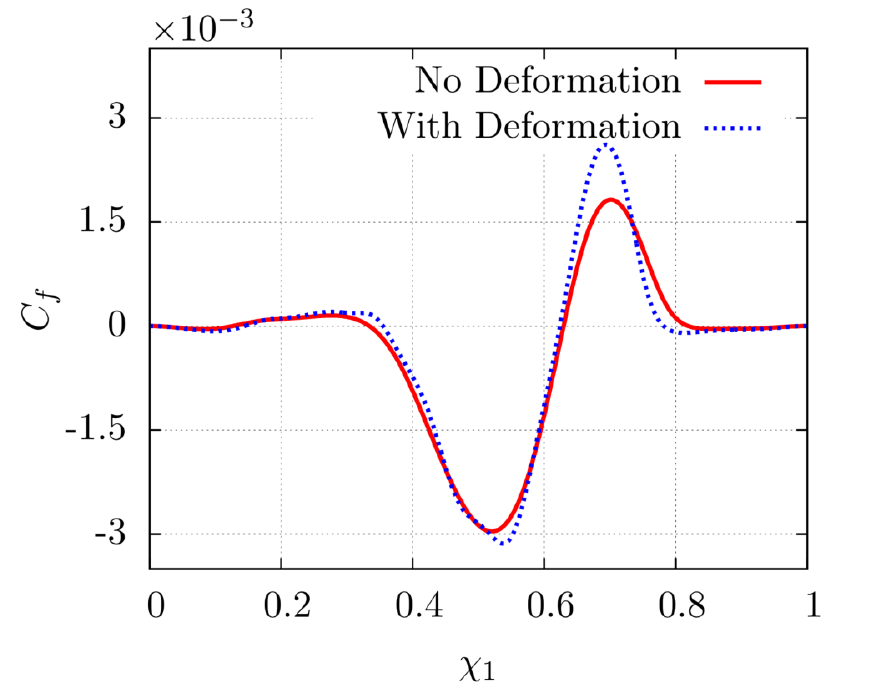}}\\(b)
\end{minipage}
\caption{Lid-driven cavity flow at $M_\infty=0.5$ and $Re_L=15{,}000$ with forced bottom surface/mesh deformation. (a) Instantaneous flow velocity magnitude with contours on a deformed domain. The inset shows a closer view of the deformed domain ($\pmb{\chi}$). (b) Skin-friction coefficient on the bottom surface with/without deformation.}
\label{fig:ldc_def}
\end{figure}

To construct a prototypical problem representing a flow with a deforming mesh,
the bottom surface of the cavity is subjected to forced deformation, keeping the other flow conditions and simulation parameters the same.
The deformation is governed by an analytical function expressed as,
\begin{eqnarray}\label{eq:mesh_def1}
\chi_1 &=& x_1 \nonumber \\
\chi_2(\tau)&=&a(1-x_2)x_1^2 (1-x_1)^2 \sin(\pi nx_1) \sin(2\pi St_f t),
\end{eqnarray}
where $a$ and $n$ are the deformation amplitude and mode number respectively. 
The non-dimensional frequency of mesh deformation is $St_f$.
The velocity of the mesh deformation, $\pmb{\mathcal{U}}^\mathtt{G}(\pmb{\mathcal{\chi}},\tau)$, is then given as,
\begin{eqnarray}\label{eq:mesh_def2}	
\frac{\partial {\chi}_1}{\partial \tau} &=& 0 \nonumber \\
\frac{\partial {\chi}_2}{\partial \tau} &=& a(1-x_2)x_1^2(1-x_1)^2\sin(\pi nx_1)\cos(2\pi St_f t) 2\pi St_f.
\end{eqnarray}
The choice of parameters, $a=0.1$, $n=10$ and $St_f=1$, is based on obtaining a case that adequately tests the LMA development.
The flow velocity in the Lagrangian (moving mesh) frame of reference can be given by,
\begin{equation}\label{eq:ul-ue-mesh}
\pmb{\mathcal{U}}(\pmb{\chi},\tau)=\pmb{u}(\mathcal{M}(\pmb{\chi},\tau))-\pmb{\mathcal{U}}^\mathtt{G}(\pmb{\chi},\tau),
\end{equation}
where $\mathcal{M}$ is the mapping of form Eq.~\ref{eq:mapping}.
{The flow solver accounts for the Eulerian-Lagrangian effect by enforcing the geometric conservation law~\citep{gordnier2002development,thomas1979geometric}, that governs spatial volume element under arbitrary mapping.}
The deformed computational domain ($\mathtt{G}_3$) at an arbitrary instant is shown in the inset of Fig.~\ref{fig:ldc_def}(a).
In addition, the figure displays the absolute velocity magnitude, $|\pmb{\mathcal{U}}(\pmb{\chi},\tau)|$, on the moving mesh at that instant.

The lid-driven cavity with bottom surface deflection exhibits many of the main flow features of the baseline (no boundary motion) flow including the large central region and smaller recirculation regions near the no-slip walls.
The domain deformation affects the motions of these flow features of course, particularly the smaller recirculation regions near the bottom wall.
In addition to the near-wall undulations on the velocity contours in Fig.~\ref{fig:ldc_def}(a), the entire flow is modified to some degree due to the surface deformation.
{Figure~\ref{fig:ldc_def}(b) displays the skin-friction coefficient (Eq.~\ref{eq:skin-friction}) on the deforming bottom wall, indicating a discernible increase at $\chi_1\approx 0.7$.}
The Lagrangian averaged flow field, estimated by accounting for the moving mesh, is largely similar to the Eulerian time-averaged flow field of Fig.~\ref{fig:ldc_Uabs}(b), where the small differences can be attributed to the mild domain deformation.

\subsection{Two-dimensional flow past a cylinder} \label{sec:CYL}


The second flow considered is that past a circular cylinder, which is also a classical problem of engineering significance.
The configuration highlights the fluid dynamics around bluff bodies, and encompasses many fundamental phenomena, including steady or unsteady separation, transition and wake vortex shedding~\citep{williamson1996vortex}, for all of which, a large body of experimental and numerical data are available for validation.
The problem is also a popular test-bed for studies on flow stability,  control, fluid-structure interaction, reduced-order modeling~\citep{shinde2016galerkin,shinde2019galerkin} and compressibility effects~\citep{canuto2015two}.

In this configuration, the flow transitions from steady state to  unsteady vortex shedding in distinct stages.
As the Reynolds number based on cylinder diameter $D$ is increased,
the initial unsteadiness is manifested for incompressible flow in $47 \lessapprox Re_D \lessapprox 178$ as periodic vortex shedding following a supercritical Hopf bifurcation at the critical Reynolds number~\citep{sreenivasan1987hopf,noack1994global}.
The initial two-dimensionality of the flow provides a suitable environment on which to demonstrate LMA.
The onset of three-dimensionality at  $Re_D\approx 178$, appears in the form of spanwise undulations~\citep{behara2010wake}.

\begin{figure}
\begin{minipage}{0.5\textwidth}
\centering  {\includegraphics[scale=0.32]{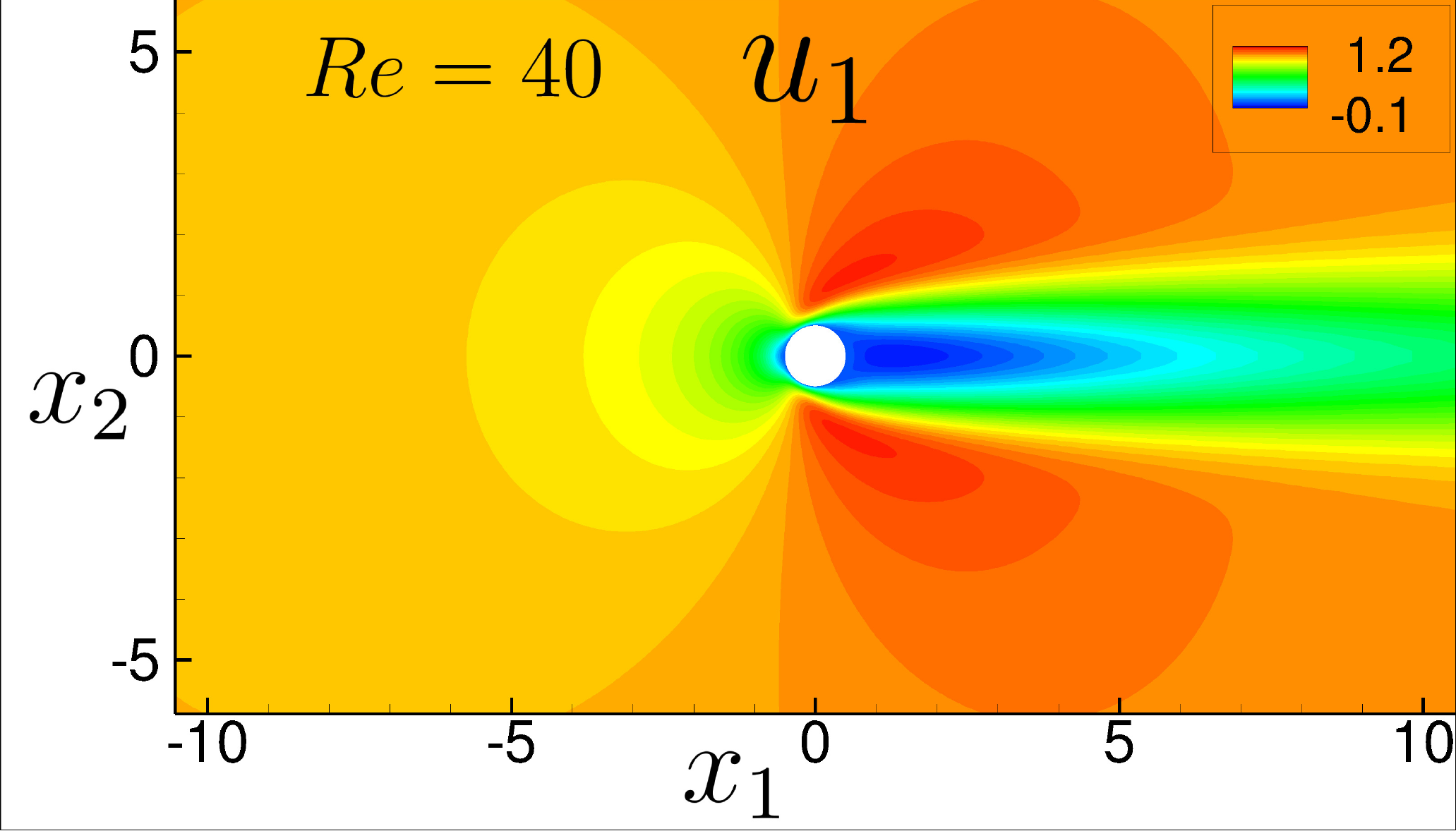}}\\(a)
\end{minipage}
\begin{minipage}{0.5\textwidth}
\centering  {\includegraphics[scale=0.8]{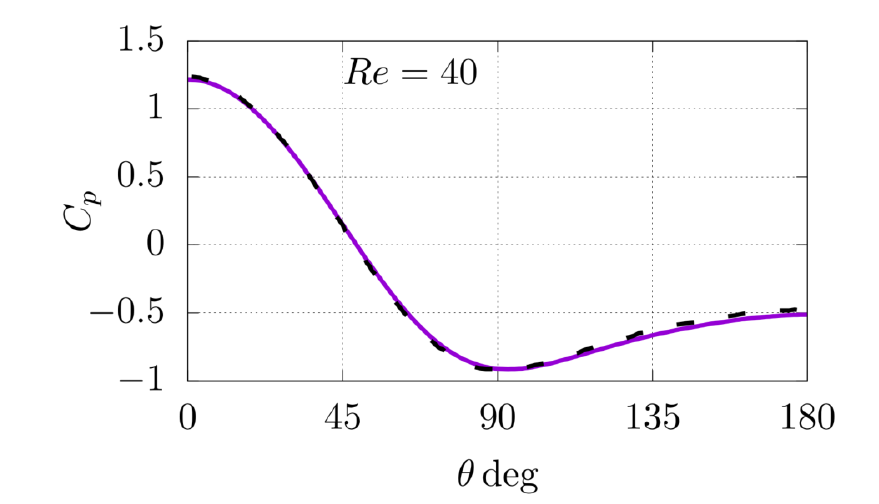}}\\(b)
\end{minipage}
\begin{minipage}{0.5\textwidth}
\centering  {\includegraphics[scale=0.32]{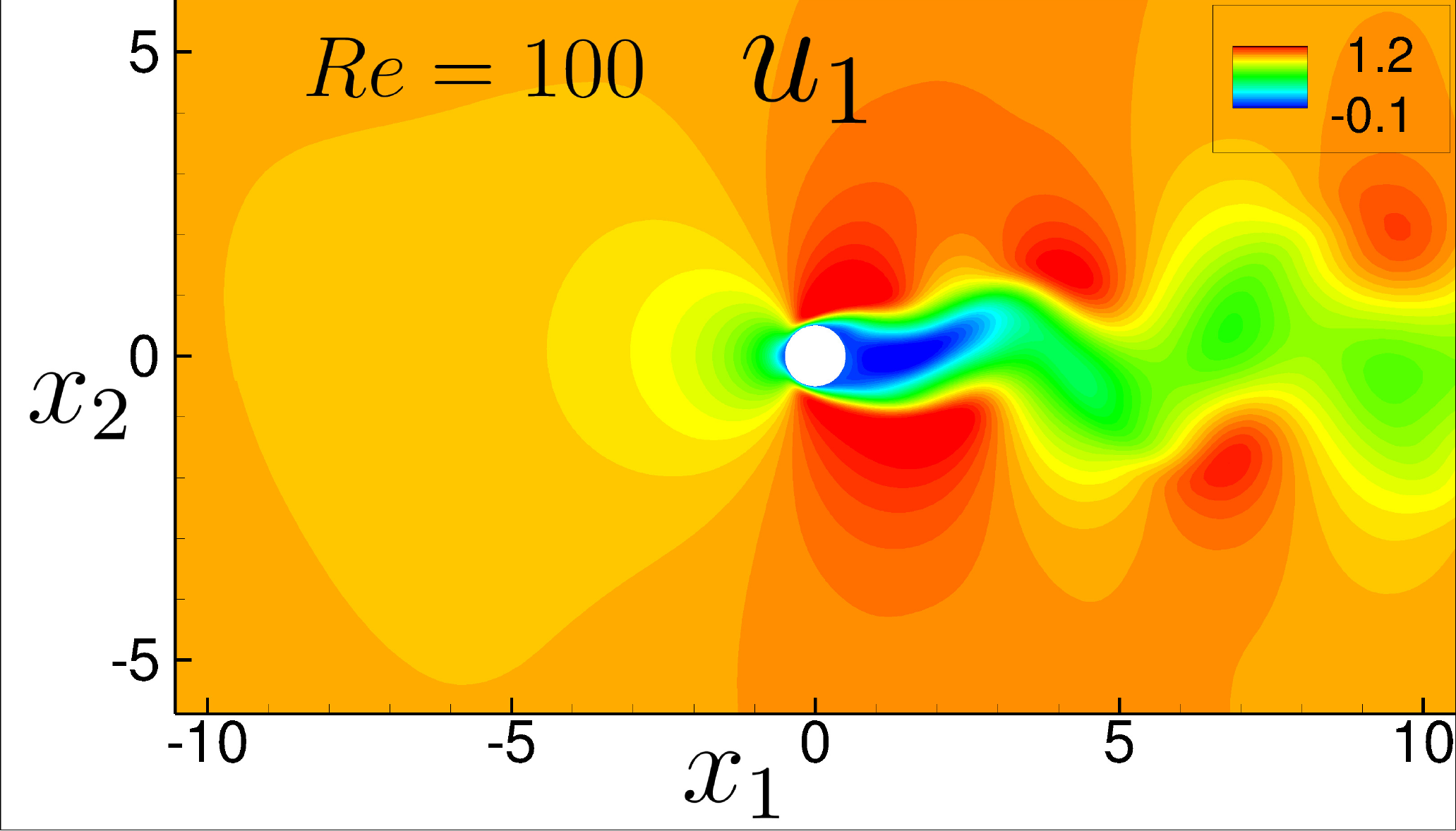}}\\(c)
\end{minipage}
\begin{minipage}{0.5\textwidth}
\centering  {\includegraphics[scale=0.32]{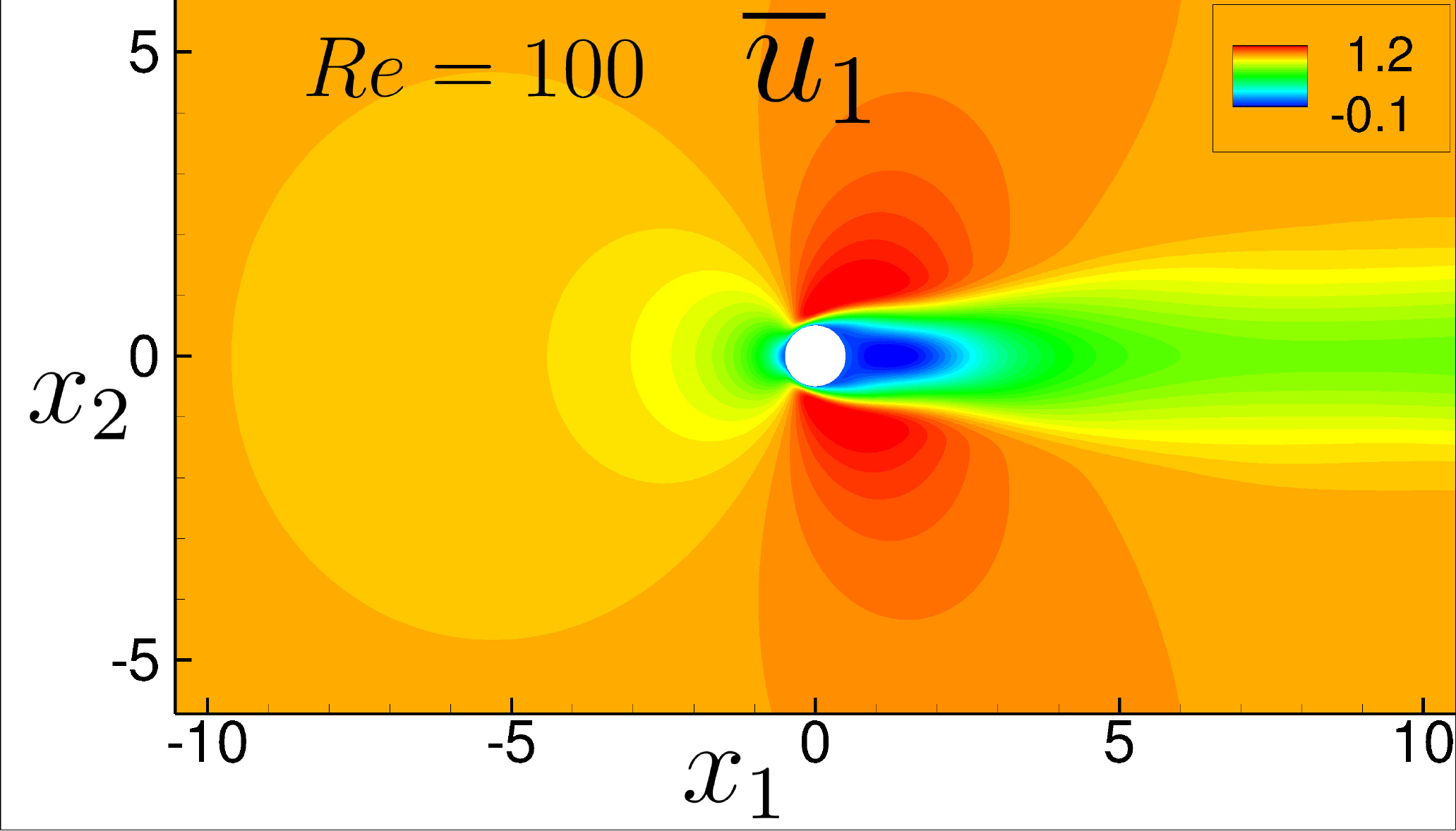}}\\(d)
\end{minipage}
\caption{Compressible flow past a cylinder, in terms of the streamwise velocity $u_1$, at Mach number $M_\infty=0.5$ and Reynolds numbers $Re_D=40$ and $Re_D=100$. (a) Steady flow at $Re_D=40$. (b) The pressure coefficient $C_p$ is compared with the DNS profile of~\cite{canuto2015two} (dashed line.) (c) Unsteady instantaneous flow at $Re_D=100$. (d) Time-mean flow at $Re_D=100$.}
\label{fig:cyl_re40_re100}
\end{figure}
Simulations are performed for a range of Reynolds number $20 \leq Re_D \leq 100$.
Appendix~\ref{sec:grid_cyl} provides details on the geometry and grid convergence study.
The flow fields are shown in Fig.~\ref{fig:cyl_re40_re100} using the normalized streamwise velocity $u_1$ for $Re_D=40$ and $Re_D=100$.
As noted earlier, the flow is steady at $Re_D=40$ (Fig.~\ref{fig:cyl_re40_re100}a).
The corresponding pressure coefficient, defined as \[C_p\equiv \frac{p-p_\infty}{\frac{1}{2}\rho_\infty u_\infty^2},\] is shown in Fig.~\ref{fig:cyl_re40_re100}(b) together with a favorable comparison with the DNS result of~\cite{canuto2015two}.
For $Re_D=100$ on the other hand, a periodic vortex shedding is observed in the wake region of the cylinder (Fig.~\ref{fig:cyl_re40_re100}c).
The time-averaged flow field for $Re_D=100$ is displayed in Fig.~\ref{fig:cyl_re40_re100}(d).
The effect of increased Reynolds number is evident when compared to the $Re_D=40$ flow field of Fig.~\ref{fig:cyl_re40_re100}(a), particularly, in the wake region, which becomes more compact for $Re_D=100$.

\section{Results and discussion} \label{sec:res_disc}
The LMA (Sec.~\ref{sec:theory}) is now applied to unsteady flow, without and with mesh deformation, as well as steady flow by considering the above lid-driven cavity and cylinder flows.
Although the focus is on LPOD and LDMD, the general procedure is applicable to all variants of these and similar decomposition approaches.
The development is performed in several steps.
The unsteady lid-driven cavity flow at $Re_L=15{,}000$ is examined first, followed by an illustration on dynamic meshes with the deforming bottom surface case.
Next, the techniques are applied to steady flow by considering both lid-driven cavity and cylinder cases at pre-critical Reynolds numbers of $Re_L=7{,}000$ and $Re_D=40$, respectively.
For the latter, the backward LMA, where the traces evolve in reversed time, in Lagrangian sense, is also examined.
Lastly, the LMA is used for the double-gyre flow pattern to illustrate the relation with FTLE.
In all cases, the modal decompositions are performed directly on the flow fields, without subtracting the averaged flow field.

\subsection{LMA on unsteady flow} \label{sec:unsteady_flows}
As noted in Sec.~\ref{sec:LDC}, the two-dimensional lid-driven cavity flow undergoes the first Hopf-bifurcation at a critical Reynolds number in $9{,}000 \lessapprox Re_L \lessapprox 11{,}000$, leading to an unsteady flow.
At $Re_L=15{,}000$, the flow undergoes successive Hopf-bifurcations, exhibiting multiple frequency peaks in the power spectral density (PSD) of the flow variables.
In particular, the PSD of the integrated streamwise force ($F_{x_1}$) on the cavity exhibits three prominent frequency peaks at Strouhal numbers $St_L=0.13$, $0.24$, and $0.37$.
The analysis is performed using $2{,}500$ solution snapshots, collected at time intervals of $0.01$, after the initial flow transients have disappeared from the simulation.
The Eulerian results from the DNS are cast in a Lagrangian frame of reference, ensuring that the time resolution is sufficient to capture phenomena with Strouhal number between $0.04\leq St_L \leq 25$.

\begin{figure}
\begin{minipage}{0.5\textwidth}
\centering \includegraphics[scale=0.09]{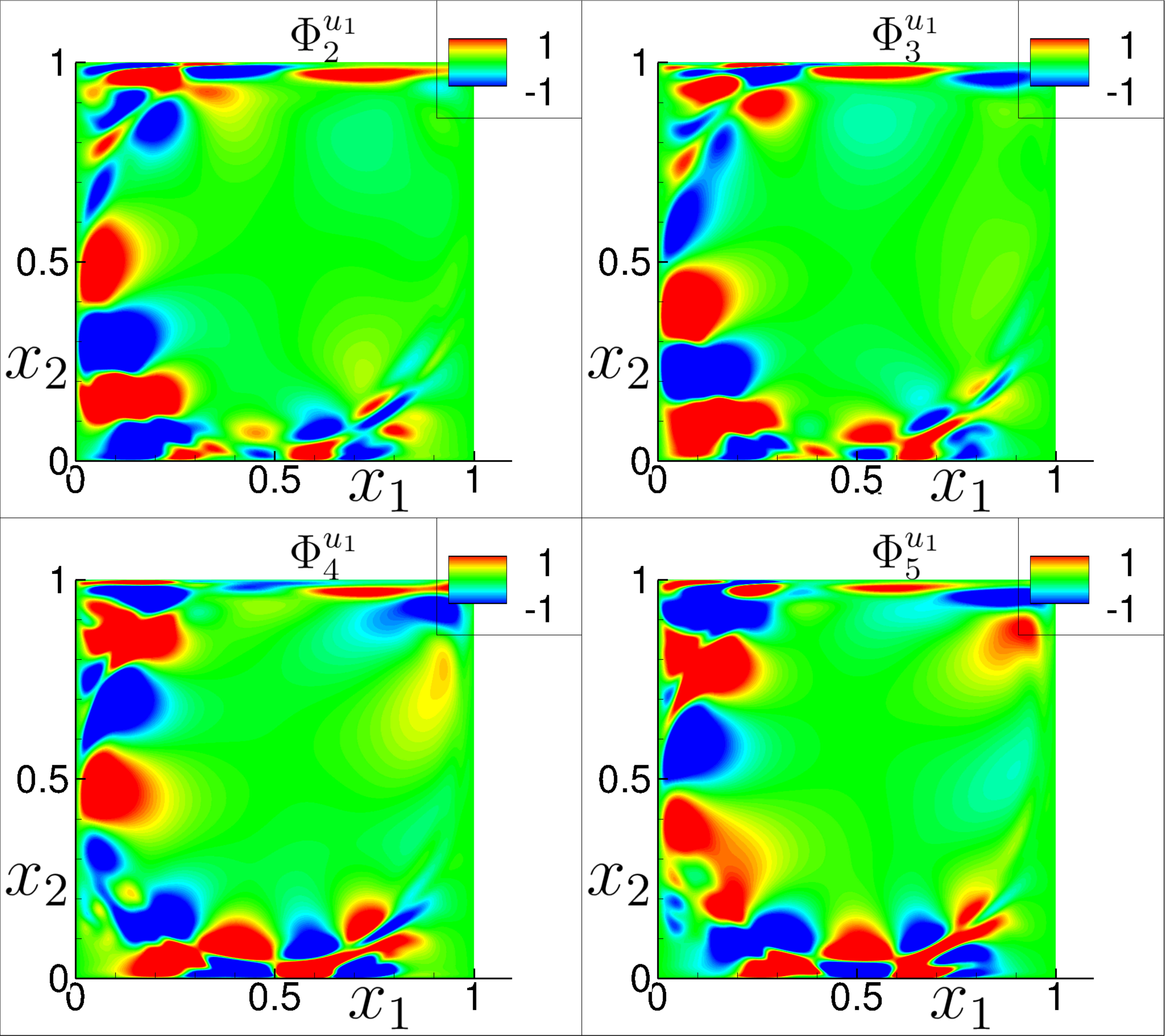}\\(a) POD modes
\end{minipage}
\begin{minipage}{0.5\textwidth}
\centering \includegraphics[scale=0.09]{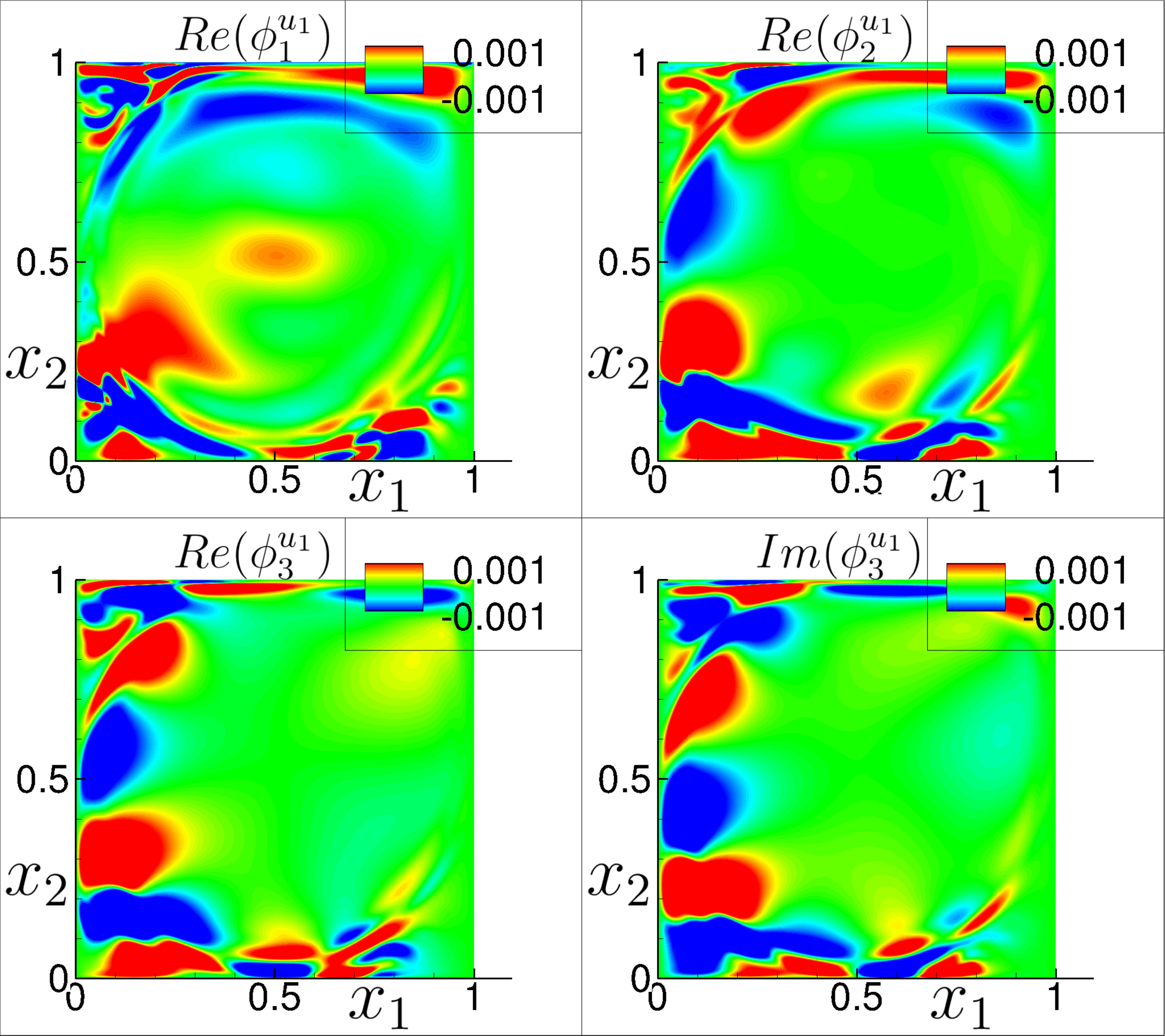}\\(b) DMD modes
\end{minipage}
\begin{minipage}{0.5\textwidth}
\centering \includegraphics[scale=0.8]{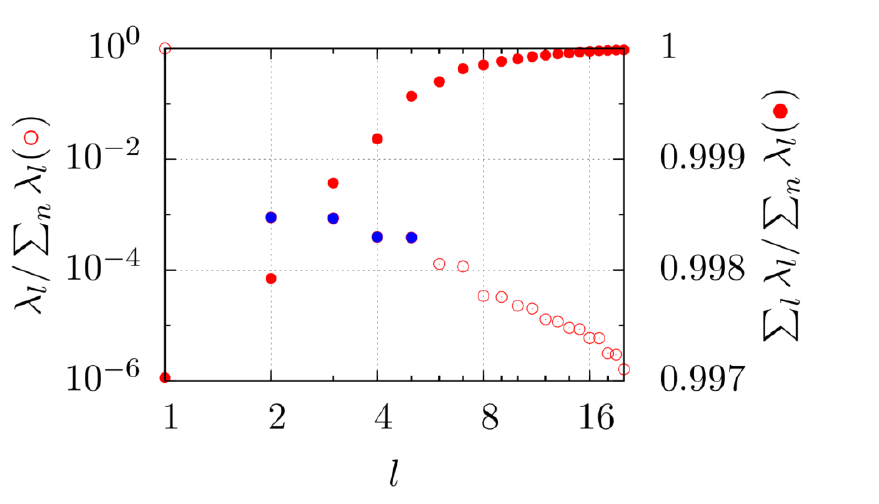}\\(c) POD eigenvalues
\end{minipage}
\begin{minipage}{0.5\textwidth}
\centering \includegraphics[scale=0.7]{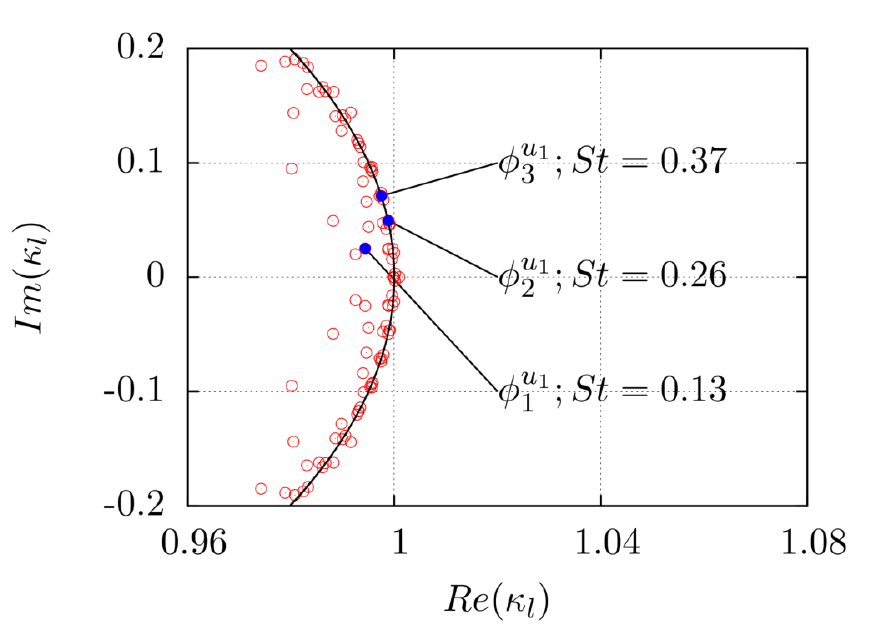}\\(d) DMD eigenvalues
\end{minipage}
\caption{Leading POD and DMD modes of the streamwise velocity for the lid-driven cavity at $Re_L=15000$ in Eulerian frame of reference. (a) POD modes $\Phi_2^{u_1}$, $\Phi_3^{u_1}$, $\Phi_4^{u_1}$, and $\Phi_5^{u_1}$. (b) DMD modes $\phi_1^{u_1}$, $\phi_2^{u_1}$, and $\phi_3^{u_1}$. (c) POD modal energies and (d) DMD eigenvalues corresponding to the spatial modes.}
\label{fig:LDC_Re15k_modes}
\end{figure}

The modal decompositions provide insights into flow organization inside the cavity, which includes a primary vortex, shear regions along cavity walls and a Couette flow region near the cavity center.
These features are evident in Figure~\ref{fig:LDC_Re15k_modes}, which  displays traditional (Eulerian) POD and DMD modes using the streamwise velocity field components.
The POD modes (Fig.~\ref{fig:LDC_Re15k_modes} a) are the energy dominant flow features;  the leading few modes, when ordered by energy content, comprise most of the flow energy, as shown in Fig.~\ref{fig:LDC_Re15k_modes}(c).
The first POD mode is non-oscillatory (not shown) and represents the time-averaged flow field, whereas the POD modes $\Phi^{u_1}_2(\pmb{x})$, $\Phi^{u_1}_3(\pmb{x})$, $\Phi^{u_1}_4(\pmb{x})$, and $\Phi^{u_1}_5(\pmb{x})$ highlight the unsteady shear regions of the lid-driven cavity.
In contrast, DMD modes are distilled based on their dynamic response (frequency) and significance, {which also accounts for the normalized magnitude of the mode}; furthermore, as noted earlier, the DMD modes and associated Ritz values are complex.
The spatial DMD modes at Strouhal numbers $St_L=0.13$, $St_L=0.24$, and $St_L=0.37$ are displayed in Fig.~\ref{fig:LDC_Re15k_modes}(b), while the associated  Ritz values are plotted in Fig.~\ref{fig:LDC_Re15k_modes}(d).

To perform the modal analysis in the Lagrangian frame of reference, the flow mapping of Eq.~\ref{eq:mapping} is used.
A pseudo-code to compute LPOD modes is provided in Appendix~\ref{sec:pseudo}.
The Eulerian coordinates ($\pmb{x},t$) are transformed into the Lagrangian coordinates ($\pmb{\chi},\tau$), starting with an identity map $\mathcal{M}(\pmb{\chi}_0,\tau_0)$, where $(\pmb{\chi}_0,\tau_0)=(\pmb{x},t)$.
The set of flow snapshots, \textit{i.e.}, the Eulerian flow fields $\pmb{u}(\pmb{x},t)$, along with Eqs.~\ref{eq:M=M-1} and~\ref{eq:ul-ue} are used to construct a set of Lagrangian flow fields with respect to the identity map, \textit{i.e.}, simply the first Eulerian snapshot.
An accurate time evolution of the flow map $\mathcal{M}$ from a reference state to a deformed geometric configuration requires higher-order time schemes and/or finer time steps.
Alternatively, for a given set of snapshots in the Eulerian frame of reference, traditional POD can be used to reconstruct a high-time-resolved flow map.
For example, the Eulerian flow fields can be reconstructed by using POD as
\begin{equation}\label{eq:epod}
\pmb{u}(\pmb{x},t_n)=\sum_{m=1}^{N_r} \sqrt{\lambda_m}\pmb{\Phi}_m(\pmb{x})\Psi_m(t_n)\text{ for } n=1,2,\dots,N_t,
\end{equation}
where $N_t$ is the number of snapshots.
$N_r$ is a reduced number of POD modes used to reconstruct the flow field $\pmb{u}$, where $N_r\leq N_t$.
The temporal coefficients of Eq.~\ref{eq:epod}, $\Psi(t)$, can be obtained at a higher time resolution over the same time duration by performing a simple interpolation procedure, which then leads to better time resolution for the flow variable as,
\begin{equation}
\widetilde{\pmb{u}}(\pmb{x},t_n)=\sum_{m=1}^{N_r} \sqrt{\lambda_m}\pmb{\Phi}_m(\pmb{x})\widetilde{\Psi_m}(t_n)\text{ for } n=1,2,\dots,N,
\end{equation}
where $N$, with $N>N_t$, is the new number of snapshots due to the time-refinement.
High-resolution reconstructions of the Eulerian flow field $\widetilde{\pmb{u}}(\pmb{x},t)$ and Eqs.~\ref{eq:mapping},~\ref{eq:M=M-1}, and~\ref{eq:ul-ue} are used to obtain the Lagrangian flow fields to the required accuracy.

\begin{figure}
\begin{minipage}{0.5\textwidth}
\centering {\includegraphics[scale=0.09]{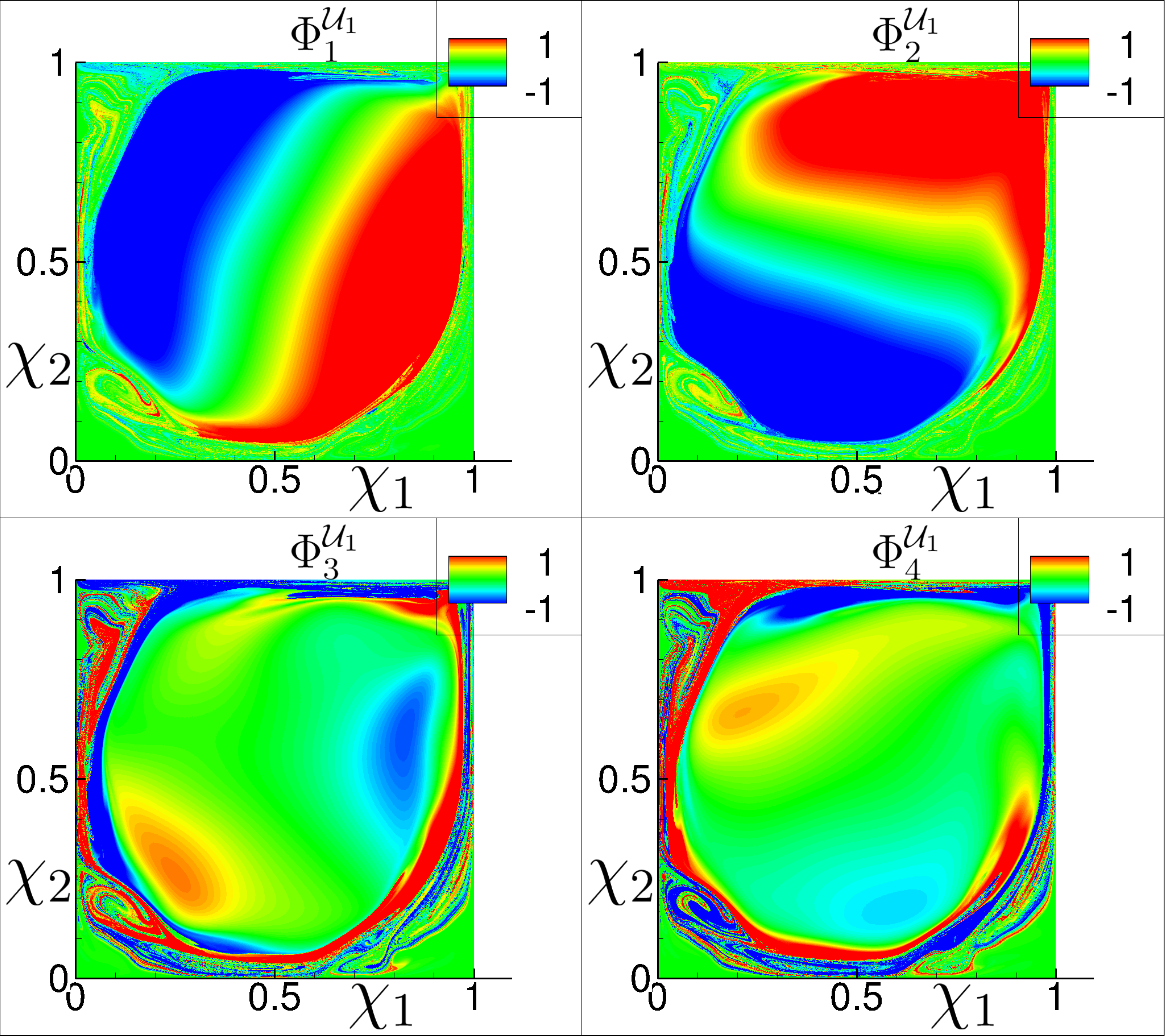}}\\(a) LPOD modes
\end{minipage}
\begin{minipage}{0.5\textwidth}
\centering {\includegraphics[scale=0.09]{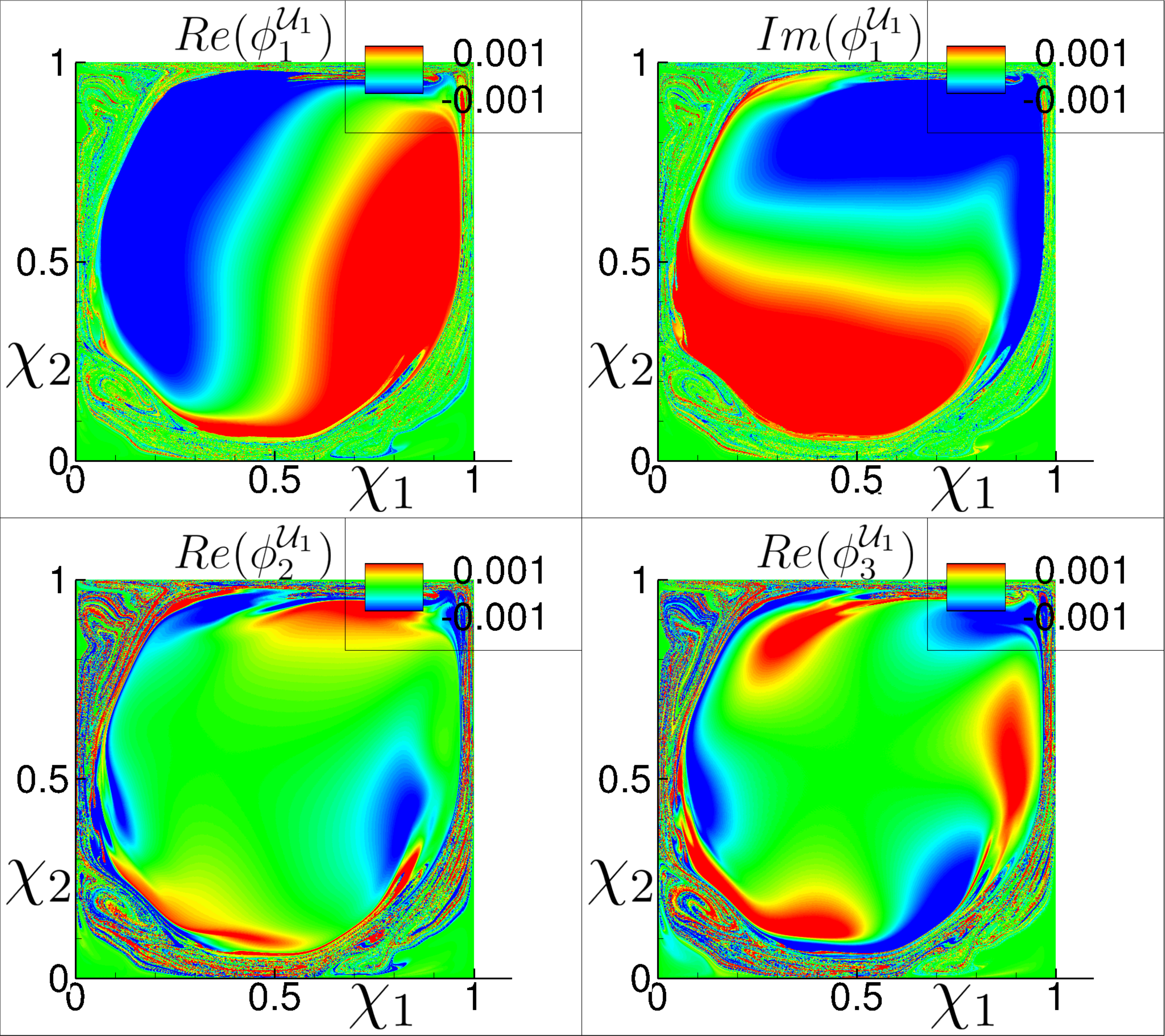}}\\(b) LDMD modes
\end{minipage}
\begin{minipage}{0.5\textwidth}
\centering {\includegraphics[scale=0.8]{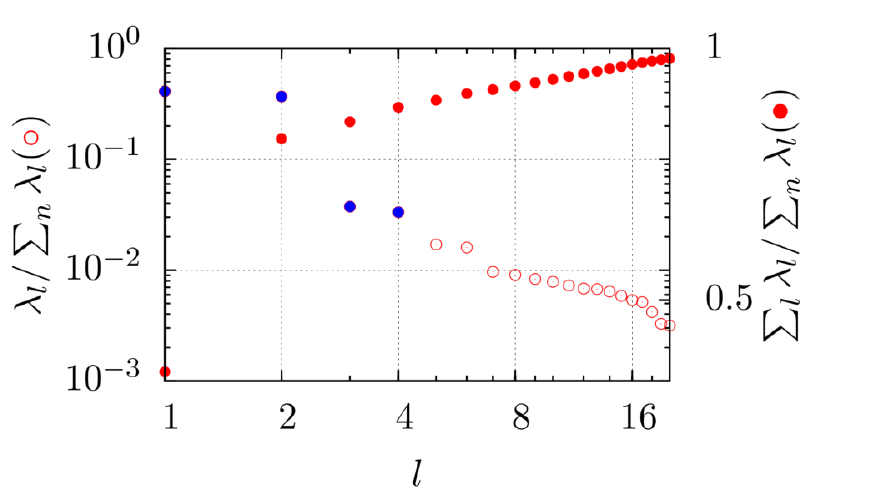}}\\(c) LPOD eigenvalues
\end{minipage}
\begin{minipage}{0.5\textwidth}
\centering {\includegraphics[scale=.7]{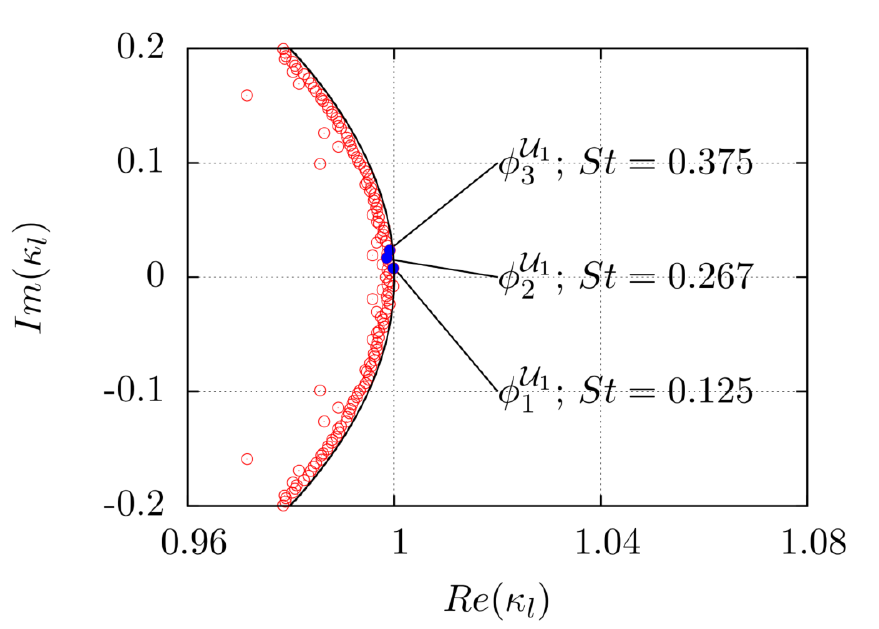}}\\(d) LDMD eigenvalues
\end{minipage}
\caption{Leading LPOD and LDMD modes of the streamwise velocity for the lid-driven cavity at $Re_L=15{,}000$ in the Lagrangian frame of reference. The modes are presented using the identity map (reference grid) $\pmb{\chi}_0(\tau_0)$ at time $\tau=\tau_0$. (a) LPOD modes $\Phi_1^{\mathcal{U}_1}$, $\Phi_2^{\mathcal{U}_1}$, $\Phi_3^{\mathcal{U}_1}$, and $\Phi_4^{\mathcal{U}_1}$. (b) LDMD modes $\phi_1^{\mathcal{U}_1}$, $\phi_2^{\mathcal{U}_1}$, and $\phi_3^{\mathcal{U}_1}$. (c) LPOD modal energies and (d) LDMD eigenvalues, indicating the values corresponding to the spatial modes.}
\label{fig:LDC_Re15k_lmodes}
\end{figure}

The corresponding LPOD and LDMD of the unsteady lid-driven cavity are obtained by using the same set of ($2{,}500$ Eulerian flow fields) snapshots.
As discussed above, the POD temporal coefficients are refined by a factor of $4$, resulting in $N=10{,}000$ Eulerian instances.
These are then transformed onto a set of $2{,}500$ Lagrangian flow fields in order to perform LPOD and LDMD, by considering the first snapshot as the identity map.
The prominent LPOD and LDMD modes of the Lagrangian streamwise flow velocity field are displayed in Figs.~\ref{fig:LDC_Re15k_lmodes}(a) and~\ref{fig:LDC_Re15k_lmodes}(b), whereas the corresponding energy contribution and frequency content are shown in Figs.~\ref{fig:LDC_Re15k_lmodes}(c) and~\ref{fig:LDC_Re15k_lmodes}(d), respectively.
The leading LPOD modes $\Phi_1^{\mathcal{U}_1}$ and $\Phi_2^{\mathcal{U}_1}$ are strikingly similar to, respectively, the real and imaginary parts of the LDMD mode $\phi_1^{\mathcal{U}_1}$, which is associated with the Strouhal number of $St_L=0.125$.
Furthermore, the higher LPOD modes $\Phi_3^{\mathcal{U}_1}$ and $\Phi_4^{\mathcal{U}_1}$ as well as the LDMD modes $\phi_2^{\mathcal{U}_1}$ and $\phi_3^{\mathcal{U}_1}$ clearly highlight the shear region of the lid-driven cavity, where the LDMD modes correspond to  unsteadiness at Strouhal numbers of $St_L=0.267$ and $St_L=0.375$, respectively.

The Lagrangian modes of Fig.~\ref{fig:LDC_Re15k_lmodes} are characteristically different from the Eulerian modes of Fig.~\ref{fig:LDC_Re15k_modes}.
The Couette flow and shear flow regions of the lid-driven cavity are distinctly exhibited by the Lagrangian modes; on the other hand, the Eulerian modes show modal shapes concentrated towards the bottom-left walls of the lid-driven cavity, focusing on the shear regions with a high contribution to the flow unsteadiness.
{In general, the decay of LPOD modal energies (Fig.~\ref{fig:LDC_Re15k_lmodes}c) for increasing number of modes, which is also shown in terms of cumulative modal energy,  appears higher for the Eulerian set of POD modes (Fig.~\ref{fig:LDC_Re15k_modes}c).}
The LDMD modal patterns (Fig.~\ref{fig:LDC_Re15k_lmodes}b) for increasing Strouhal number are more intelligible compared to the DMD modal patterns of Fig.~\ref{fig:LDC_Re15k_modes}(b) for increasing Strouhal number.

{
The notion of non-uniqueness of the hyperbolic trajectories and Lyapunov exponents due to the finite time applies to the LMA ansatz as well, in the sense that the set of LMA coherent structures (\textit{e.g.} LPOD/LDMD modes) changes with the identity map and time duration $\mathcal{T}$.
However, the uniqueness of the LMA modes can be ensured by considering a sufficiently long time duration and/or an appropriate flow region, where the issue naturally relates to the spatio-temporal resolution of the flow.
The Ritz eigenvalues associated with DMD modes provide a measure of flow stationarity/convergence in terms of the non-growing/non-decaying global modes that lie on the unit circle.
For instance, Fig.~\ref{fig:LDC_Re15k_lmodes}(d) displays the Ritz eigenvalues for LDMD modes of the lid-driven cavity at $Re_L=15{,}000$, where except for few outliers most of the eigenvalues are along the unit circle.
The corresponding Eulerian DMD eigenvalues of Fig.~\ref{fig:LDC_Re15k_modes}(d) also lie along the unit circle with some outliers.
However, the eigenvalues are not perfectly on the unit circle, indicating stable (for inside the circle) and unstable (for outside the circle) tendencies of the modes.
For instance, \cite{chen2012variants,towne2018spectral} discuss the equivalence between DMD and Fourier modes for zero-centered data, which ensure the zero growth/decay rate of the modes, \textit{i.e}, the eigenvalues strictly lie on the unit circle.
Nonetheless, LMA employs unaltered flow map data in the context of finite-time unsteady/transient dynamics, analogous to the finite-time hyperbolic trajectories and Lyapunov exponents.
Notably, the Ritz values for the LDMD modes are well aligned with the unit circle (\textit{e.g.} Fig.~\ref{fig:LDC_Re15k_lmodes}d), as opposed to the Eulerian counterpart (\textit{e.g.} Fig.~\ref{fig:LDC_Re15k_modes}d).
}

\begin{figure}
\begin{minipage}{0.24\textwidth}
\centering {\includegraphics[scale=0.045]{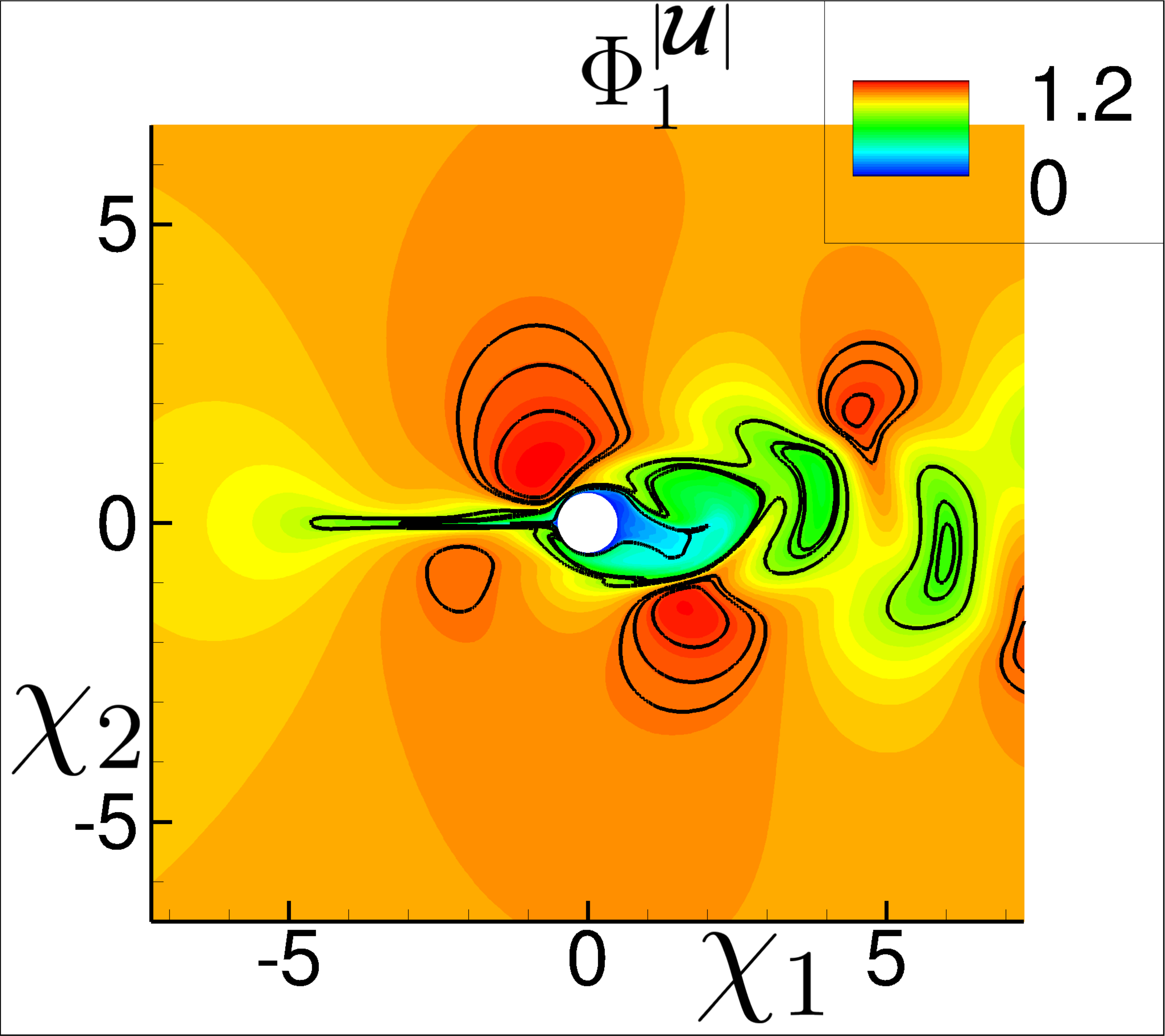}}\\(a) 
\end{minipage}
\begin{minipage}{0.24\textwidth}
\centering {\includegraphics[scale=0.045]{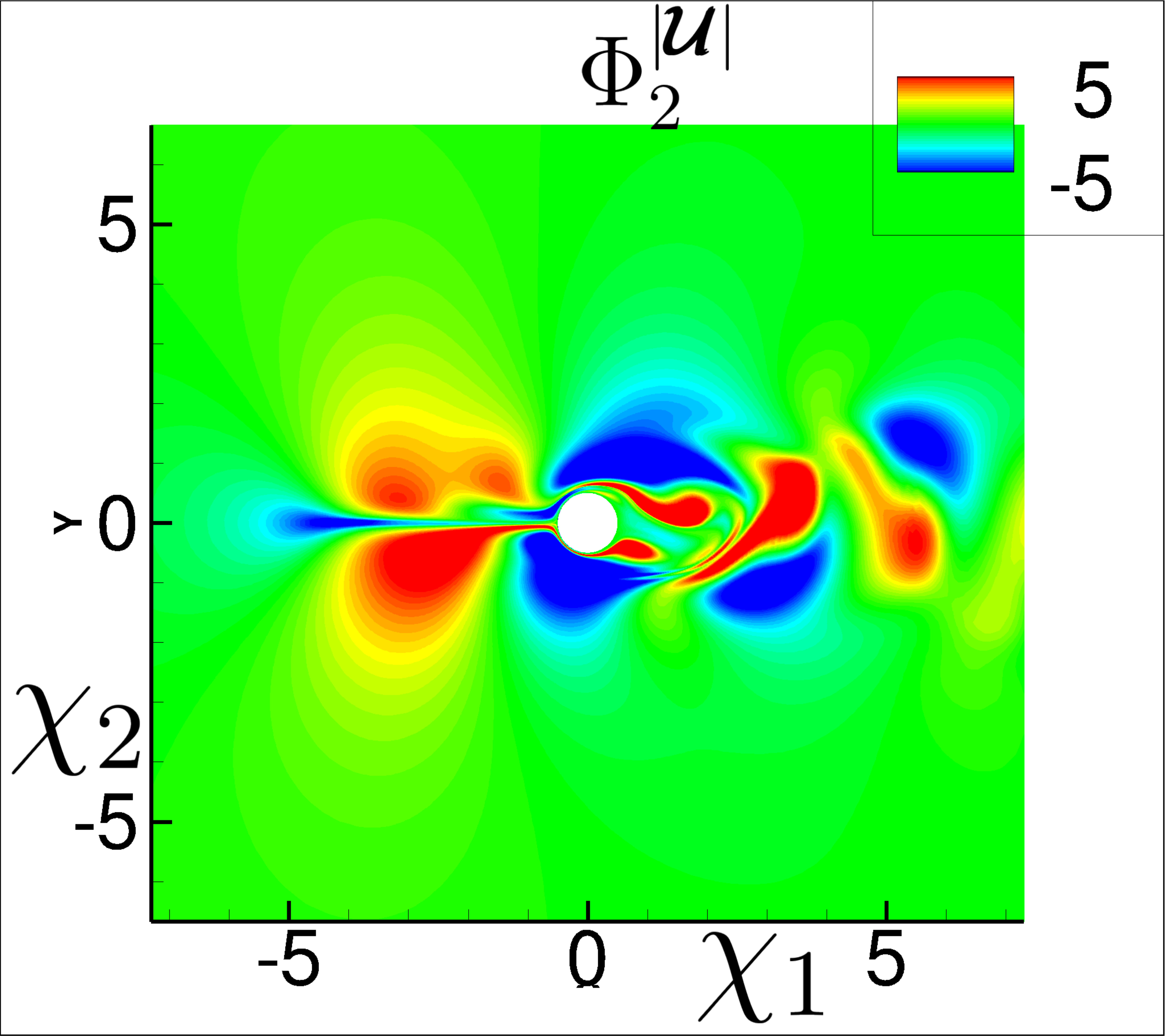}}\\(b) 
\end{minipage}\hfill
\begin{minipage}{0.24\textwidth}
\centering {\includegraphics[scale=0.045]{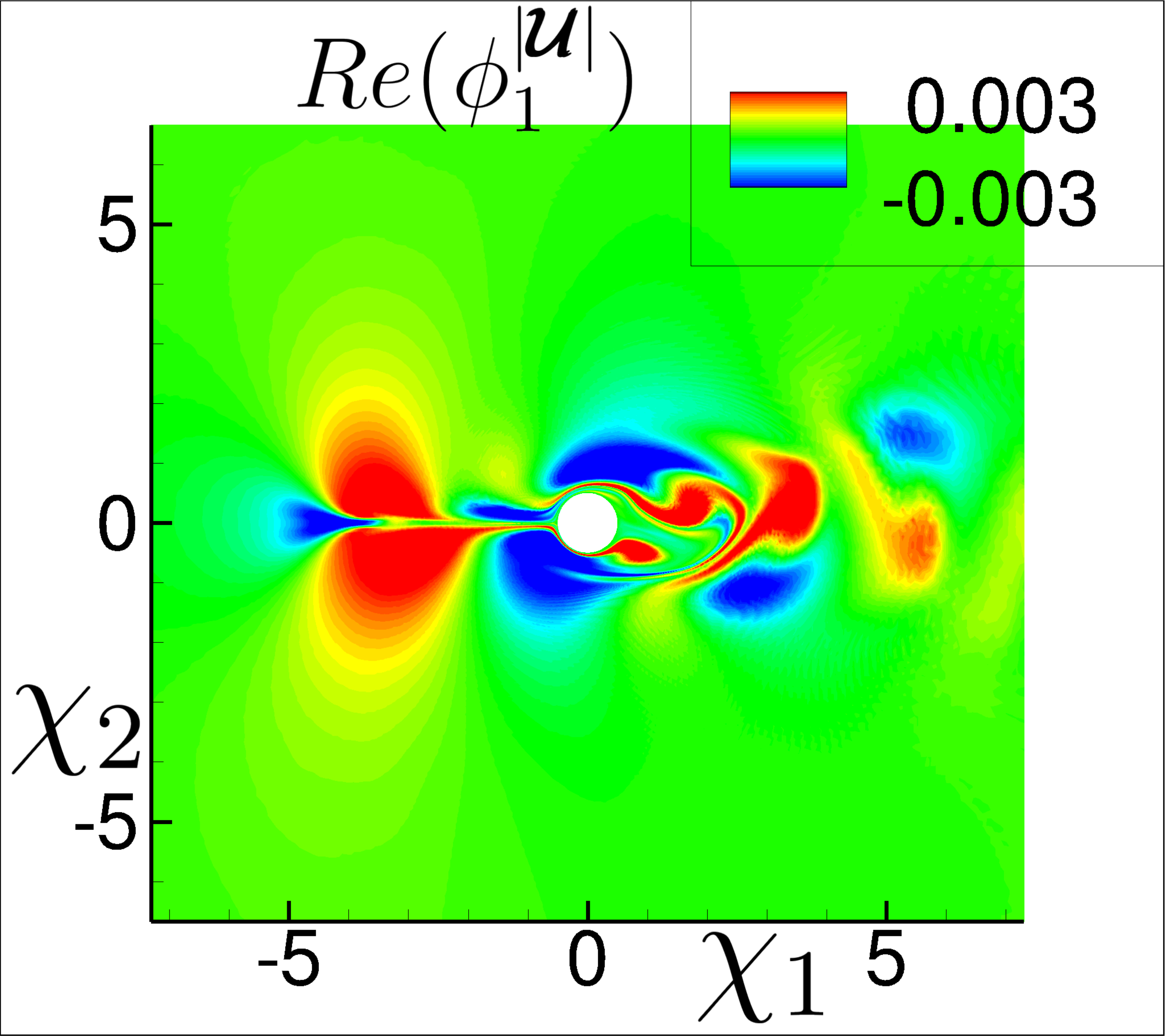}}\\(c) 
\end{minipage}
\begin{minipage}{0.24\textwidth}
\centering {\includegraphics[scale=0.045]{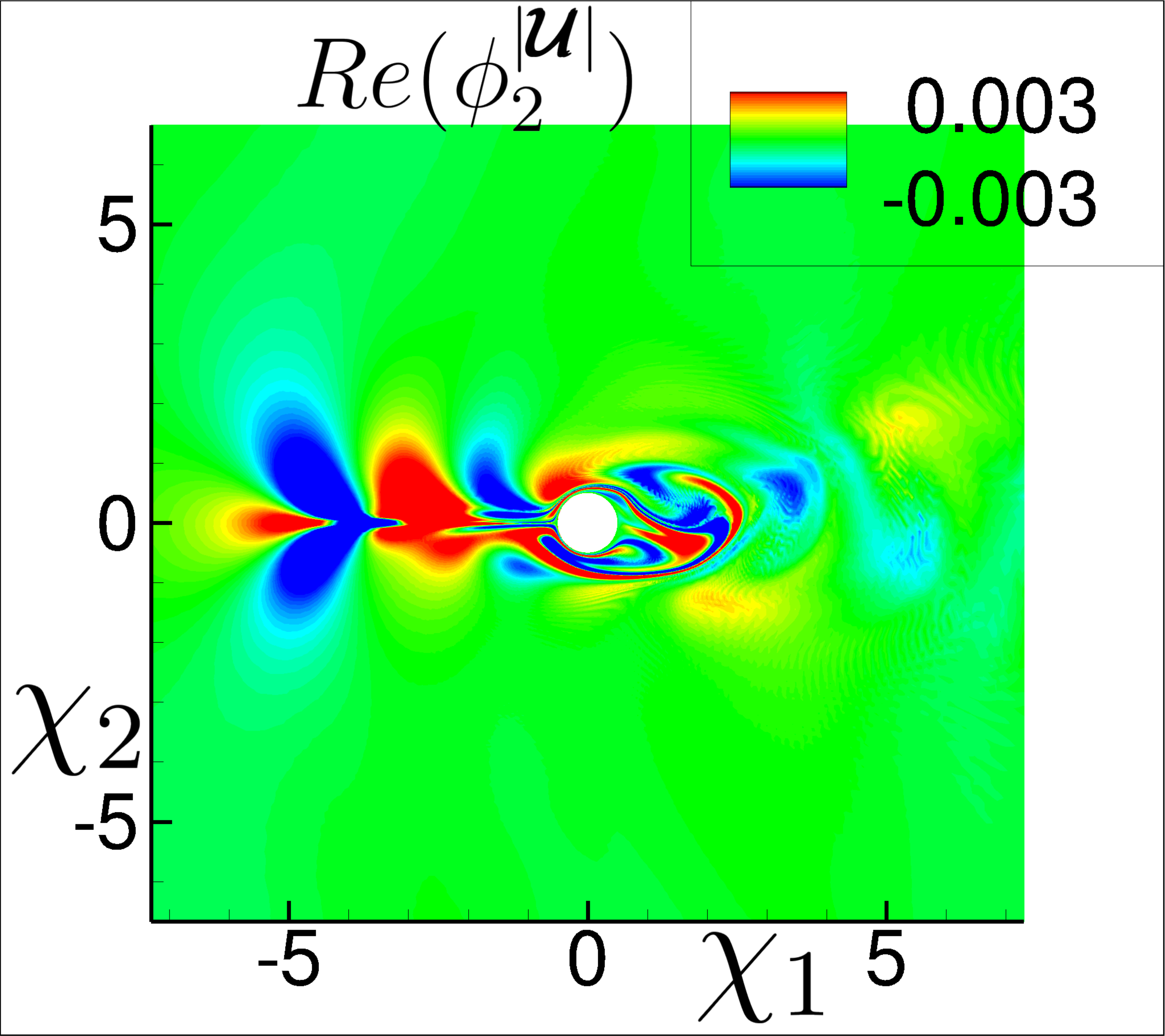}}\\(d)
\end{minipage}
\begin{minipage}{0.24\textwidth}
\centering {\includegraphics[scale=0.045]{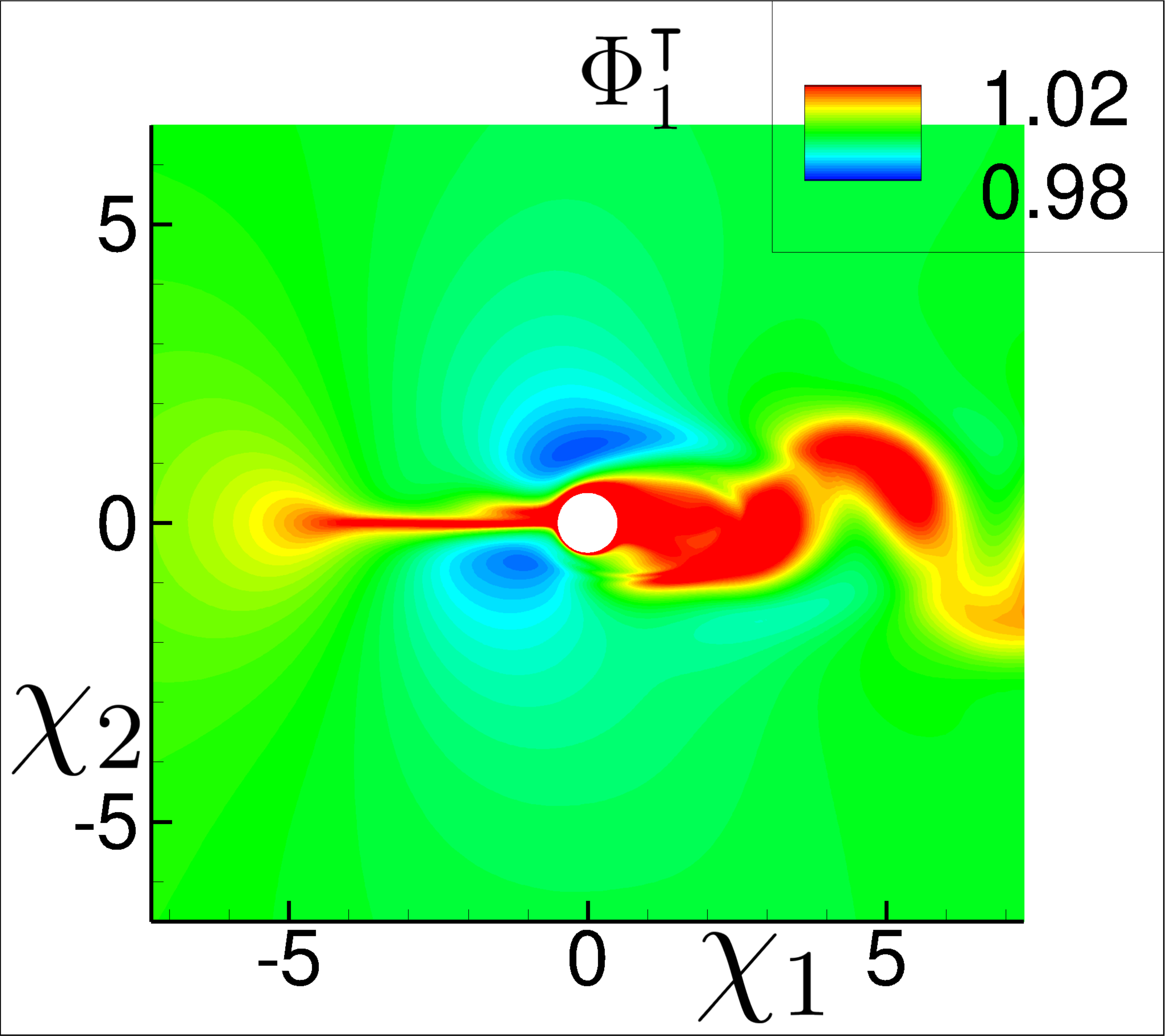}}\\(e)
\end{minipage}
\begin{minipage}{0.24\textwidth}
\centering {\includegraphics[scale=0.045]{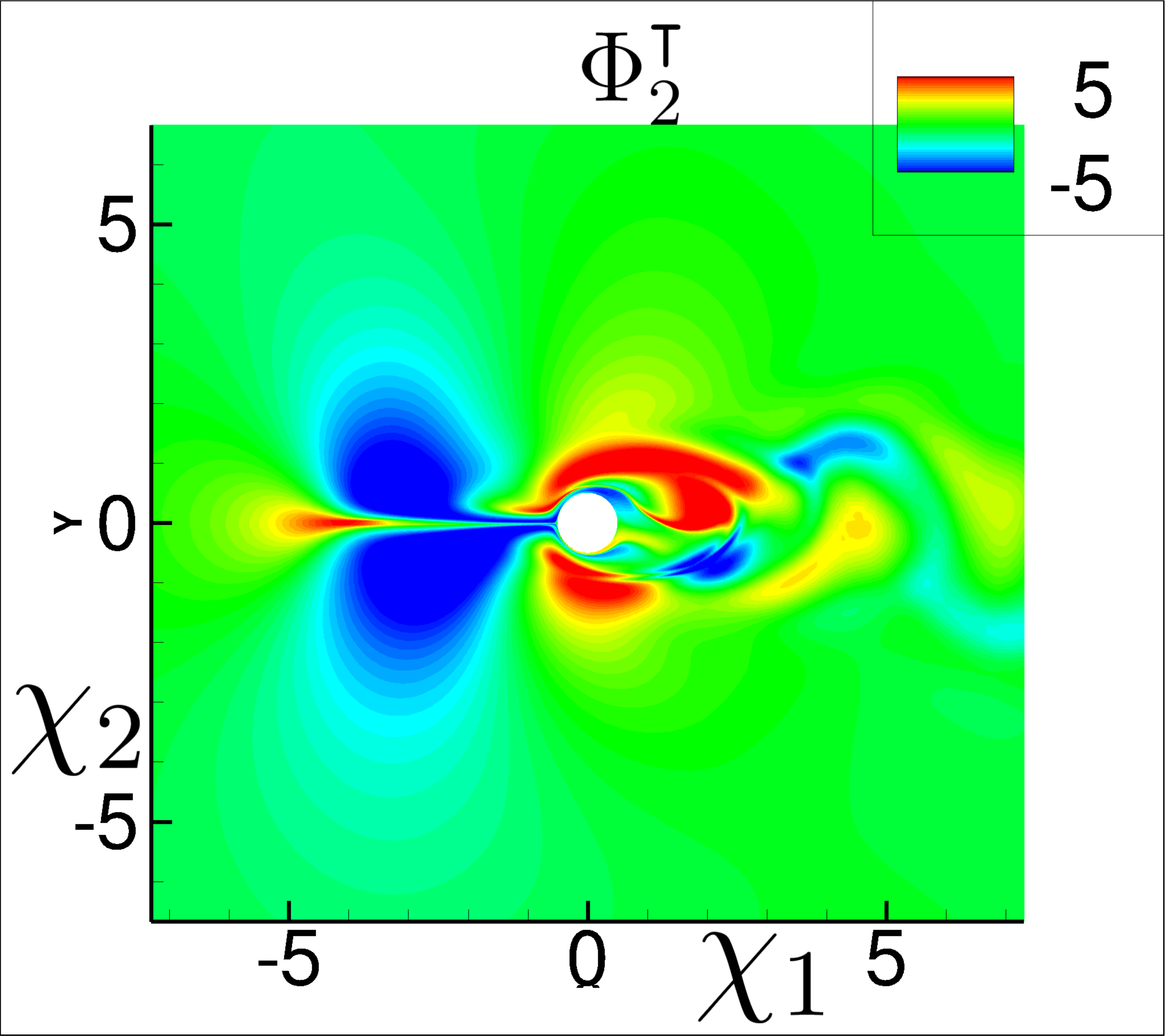}}\\(f)
\end{minipage}\hfill
\begin{minipage}{0.24\textwidth}
\centering {\includegraphics[scale=0.045]{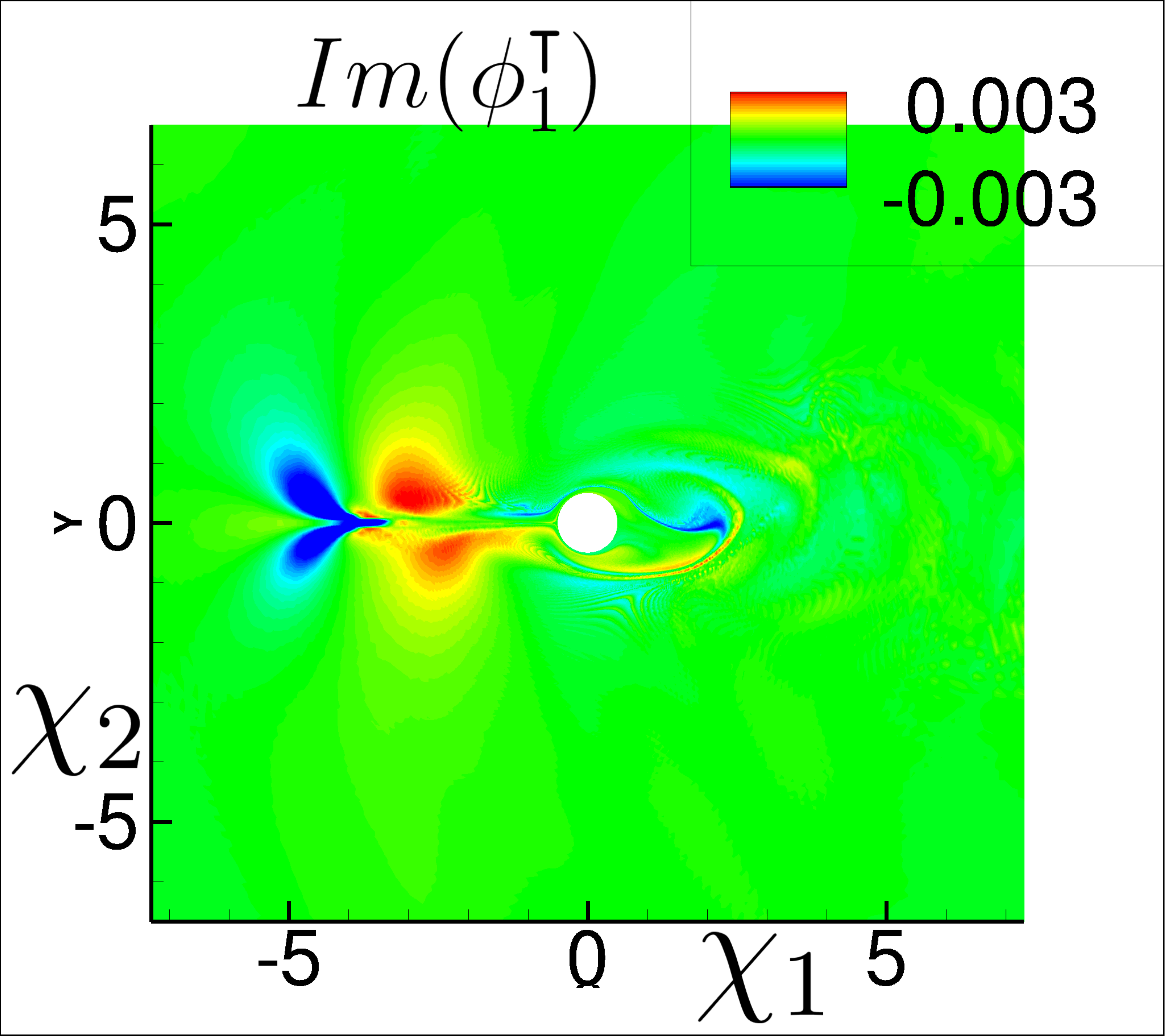}}\\(g) 
\end{minipage}
\begin{minipage}{0.24\textwidth}
\centering {\includegraphics[scale=0.045]{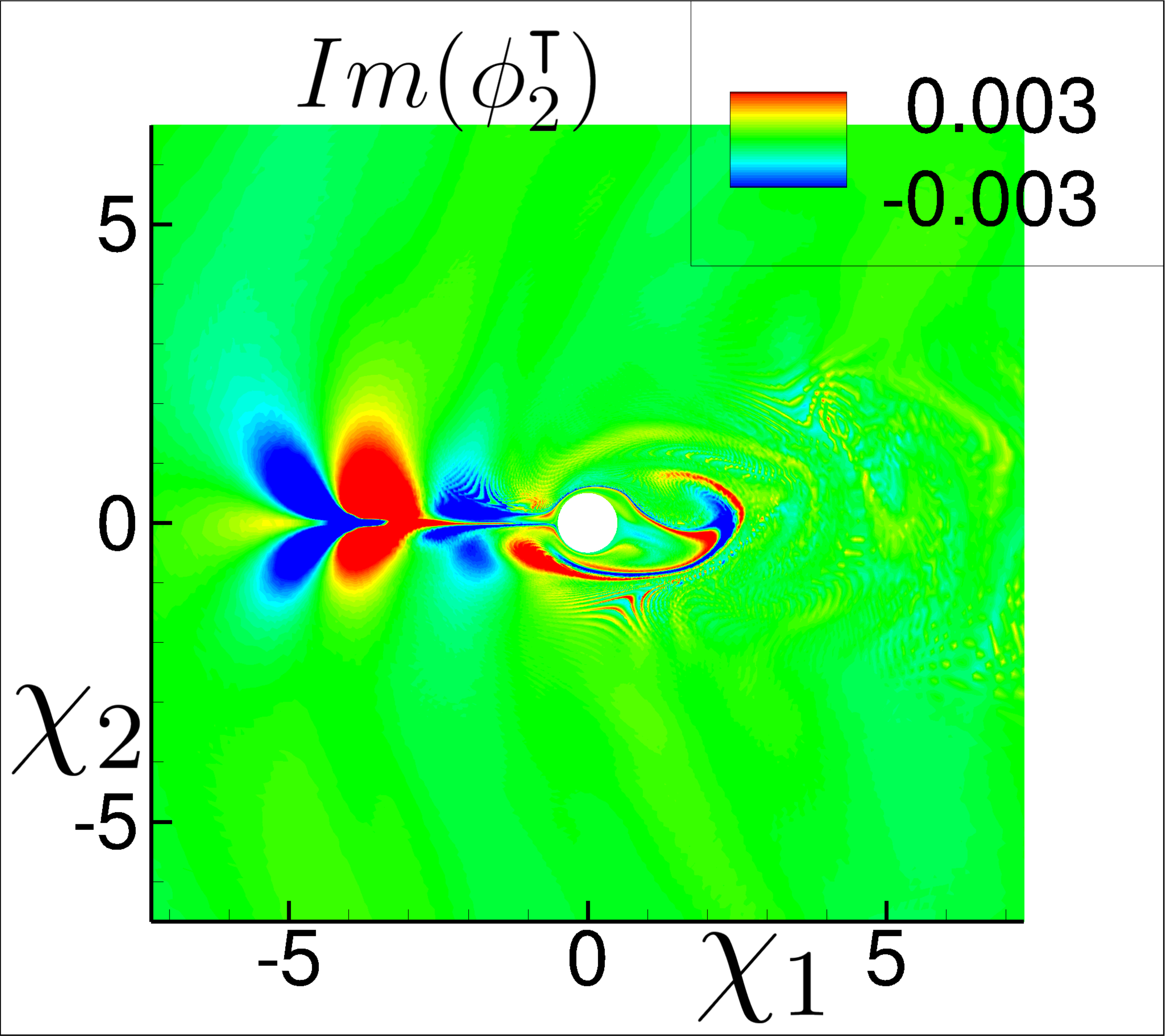}}\\(h)
\end{minipage}
\caption{Leading LPOD and LDMD modes of the absolute velocity and temperature for flow past cylinder at $Re_D=100$ and $M_\infty=0.5$. The modes are displayed on the identity map (reference grid) $\pmb{\chi}_0(\tau_0)$ at time $\tau=\tau_0$. (a) First LPOD mode $\Phi_1^{|\pmb{\mathcal{U}}|}$ and (b) second LPOD mode $\Phi_2^{|\pmb{\mathcal{U}}|}$ of the absolute velocity; (c) real part of LDMD mode $Re(\phi_1^{|\pmb{\mathcal{U}}|})$ with $St_D=0.13$ and (d) real part of LDMD mode $Re(\phi_2^{|\pmb{\mathcal{U}}|})$ with $St_D=0.33$ for the absolute velocity. (e) First LPOD mode $\Phi_1^\intercal$ and (f) second LPOD mode $\Phi_2^\intercal$ of the temperature; (g) imaginary part of LDMD mode $Im(\phi_1^{\intercal})$ with $St_D=0.13$ and (h) imaginary part of LDMD mode $Im(\phi_2^{\intercal})$ with $St_D=0.33$ for the temperature.}
\label{fig:CYL_Re100_lmodes_TabsU}
\end{figure}
{
In general, Lagrangian transport, mixing, and chaos are studied using finite-time stable and unstable manifolds of hyperbolic trajectories via measures such as the finite-time Lyapunov exponents, in order to identify and monitor the transport and mixing barriers in the flow, which are the material lines and surfaces on these attracting and repelling manifolds~\citep{branicki2010finite,haller2015lagrangian,balasuriya2016hyperbolic}.
As noted earlier, Lyapunov exponents represent the growth of small separation between trajectories with time, leading to $d$ number of exponents for a $d$-dimensional state space.
A positive value of the exponent indicates an unstable trajectory, where the largest Lyapunov exponent governs the dynamics, albeit locally~\citep{ottino1989kinematics}.
On the other hand, the LMA ansatz acts on the entire material surface (or more generally material volume) for its global dynamics, comprising the locally stable and unstable manifolds over the considered time duration from the initial flow map.
The spatio-temporal material surface/volume may exhibit linear/non-linear, steady/unsteady and/or chaotic dynamics, dictating the transport and mixing process, whose dynamics can be conveniently examined by means of suitable LMA techniques, \textit{e.g.} LPOD, LDMD and their variants.
}

{
The link between the FTLE and LMA is formulated in Sec.~\ref{sec:LE}, elucidating the equivalence between the dominant LPOD mode of the velocity magnitude and maximum FTLE.
Figure~\ref{fig:CYL_Re100_lmodes_TabsU}(a) displays the first LPOD mode of the absolute velocity for the flow past a cylinder at $Re_D=100$ and $M_\infty=0.5$ (see Sec.~\ref{sec:CYL} and Appendix~\ref{sec:grid_cyl}), where the regions of high/low modal amplitudes are also delineated in terms of selected contours.
This LMA analysis 
is performed over time $\tau \in [0,\mathcal{T}=5]$, utilizing $n=500$ material (Lagrangian) snapshots.
The maximum FTLE fields at other parameters
by, for example, \citet{kasten2009localized,finn2013integrated} share several structural similarities with the LPOD mode of Fig.~\ref{fig:CYL_Re100_lmodes_TabsU}(a), in particular the lower amplitude contours (on the green region) in the cylinder wake and along the upstream centerline ($\chi_2=0$).
These regions of the flow map (material surface) are at lower absolute velocities, in the Lagrangian mean sense, over the considered finite time duration from an initial state, as opposed to the higher amplitude regions (in red).
The second LPOD mode of Fig.~\ref{fig:CYL_Re100_lmodes_TabsU}(b) displays the flow map regions with the largest contribution to the variance of the absolute velocity about the first LPOD mode.
These flow map regions are also associated with the von-Karman vortex shedding, which becomes evident in terms of the corresponding LDMD mode with $St_D=0.13$ shown in Fig.~\ref{fig:CYL_Re100_lmodes_TabsU}(c).
Clearly, the LPOD and LDMD modes of Figs.~\ref{fig:CYL_Re100_lmodes_TabsU}(b) and (c), respectively, indicate the prominent regions of the flow map exhibiting the von-Karman vortex shedding.
In addition, an LDMD mode with $St_D=0.33$ is displayed in Fig.~\ref{fig:CYL_Re100_lmodes_TabsU}(d), where the flow map regions are associated with the first higher harmonic of the vortex shedding frequency ($St_D=0.16$).
The LPOD and LDMD modes of Figs.~\ref{fig:CYL_Re100_lmodes_TabsU}(a), (b), (c), and (d), along with the complete sets of LPOD/LDMD modes showcase spatio-temporal dynamics of the material surface, providing insights into its transport and chaotic mixing characteristics.

In many practical situations, it is desired to obtain Lagrangian coherence of quantities other than the primary flow variables, namely, velocity, pressure, and density~\citep{balasuriya2018generalized}.
For example, the flow temperature, species concentration, and in general the derived quantities from the primary variables are relevant to many multi-physics processes including, among others, turbulent combustion, acoustics, and magnetohydrodynamics.
In the LMA ansatz, the non-primary flow variables can be also subjected to 
many other modal decomposition techniques.
To demonstrate this, we consider LPOD and LDMD of the temperature field that is associated with the Lagrangian flow map of the flow past cylinder case (of Sec.~\ref{sec:CYL}), where again the LMA analysis is performed over time $\tau \in [0,\mathcal{T}=5]$, utilizing $n=500$ material (Lagrangian) snapshots.
The first two LPOD modes of the temperature field (Fig.~\ref{fig:CYL_Re100_lmodes_TabsU}(e) and (f)) exhibit coherent regions of the flow map, in terms of the Lagrangian mean and a greater part of the variance of the temperature, respectively.
The LDMD modes of the temperature of Fig.~\ref{fig:CYL_Re100_lmodes_TabsU}(g) and (h) manifest the temporal dynamics of the Lagrangian thermal field with the unsteadiness of $St_D=0.12$ and $St_D=0.33$, respectively.
The temperature field in the Eulerian simulation relates to the absolute velocity via the pressure and density; there is a noticeable dependence of the temperature modes on the corresponding velocity magnitude modes in Fig.~\ref{fig:CYL_Re100_lmodes_TabsU}.

The Lagrangian approach in fluid turbulence provides a systematic description of particles and field statistics, educing chaos and exponentially separating trajectories~\citep{falkovich2001particles,yeung2002lagrangian}.
In addition to remarkable physical insights into the transport, mixing and dispersion, the Lagrangian description of turbulence is useful in, for example, stochastic modeling and probability density functions~\citep{yeung1989lagrangian,pope1994lagrangian}, turbulence modeling in large eddy simulations~\citep{meneveau1996lagrangian}, and reduced-order modeling~\citep{lu2020lagrangian,xie2020lagrangian}.
However such studies are limited in number compared to the Eulerian counterpart.
The plethora of modal analysis techniques, in Eulerian reference frame, inherently connect to the spatial-temporal scales of turbulence, educing coherent flow structures, where these prominent flow features are utilized in, among others, flow control and optimization, reduced order modeling, and turbulence modeling~\citep{rowley2017model,shinde2020proper}.
Thus, the LMA provides a means to transform the Eulerian modal analysis techniques and their applications to the Lagrangian flow maps (material surfaces/volumes).
}

\subsection{LMA on flow with mesh deformation} \label{sec:mesh_deformation}
The difficulties of applying modal decompositions to deforming meshes and associated domains in the Eulerian frame are overcome by recasting such problems in the Lagrangian formulation.
{The LMA on the flow with deforming meshes can be performed at least in two settings: in the first case, the (material) flow map is considered to be the frame of reference, whereas in the second case the moving/deforming mesh is taken as the frame of reference.
In either case, the modal analysis is not restricted because of the moving/deforming domain.
Here we demonstrate the second possibility of performing LMA, where the Lagrangian frame of reference is the deforming mesh.
Two major advantages of this approach are: 1. the flow fields for the LMA, in general, already account for the deforming mesh in the Eulerian-Lagrangian computations, and 2. the LMA distills out modes that are associated with the mesh deformation from the modes of the flow fields, as illustrated below.}

The two-dimensional lid-driven cavity with mesh deformation (see Sec.~\ref{sec:mesh_def_2D-LDC}) is considered.
As detailed in Sec.~\ref{sec:mesh_def_2D-LDC}, the bottom wall of the lid-driven cavity is subjected to forced sinusoidal deformation (Eqs.~\ref{eq:mesh_def1} and~\ref{eq:mesh_def2}) at a Strouhal  number, $St_L=1$,  which affects the mesh in the entire domain.
The resulting flow fields are always unsteady at all Reynolds numbers due to the time-dependent domain deformation.
\begin{figure}
\begin{minipage}{0.5\textwidth}
\centering {\includegraphics[scale=0.09]{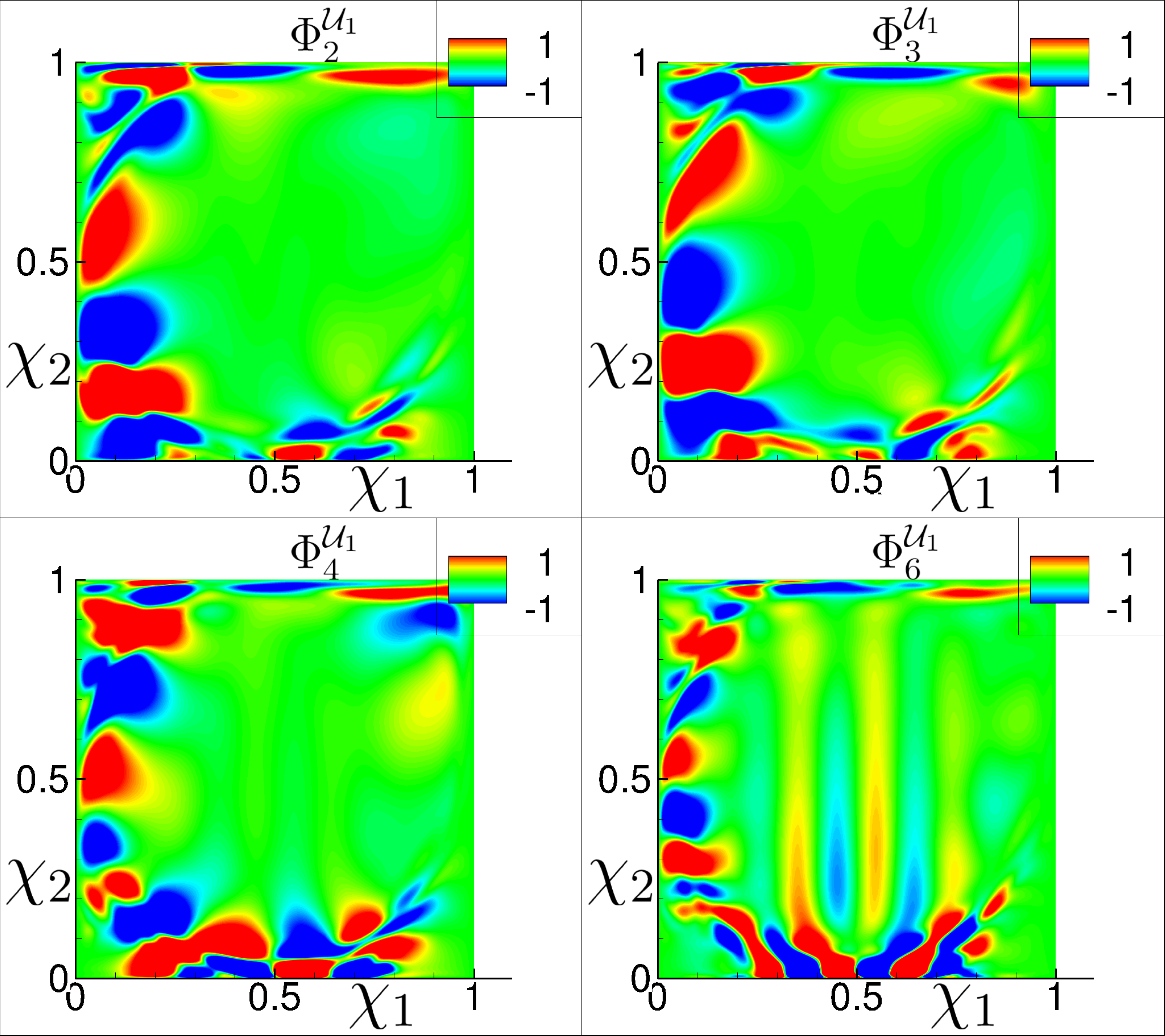}}\\(a) LPOD modes
\end{minipage}
\begin{minipage}{0.5\textwidth}
\centering {\includegraphics[scale=0.09]{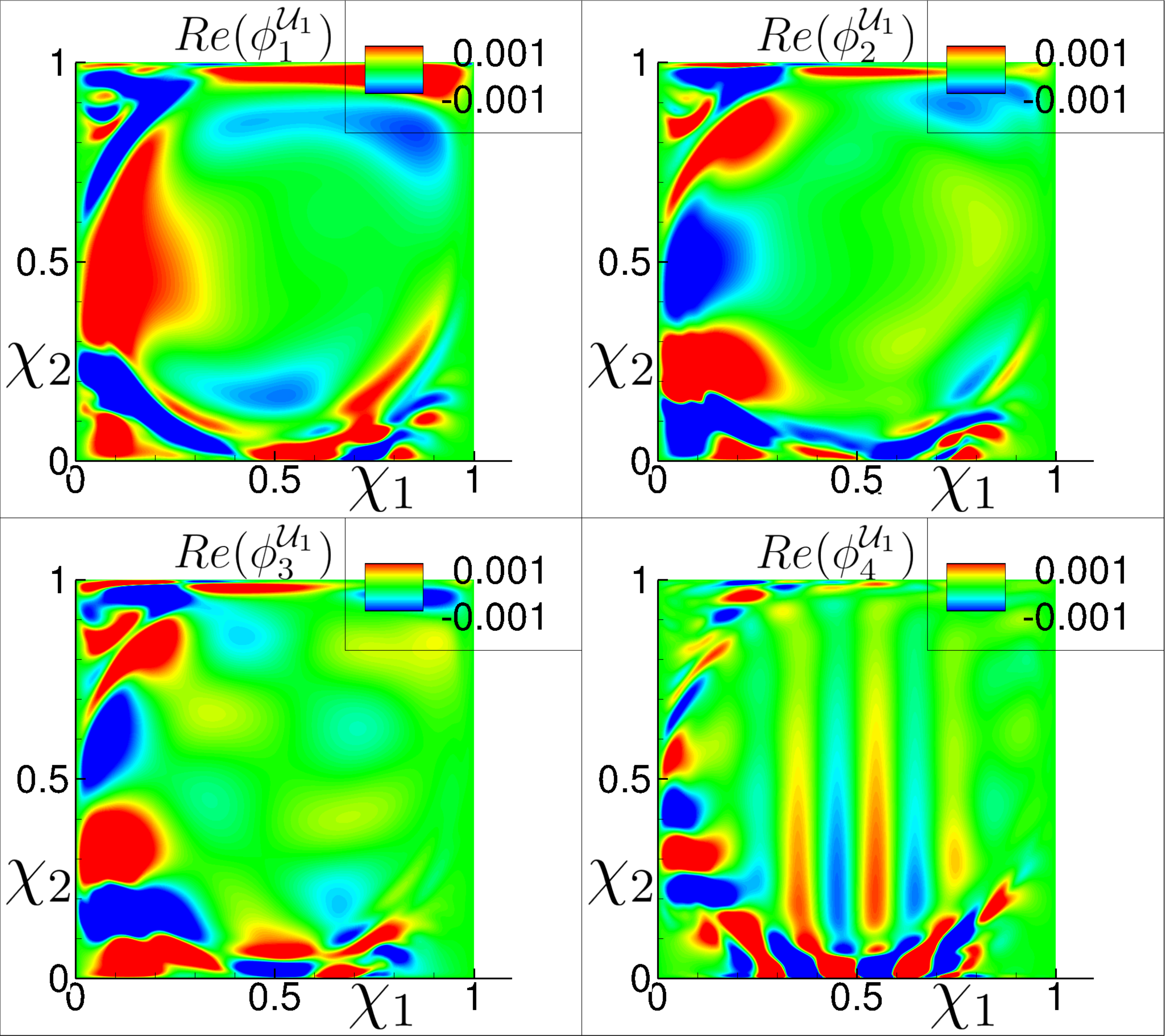}}\\(b) LDMD modes
\end{minipage}
\begin{minipage}{0.5\textwidth}
\centering {\includegraphics[scale=0.8]{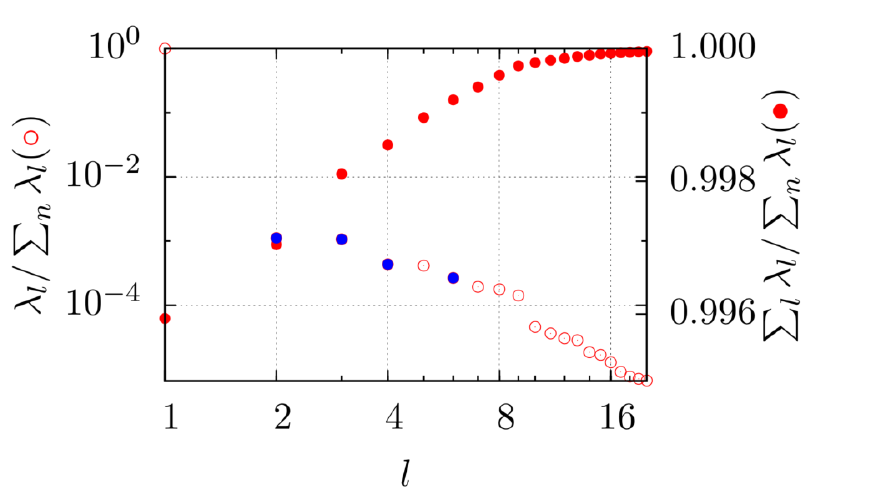}}\\(c) LPOD eigenvalues
\end{minipage}
\begin{minipage}{0.5\textwidth}
\centering {\includegraphics[scale=.7]{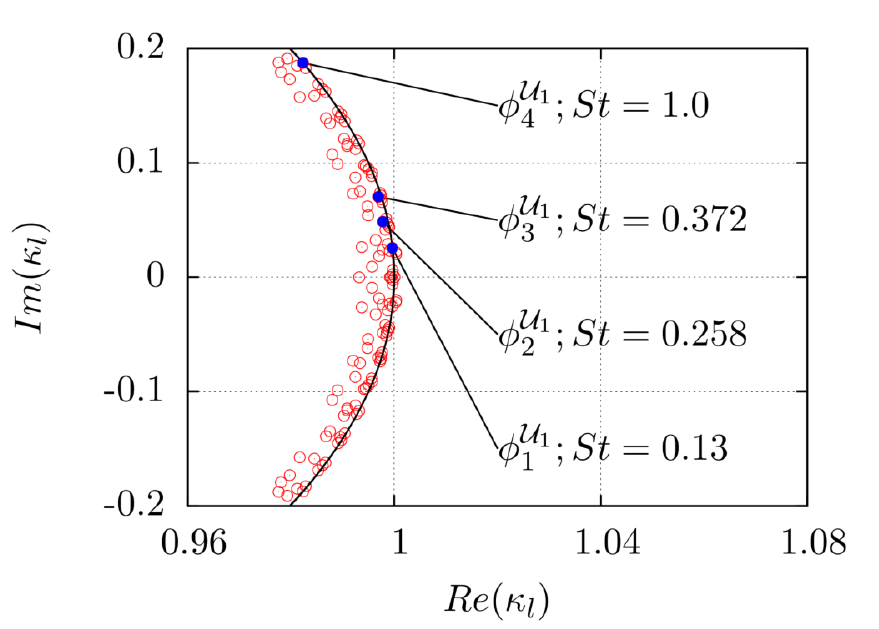}}\\(d) LDMD eigenvalues
\end{minipage}
\caption{Leading LPOD and LDMD modes of the streamwise velocity for the lid-driven cavity at $Re_L=15{,}000$  subjected to forced bottom surface deformation at $St_L=1.0$. 
The modes are presented using the identity map (reference grid) $\pmb{\chi}_0(\tau_0)$ at time $\tau=\tau_0$. (a) LPOD modes $\Phi_2^{\mathcal{U}_1}$, $\Phi_3^{\mathcal{U}_1}$, $\Phi_4^{\mathcal{U}_1}$, and $\Phi_6^{\mathcal{U}_1}$. (b) LDMD modes $\phi_1^{\mathcal{U}_1}$, $\phi_2^{\mathcal{U}_1}$, $\phi_3^{\mathcal{U}_1}$ and $\phi_4^{\mathcal{U}_1}$. (c) LPOD modal energies and (d) LDMD eigenvalues, indicating the values corresponding to the spatial modes.}
\label{fig:LDC_Re15k_fsi_k10_modes}
\end{figure}
As before, $2{,}500$ snapshots are collected at time intervals of $0.01$, encompassing Strouhal numbers in $0.04\leq St \leq 25$.
The first Eulerian snapshot is considered as the identity map for the Lagrangian transformation of Eq.~\ref{eq:mapping}, where the Lagrangian coordinates $(\pmb{\chi}_0,\tau_0)=(\pmb{x},t)$.

The leading LPOD and LDMD modes of the streamwise flow velocity are displayed in Fig.~\ref{fig:LDC_Re15k_fsi_k10_modes} together with their energy and frequency content data, respectively.
Sub-figures, Figs.~\ref{fig:LDC_Re15k_fsi_k10_modes}(a) and~\ref{fig:LDC_Re15k_fsi_k10_modes}(b), use the Eulerian/Lagrangian coordinates at the reference state (identity map).
The leading spatial LPOD and LDMD modes of Fig.~\ref{fig:LDC_Re15k_fsi_k10_modes} closely resemble the Eulerian POD and DMD modes of Fig.~\ref{fig:LDC_Re15k_modes}, in terms of the spatial structure, modal energy and frequency content.
This is consistent with the findings of~\cite{menon2020dynamic}, where the DMD modes at frequencies other than the rigid body motion were shown to be unmodified.
However, the spatial modes of Fig.~\ref{fig:LDC_Re15k_fsi_k10_modes} display differences compared to those of Fig.~\ref{fig:LDC_Re15k_modes}, mainly in regions of large mesh deformation, \textit{i.e}, near the bottom wall of the lid-driven cavity.
These differences between the Eulerian and Lagrangian modes are congruent with those between the corresponding flow fields of Fig.~\ref{fig:ldc_Uabs} and Fig.~\ref{fig:ldc_def}, respectively.

Furthermore, Fig.~\ref{fig:LDC_Re15k_fsi_k10_modes}(a) and Fig.~\ref{fig:LDC_Re15k_fsi_k10_modes}(b) include an LPOD and LDMD mode, respectively, which correspond to the forced mesh deformation, where the LDMD mode ($\phi_4^{\mathcal{U}_1}$) exhibits a modal frequency equal to the forcing frequency of $St=1$.
The Lagrangian modes that correspond to the domain deformation, $\Phi_6^{\mathcal{U}_1}$ (LPOD) and $Re(\phi_4^{\mathcal{U}_1})$ (LDMD), precisely indicate the regions of the flow that are affected by the bottom surface deflection.
This is of significant practical importance for the flow control using surface morphing~\citep{bruce2015review,shinde2020control,shinde2021supersonic}, where the effect of control surface deformation on the flow fields can be identified using LMA for the efficacy of control.

\subsection{LMA on steady flows} \label{sec:steady_flows}
In the Eulerian description, a steady flow is characterized by  time-independent flow variables.
From the stability point of view, a base flow or fixed point is a flow state where all the solutions to an initial value problem converge monotonically.
The stability properties of the flow can be characterized based on the temporal/spatial evolution of the external perturbations superimposed on the steady base flow state.
In addition to stability analysis, the steady base flow can be subjected to different analyses that are of practical interest, such as for instance, resolvent analysis~\citep{schmid2014analysis,sharma2013coherent}.
A steady non-uniform flow in the Eulerian frame of reference transforms to an unsteady flow in the Lagrangian frame of reference, enabling the use of Lagrangian modal decompositions (of Sec.~\ref{sec:theory}) directly for the steady flows.
The Eulerian steady base flow/fixed point again serves as the identity map of Eq.~\ref{eq:mapping} for the Lagrangian flow diffeomorphism.
A set of flow snapshots may be extracted from the phase portrait of the base flow, starting at the identity map and tracing the diffeomorphism of the Lagrangian domain for increasing time.
Alternatively, the time direction may be reversed by tracing the diffeomorphism of the Lagrangian domain in the negative (backward) time direction.
The forward and adjoint (backward) formulations to perform the Lagrangian modal analysis of a steady base flow are discussed in this section.

\subsubsection{Forward Lagrangian approach} \label{sec:forward}
The lid-driven cavity flow at the pre-critical Reynolds number of $Re_L=7{,}000$ (see Sec.~\ref{sec:LDC}, Fig.~\ref{fig:ldc_Uabs}a) is considered here for the Lagrangian modal analysis.
The steady base flow in the Eulerian frame of reference is then represented in the Lagrangian frame of reference according to Eqs.~\ref{eq:mapping},~\ref{eq:M=M-1}, and~\ref{eq:ul-ue}.
The Lagrangian flow map is allowed to evolve in the positive/forward time direction from the reference steady state by solving an initial value problem.
A set of $2{,}500$ snapshots of the Lagrangian flow fields at time interval of $0.01$ are collected to perform the modal analysis, while a much smaller time-step of $\delta t = 0.001$ is used to accurately track the Lagrangian flow map.

\begin{figure}
\begin{minipage}{0.5\textwidth}
\centering {\includegraphics[scale=0.09]{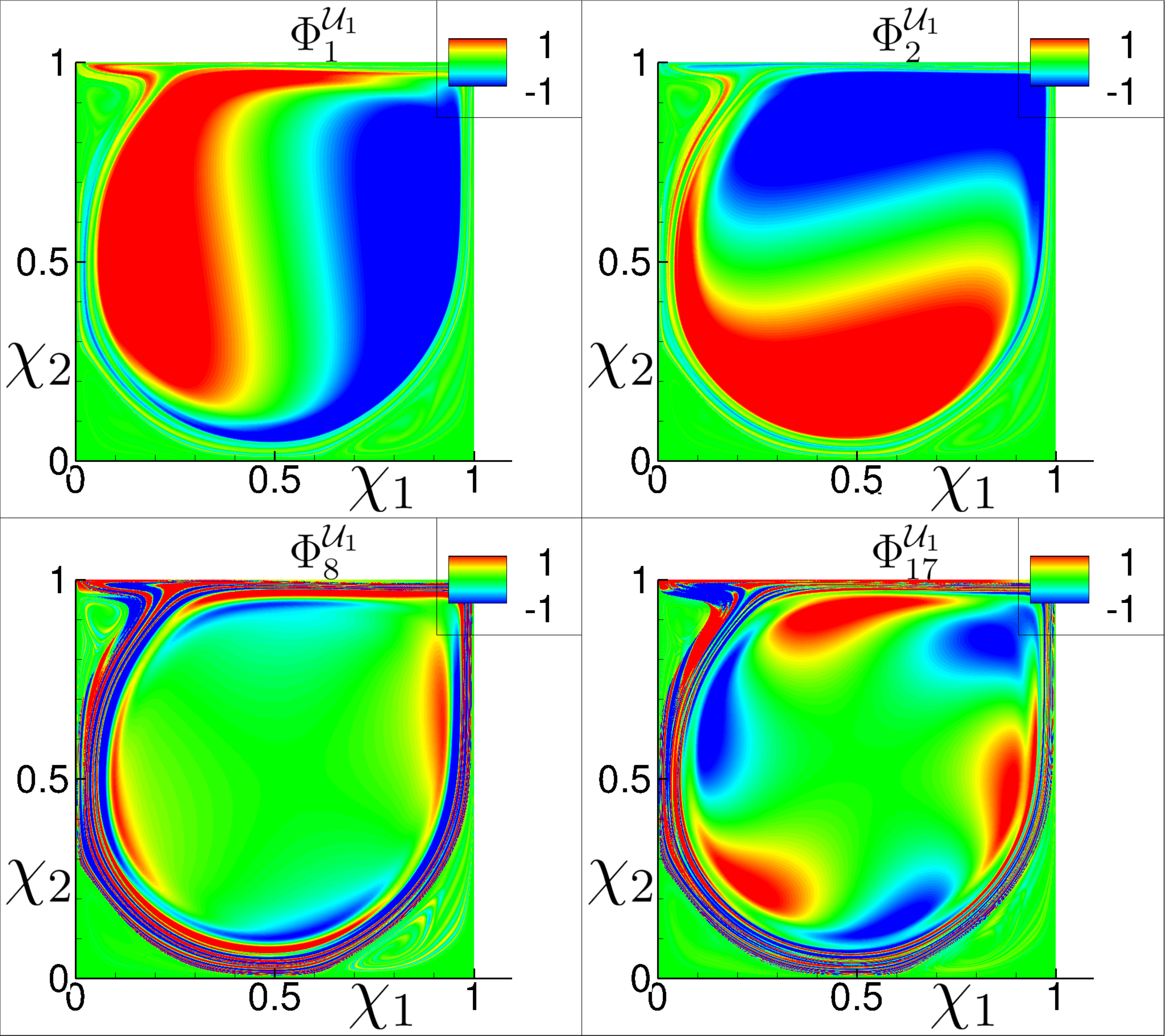}}\\(a) LPOD modes
\end{minipage}
\begin{minipage}{0.5\textwidth}
\centering {\includegraphics[scale=0.09]{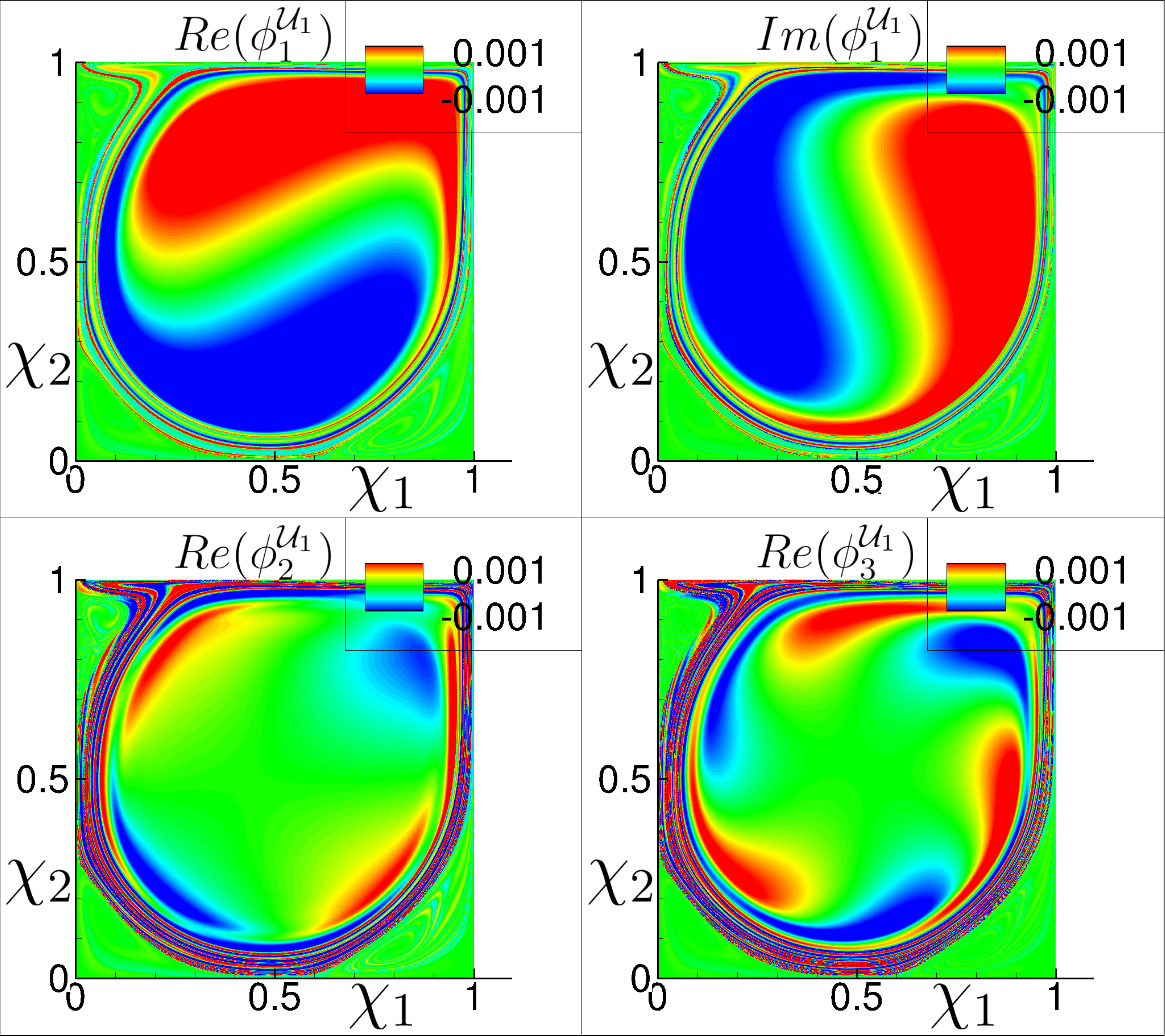}}\\(b) LDMD modes
\end{minipage}
\begin{minipage}{0.5\textwidth}
\centering {\includegraphics[scale=0.8]{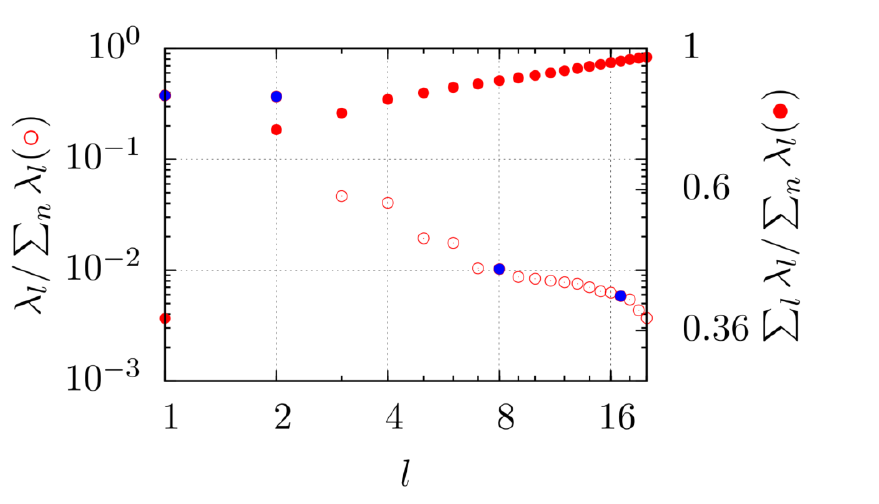}}\\(c) LPOD eigenvalues
\end{minipage}
\begin{minipage}{0.5\textwidth}
\centering {\includegraphics[scale=.7]{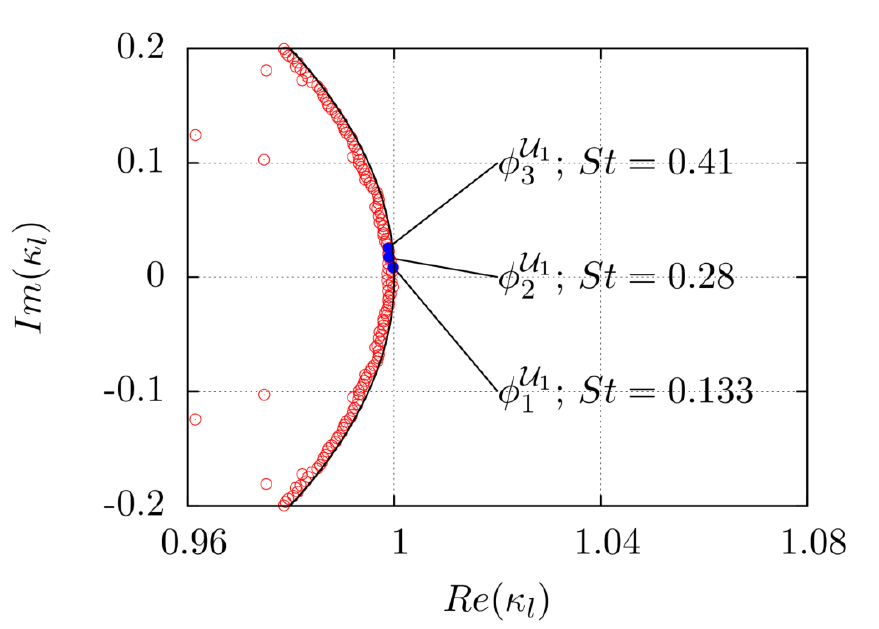}}\\(d) LDMD eigenvalues
\end{minipage}
\caption{Leading modes of the steady state streamwise flow velocity for the lid-driven cavity at $Re_L=7{,}000$ in the Lagrangian frame of reference. (a) LPOD modes $\Phi_1^{\mathcal{U}_1}$, $\Phi_2^{\mathcal{U}_1}$, $\Phi_8^{\mathcal{U}_1}$, and $\Phi_{17}^{\mathcal{U}_1}$. (b) LDMD modes $\phi_1^{\mathcal{U}_1}$, $\phi_2^{\mathcal{U}_1}$ and $\phi_3^{\mathcal{U}_1}$. (c) LPOD modal energies and (d) LDMD eigenvalues, indicating the values corresponding to the spatial modes.}
\label{fig:LDC_Re7k_lmodes}
\end{figure}

The leading LPOD and LDMD modes of the streamwise flow velocity for the steady lid-driven cavity at $Re_L=7{,}000$ are displayed in Figs.~\ref{fig:LDC_Re7k_lmodes}(a) and~\ref{fig:LDC_Re7k_lmodes}(b), while the corresponding modal energies of LPOD and modal frequencies of LDMD modes are shown in Figs.~\ref{fig:LDC_Re7k_lmodes}(c) and~\ref{fig:LDC_Re7k_lmodes}(d), respectively.
The energy dominant LPOD modes $\Phi_1^{\mathcal{U}_1}$ and $\Phi_2^{\mathcal{U}_1}$ are nearly identical to the real and imaginary components of the LDMD mode $\phi_1^{\mathcal{U}_1}$, respectively.
Notably, although the flow is steady in the Eulerian sense, the LDMD mode $\phi_1^{\mathcal{U}_1}$ corresponds to the unsteadiness of $St_L=0.133$, as shown in Fig.~\ref{fig:LDC_Re7k_lmodes}(d).
Furthermore, the LDMD modes $\phi_2^{\mathcal{U}_1}$ and $\phi_3^{\mathcal{U}_1}$ of Fig.~\ref{fig:LDC_Re7k_lmodes}(b) are associated with the Strouhal numbers $St_L=0.28$ and $St_L=0.41$, respectively, from the Lagrangian point of view.
The LPOD modes $\Phi_5^{\mathcal{U}_1}$ and $\Phi_7^{\mathcal{U}_1}$ (of Fig.~\ref{fig:LDC_Re7k_lmodes}a) correspond to the LDMD modes $\phi_2^{\mathcal{U}_1}$ and $\phi_3^{\mathcal{U}_1}$, respectively.

As discussed in Sec.~\ref{sec:LDC}, the lid-driven cavity flow becomes unstable beyond the critical Reynolds number of $Re_c \approx 10{,}500$; furthermore, at the post-critical Reynolds number of $Re_L=15{,}000$, the flow exhibits dominant frequency peaks at $St_L=0.13$, $St_L=0.25$, and $St_L=0.39$.
For the steady lid-driven cavity flow at $Re_L=7{,}000$, the LDMD modes at these frequencies closely resemble with the LDMD modes at $Re_L=15{,}000$ (Fig.~\ref{fig:LDC_Re15k_lmodes}b), which also represent energetically significant LPOD modes.
Importantly, the Lagrangian approach of modal decomposition provides a means to extract the dynamically significant flow features that are embedded in the steady base flow.
This feature of LMA is of significant importance from the base flow stability point of view.

The Eulerian to Lagrangian flow mapping of Eq.~\ref{eq:mapping} for the flow past a cylinder (of Sec.~\ref{sec:CYL}) needs a special treatment due to the convective/open nature of the flow; this is different from the lid-driven cavity flow (of Sec.~\ref{sec:LDC}), where the diffeomorphism $\textit{SDiff}(\mathcal{D})$ is always confined to the original Eulerian flow domain $\mathsf{D}$.
The time evolution of the diffeomorphism $\textit{SDiff}(\mathcal{D})$ for the flow past a cylinder is not confined to the original Eulerian flow domain $\mathsf{D}$ or the flow domain $\mathcal{D}$ of the identity map $\mathcal{M}(\pmb{\chi}_0,\tau_0)$.
Note that this applies to the unsteady lid-driven cavity and cylinder flows discussed in Sec.~\ref{sec:unsteady_flows}.
A simple remedy to this problem is to use the (Eulerian) flow conditions at the outlet boundary for the (Lagrangian) flow that has left the computational domain, and considering the diffeomorphism for a limited time for a meaningful analysis.

\begin{figure}
\begin{minipage}{0.5\textwidth}
\centering {\includegraphics[scale=0.09]{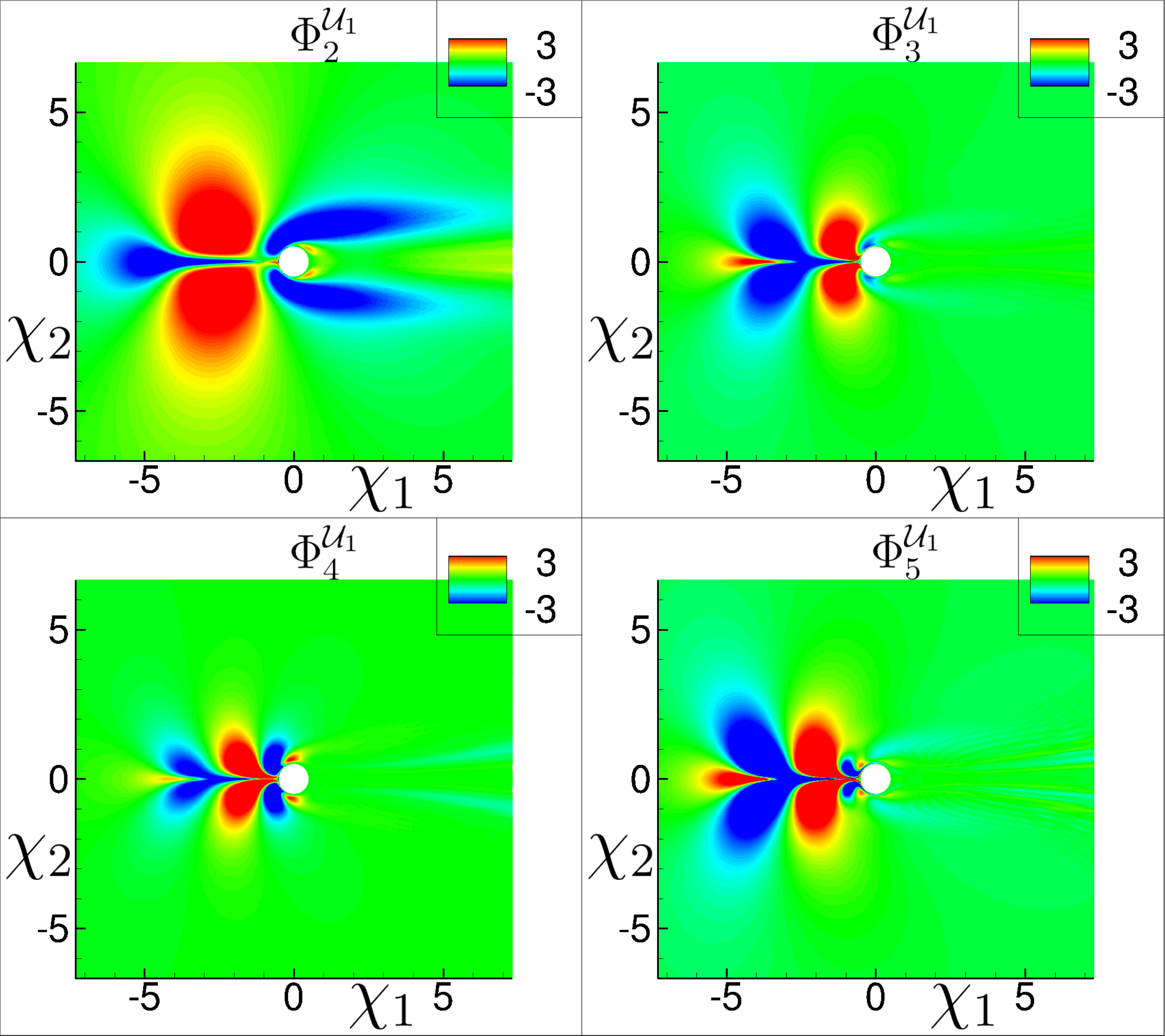}}\\(a) LPOD modes
\end{minipage}
\begin{minipage}{0.5\textwidth}
\centering {\includegraphics[scale=0.09]{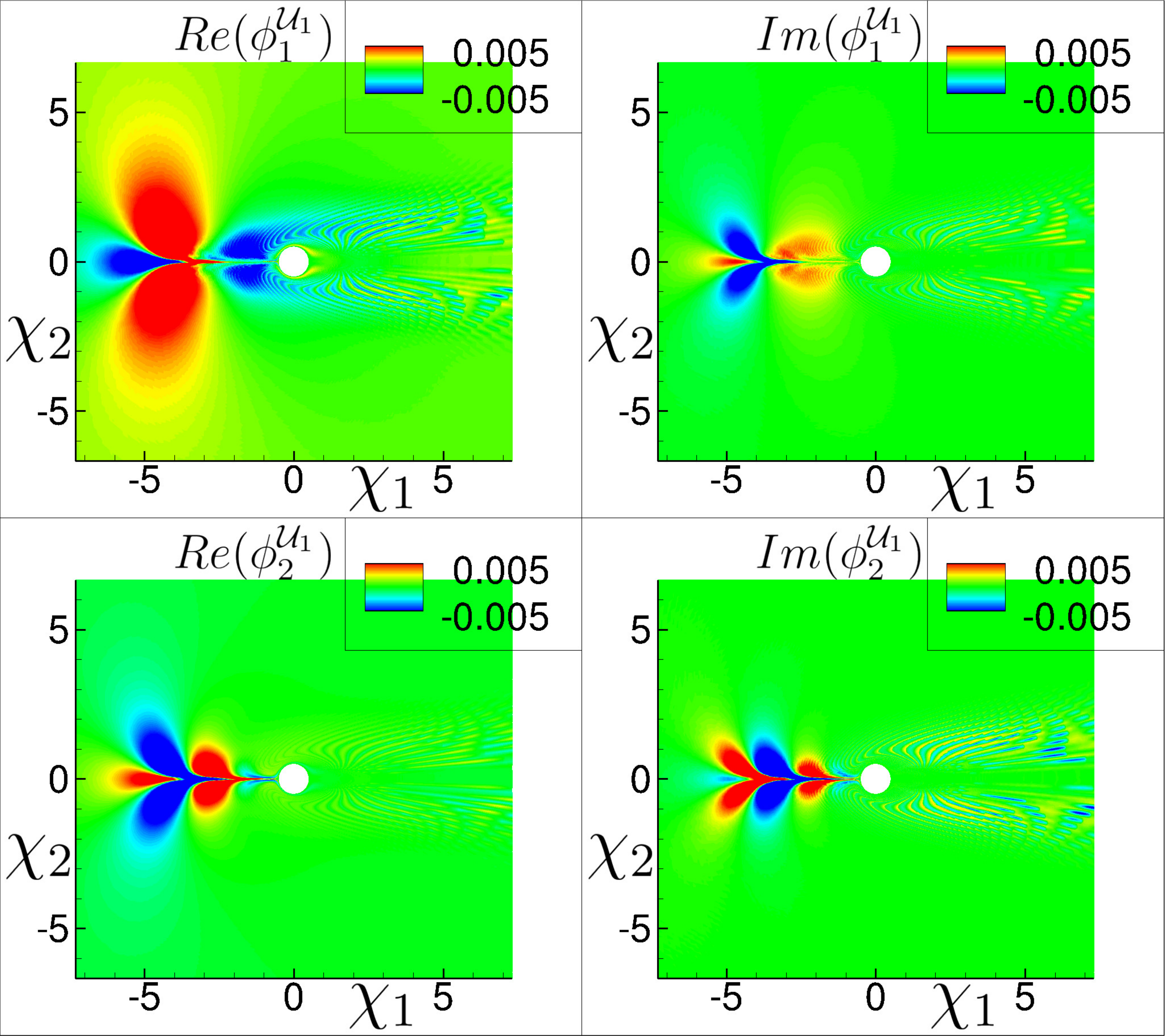}}\\(b) LDMD modes
\end{minipage}
\begin{minipage}{0.5\textwidth}
\centering {\includegraphics[scale=0.8]{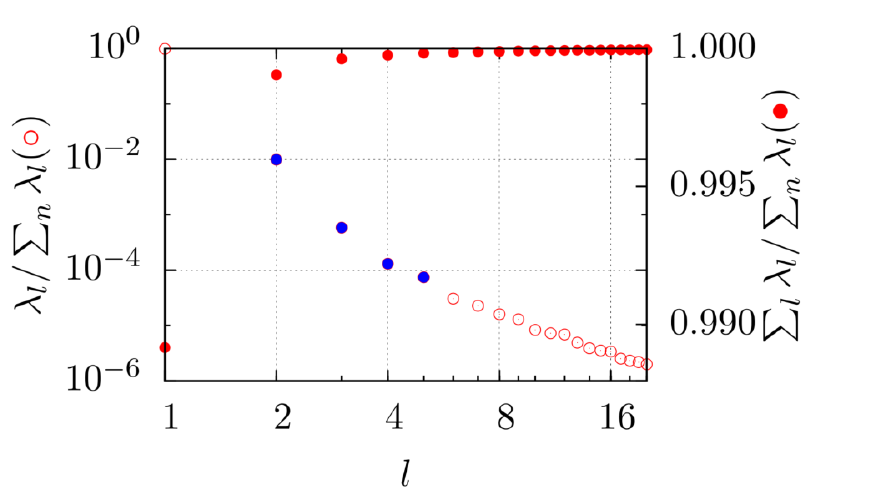}}\\(c) LPOD eigenvalues
\end{minipage}
\begin{minipage}{0.5\textwidth}
\centering {\includegraphics[scale=0.7]{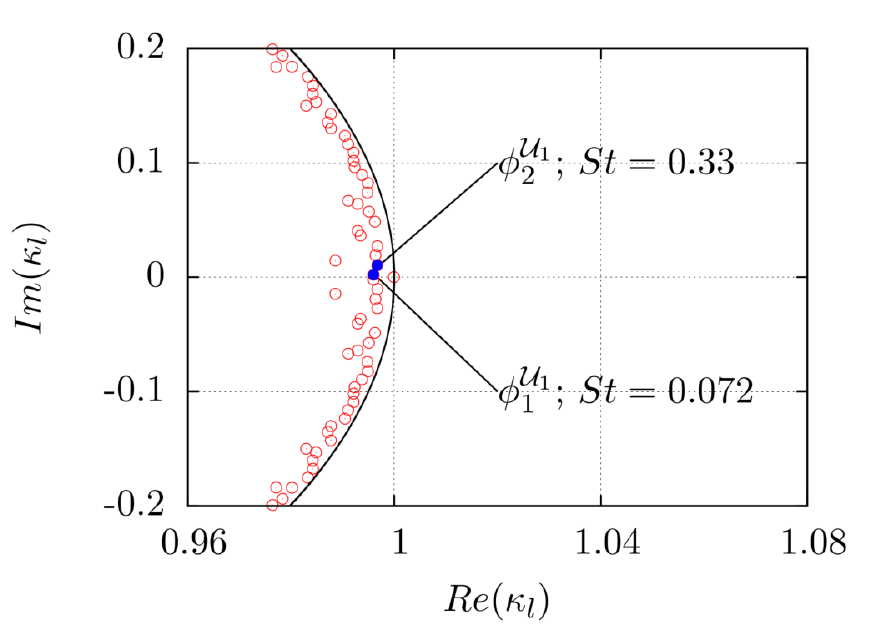}}\\(d) LDMD eigenvalues
\end{minipage}
\caption{Leading modes of the steady state streamwise flow velocity for the flow past a cylinder at $Re_D=40$ in the Lagrangian frame of reference. (a) LPOD modes $\Phi_2^{\mathcal{U}_1}$, $\Phi_3^{\mathcal{U}_1}$, $\Phi_4^{\mathcal{U}_1}$, and $\Phi_5^{\mathcal{U}_1}$. (b) LDMD modes $\phi_1^{\mathcal{U}_1}$ and $\phi_2^{\mathcal{U}_1}$. (c) LPOD modal energies and (d) LDMD eigenvalues, indicating the values corresponding to the spatial modes.}
\label{fig:CYL_Re40_lmodes}
\end{figure}

The LMA is also performed for the steady flow past a cylinder at $Re_D=40$.
The steady base flow serves as the reference state for the Lagrangian mapping of Eq.~\ref{eq:mapping}, while Eq.~\ref{eq:M=M-1} and Eq.~\ref{eq:ul-ue} are used to extract the time-dependent Lagrangian flow fields. 
$1{,}000$ snapshots are collected at time intervals of $0.005$, and the Lagrangian flow  map is again obtained at a much smaller time step of $\delta t=0.001$ for accuracy.
Thus, the finite-time duration for the LMA is $\tau\in[0,\mathcal{T}=5]$, same as for the LMA on unsteady flow past cylinder at $Re_D=100$ discussed before in Sec.~\ref{sec:unsteady_flows}.
The significant LPOD and LDMD modes of the streamwise flow velocity are depicted in Figs.~\ref{fig:CYL_Re40_lmodes}(a) and~\ref{fig:CYL_Re40_lmodes}(b) with their energy and frequency content in Figs.~\ref{fig:CYL_Re40_lmodes}(c) and~\ref{fig:CYL_Re40_lmodes}(d), respectively.
Figure~\ref{fig:CYL_Re40_lmodes}(b) displays the real and imaginary parts of the LDMD modes $\phi_1^{\mathcal{U}_1}$ and $\phi_2^{\mathcal{U}_1}$ that exhibit Strouhal numbers of $St_D=0.072$ and $St_D=0.33$, respectively.
The LPOD mode $\Phi_2^{\mathcal{U}_1}$ resembles the LDMD mode $\phi_2^{\mathcal{U}_1}$, which is associated with the Strouhal number of $0.072$; furthermore, the higher LPOD modes ($\Phi_3^{\mathcal{U}_1}$, $\Phi_4^{\mathcal{U}_1}$, and $\Phi_5^{\mathcal{U}_1}$), in general, show the structural similarity with the LDMD modes.

The two-dimensional flow past a cylinder in the post-critical regime ($50 \lessapprox Re_D \lessapprox 190$) exhibits a distinct unsteadiness at Strouhal number of $St_D=0.16$, which is associated with the shedding of vortices in the wake region (refer Sec.~\ref{sec:CYL}).
The LDMD modes $\phi_1^{\mathcal{U}_1}$ and $\phi_2^{\mathcal{U}_1}$ exhibit Strouhal numbers of $St_D=0.072$ and $St_D=0.33$, while the LDMD modes of the flow normal velocity component $\phi_1^{\mathcal{U}_2}$ and $\phi_2^{\mathcal{U}_2}$ (not shown) exhibit Strouhal numbers of $St_D=0.06$ and $St_D=0.18$.
The LPOD mode $\Phi_2^{\mathcal{U}_1}$ of Fig.~\ref{fig:CYL_Re40_lmodes}(a) and LDMD mode $Re(\phi_2^{\mathcal{U}_1})$ of Fig.~\ref{fig:CYL_Re40_lmodes}(a) exhibit structural similarities with the LPOD mode of Fig.~\ref{fig:CYL_Re100_lmodes_TabsU}(b) and LDMD mode of Fig.~\ref{fig:CYL_Re100_lmodes_TabsU}(d), respectively, particularly in the upstream region of the cylinder.
Note that the steady case ($Re_D=40$; Fig.~\ref{fig:CYL_Re40_lmodes}) displays LMA of the streamwise velocity, whereas the unsteady case ($Re_D=100$; Fig.~\ref{fig:CYL_Re100_lmodes_TabsU}) displays LMA of the absolute velocity.
Nonetheless, the breakdown of the modal symmetry about $\chi_2=0$ for increased Reynolds number is evident, particularly in the wake region, clearly manifesting the unsteady effects of the von Karman shedding in terms of the considered flow maps (material surfaces).
Importantly, the LMA of the steady base flow can extract flow features that, in general, correspond to the unsteady flow dynamics from the post-bifurcation regime.

The LPOD and LDMD modes (of Fig.~\ref{fig:CYL_Re40_lmodes}(a) and~\ref{fig:CYL_Re40_lmodes}(b), respectively) are displayed on the reference flow state or identity map, which is the initial condition for the Lagrangian flow map.
Interestingly, the LPOD and LDMD modes indicate upstream regions of the cylinder that correspond to the unsteady flow dynamics associated with the downstream/wake region of the cylinder.
In other words, the procedure tracks the Lagrangian flow map in the future time/space and then traces back the flow variations to the identity map in terms of the modal decomposition.
Alternatively, the analysis can be performed by reversing the time direction, in which the Lagrangian flow map is traced in prior time/space and the flow variations can then be tracked forward to the identity map.
The latter approach is referred to as the adjoint Lagrangian approach of modal decomposition, which we discuss in the following section (Sec.~\ref{sec:adjoint}).

\subsubsection{Adjoint Lagrangian approach} \label{sec:adjoint}
Adjoint-based analyses find utility in the study of flow receptivity, sensitivity, and stability; in addition, these procedures are employed  in design and optimization~\citep{hill1995adjoint,schmid2007nonmodal,luchini2014adjoint,browne2014sensitivity,iorio2014direct}.
Briefly, the adjoint equations of a linear or nonlinear system of equations are effectively solved by reversing the direction of time~\citep{chandler2012adjoint}.
In the similar manner, the Lagrangian flow map of Eq.~\ref{eq:mapping} can be obtained for negative direction of time $t$ as:
\begin{eqnarray} \label{eq:mapping_adjoint}
\mathcal{M}:\mathcal{D}\times [0,\mathcal{T}] \rightarrow \textit{SDiff}(\mathcal{D})\subseteq \mathsf{E}=\mathbb{R}^3: (\pmb{\chi},\tau) \mapsto \mathcal{M}(\pmb{\chi},\tau)=(\pmb{x},-t), \text{ and }  \nonumber \\
\mathcal{M}(\pmb{\chi}_0,\tau_0)=\text{identity map},
\end{eqnarray}
where the identity map is now the terminal condition, from the Eulerian point of view.
In addition, Eqs.~\ref{eq:M=M-1} and~\ref{eq:ul-ue} also consider the negative time direction, while the Eulerian flow fields are transformed to the Lagrangian flow fields.

{
In matrix form, let $\pmb{X}\in \mathbb{R}^{m\times n}$ and $\pmb{Y}\in \mathbb{R}^{m\times n}$ be the discrete sets of Lagrangian flow fields that are gathered in the forward and backward time direction.
The real symmetric tensors $\pmb{X}^T\pmb{X}$ and $\pmb{Y}^T\pmb{Y}$ are self-adjoint by definition.
For reversed initial and terminal conditions, the corresponding eigenvalue problems for the two tensors result in an identical set of orthonormal eigenfunctions ($\pmb{\Phi}$) and real eigenvalues ($\pmb{\Lambda}$), indicating that the two symmetric tensors are adjoint and normal~\citep{chandler2012adjoint}.
Following the LPOD/LDMD framework (of Sec.~\ref{sec:theory}), the matrices can be expressed as,
\begin{equation}
\pmb{X}=\pmb{\Phi}\pmb{\Lambda}^{\frac{1}{2}}\pmb{\Psi}_{\textsc{x}}^T \hspace{2mm} \text{and} \hspace{2mm} \pmb{Y}=\pmb{\Phi}\pmb{\Lambda}^{\frac{1}{2}}\pmb{\Psi}_{\textsc{y}}^T,
\end{equation}
which leads to
\begin{equation}
\pmb{X}\pmb{\Psi}_{\textsc{x}}=\pmb{Y}\pmb{\Psi}_{\textsc{y}},
\end{equation}
satisfying an adjoint formalism
\begin{equation}
\pmb{X}=\pmb{Y}\pmb{\Psi}_\textsc{y}\pmb{\Psi}_\textsc{x}^T \hspace{2mm} \text{and} \hspace{2mm}
\pmb{Y}=\pmb{X}\pmb{\Psi}_\textsc{x}\pmb{\Psi}_\textsc{y}^T
\end{equation}
with an adjoint operator $\left(\pmb{\Psi}_\textsc{y}\pmb{\Psi}_\textsc{x}^T\right) \in \mathbb{R}^{n\times n}$.
However, for the same initial condition (or the identity map), the data matrices $\pmb{X}$ and $\pmb{Y}$ may not lead to the exact same set of eigenfunctions and eigenvalues due to non-normality of the flow fields with advection~\citep{chomaz2005global,marquet2009direct}; on the other hand, the forward and backward data matrices can lead to normal operators depending on flow symmetries, for instance those in the absolutely unstable flows~\citep{symon2018non}.
}

To compute the backward/adjoint modes, similar to the forward Lagrangian modal decomposition of the flow past a cylinder in Sec.~\ref{sec:forward},  $1{,}000$ snapshots are collected at time intervals of $0.01$, \textit{i.e.}, $-0.01$ from the Eulerian point of view, while the Lagrangian flow map evolves with a finer time step of $\delta t =0.001$ for accuracy.
\begin{figure}
\begin{minipage}{0.5\textwidth}
\centering {\includegraphics[scale=0.09]{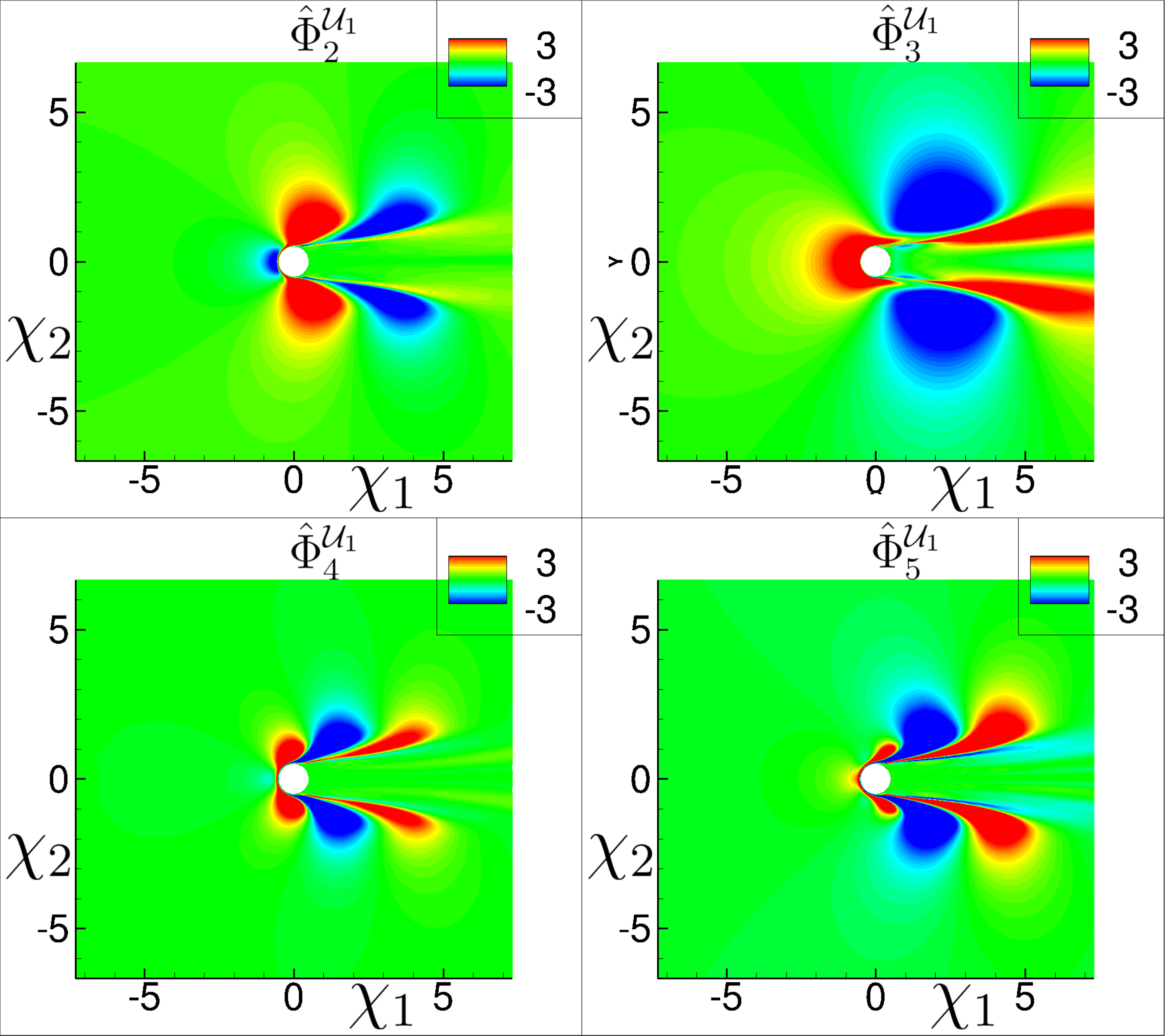}}\\(a) Adjoint LPOD modes
\end{minipage}
\begin{minipage}{0.5\textwidth}
\centering {\includegraphics[scale=0.09]{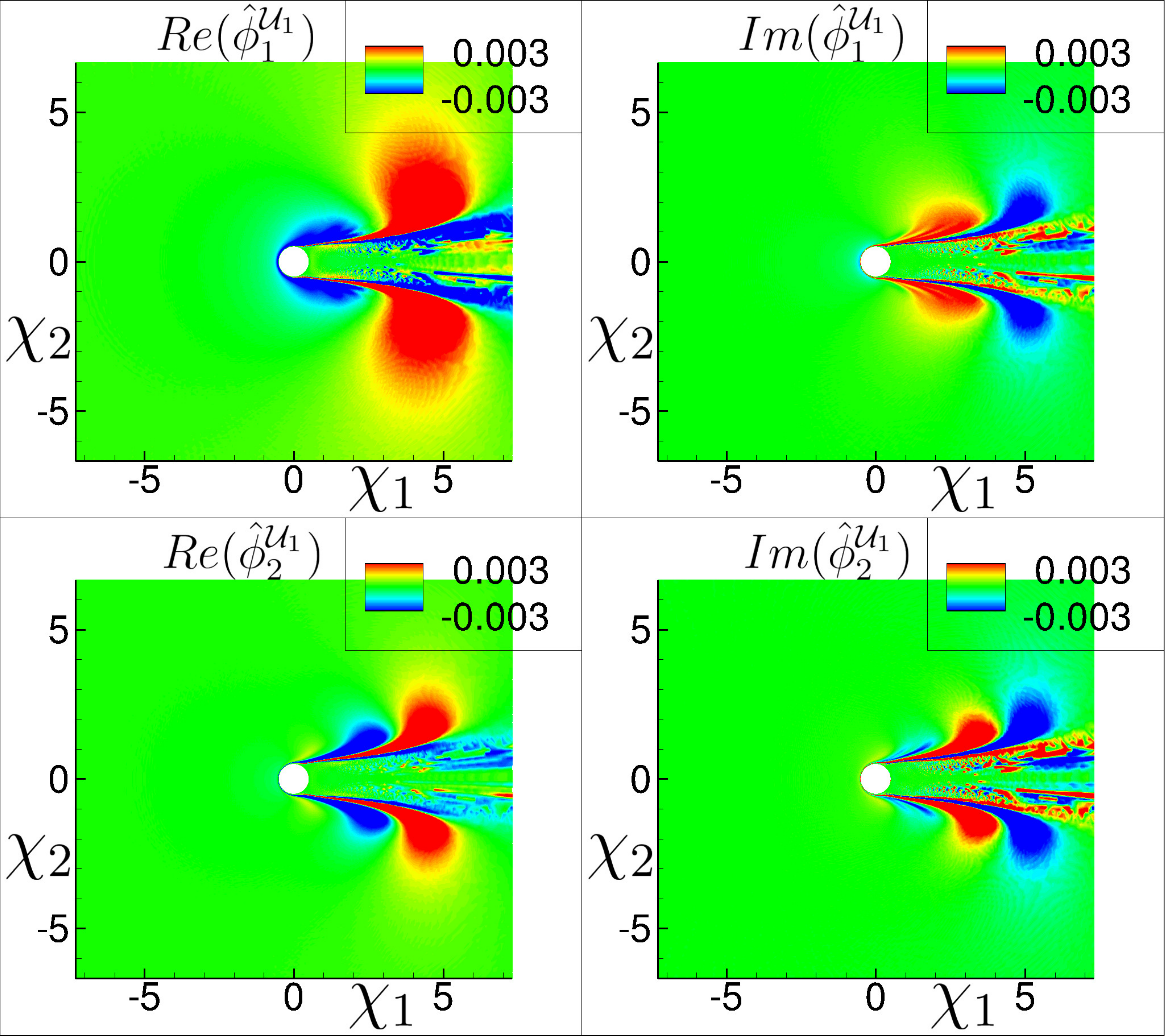}}\\(b) Adjoint LDMD modes
\end{minipage}
\begin{minipage}{0.5\textwidth}
\centering {\includegraphics[scale=0.8]{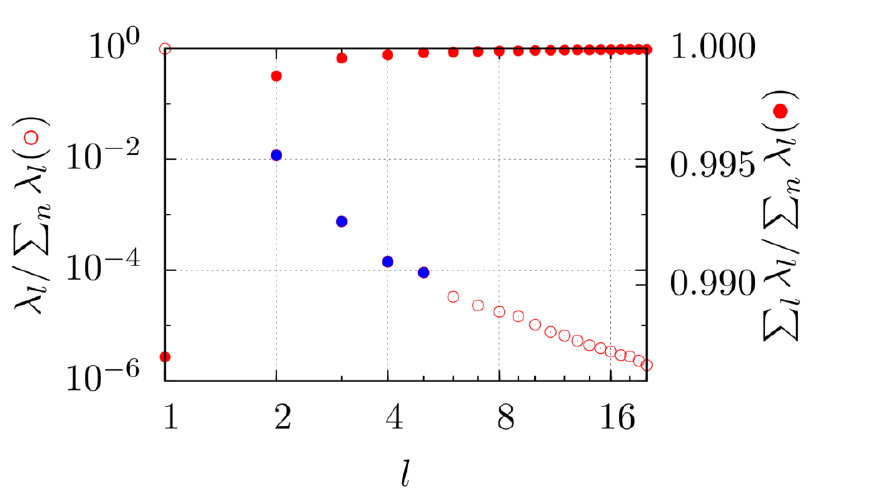}}\\(c) Adjoint LPOD eigenvalues
\end{minipage}
\begin{minipage}{0.5\textwidth}
\centering {\includegraphics[scale=0.7]{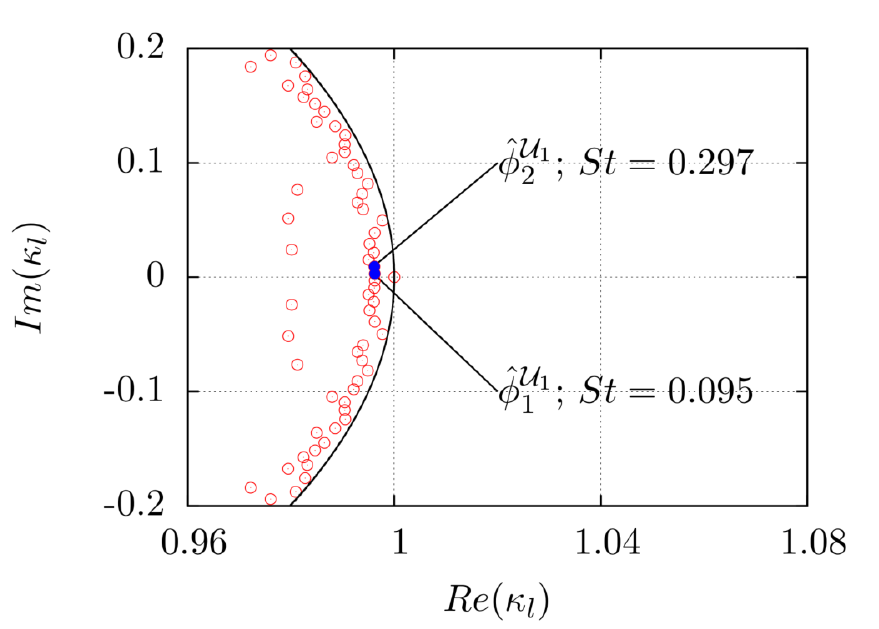}}\\(d) Adjoint LDMD eigenvalues
\end{minipage}
\caption{Leading \underline{adjoint} modes of the steady state streamwise flow velocity for the flow past a cylinder at $Re_D=40$ in the Lagrangian frame of reference. (a) Adjoint LPOD modes $\hat{\Phi}_2^{\mathcal{U}_1}$, $\hat{\Phi}_3^{\mathcal{U}_1}$, $\hat{\Phi}_4^{\mathcal{U}_1}$, and $\hat{\Phi}_5^{\mathcal{U}_1}$. (b) Adjoint LDMD modes $\hat{\phi}_1^{\mathcal{U}_1}$ and $\hat{\phi}_2^{\mathcal{U}_1}$. (c) Adjoint LPOD modal energies and (d) adjoint LDMD eigenvalues, indicating frequencies associated with the spatial modes.}
\label{fig:CYL_Re40_almodes}
\end{figure}
The leading adjoint LPOD and LDMD modes of the streamwise flow velocity are shown in Figs.~\ref{fig:CYL_Re40_almodes}(a) and~\ref{fig:CYL_Re40_almodes}(b) together with their energy and frequency content in Figs.~\ref{fig:CYL_Re40_almodes}(c) and~\ref{fig:CYL_Re40_almodes}(d), respectively.
As opposed to the forward Lagrangian modes of Figs.~\ref{fig:CYL_Re40_lmodes}(a) and~\ref{fig:CYL_Re40_lmodes}(b), the adjoint modes appear in the downstream region of the flow.
These  modes indicate synchronized regions of the flow (\textit{i.e}, the streamwise velocity), which were subjected to similar variations of the magnitude and wavelength over the considered time in the past.
Like for the forward Lagrangian modes (of Fig.~\ref{fig:CYL_Re40_lmodes}), the adjoint modes are symmetric about $\chi_2=0$ line, mainly indicating the shear layer regions of the flow.
Notably, similar to the forward LDMD, the adjoint LDMD modes $\hat{\phi}_1^{\mathcal{U}_1}$ and $\hat{\phi}_2^{\mathcal{U}_1}$ of Fig.~\ref{fig:CYL_Re40_almodes}(b), exhibit unsteadiness at Strouhal numbers of $St_D=0.095$ and $St_D=0.297$, respectively, relevant to the post-bifurcation dynamics.

\subsection{Double-gyre flow} \label{sec:gyre}
As noted earlier, the (maximum) finite time Lyapunov exponent is extensively used in understanding the flow chaos, mixing and transport, which play an important role in the geophysical flows~\citep{garaboa2015lagrangian,bozorgmagham2015atmospheric,allshouse2015lagrangian,d2004mixing}; in addition, the FTLE finds application in biological~\citep{green2010using,shadden2015lagrangian} and several industrial flows~\citep{dauch2019analyzing,gonzalez2016finite}.
To elaborate on the relation between the FTLE and LMA, we consider a double-gyre flow pattern of interest in geophysical flows.
As presented by~\cite{shadden2005definition}, a simple potential flow leads to an oscillating double-gyre flow pattern for a non-zero perturbation parameter, exhibiting an engagement between the classical stable/unstable manifolds~\citep{guckenheimer2013nonlinear,rom1990analytical}.

The periodically varying double-gyre flow is described by a stream function, in Eulerian frame of reference, as
\begin{equation}\label{eq:streamF}
\uppsi(x_1,x_2,t) = \textsc{a} \sin(\pi f(x_1,t)) \sin(\pi x_2),
\end{equation}
where
\begin{eqnarray}
f(x_1,t) &=& a(t) x_1^2 + b(t) x_1  \\ 
a(t) &=& \epsilon \sin(2\pi St_f t) \\
b(t) &=& 1 - 2\epsilon \sin(2\pi St_f t)
\end{eqnarray}
over a domain $\mathsf{D} \times [0,\mathsf{T}]$ such that $x_1 \in [0,2]$, $x_2\in [0,1]$ and $t\in[0,15]$.
This simple mathematical model produces two counter-rotating flow cells.
A non-zero value of the parameter $\epsilon$ leads to the time-dependent double-gyre flow~\citep{shadden2005definition}.
The two components of the velocity field are
\begin{eqnarray} \label{eq:gyre_uv}
u_1 &=& -\frac{\partial \uppsi}{\partial x_2} = -\pi \textsc{a} \sin(\pi f(x_1,t)) \cos(\pi x_2), \\
u_2 &=& \frac{\partial \uppsi}{\partial x_1} = \pi \textsc{a} \cos(\pi f(x_1,t)) \frac{d f(x_1,t)}{dx_1} \sin(\pi x_2).
\end{eqnarray}
The parameter $\textsc{a}$ determines the magnitude of velocity, while $St$ is the Strouhal number associated with the flow unsteadiness.
The Eulerian flow fields are generated for $\textsc{a}=0.1$, $St=0.1$, and $\epsilon=0.1$ over the computational domain discretized by $n_{x_1}\times n_{x_2} = 601 \times 301$ grid points, while the time-step value of $\delta t=0.05$ results in a total number of $301$ snapshots.

\begin{figure}
\begin{minipage}{0.3\textwidth}
\centering \includegraphics[scale=0.9]{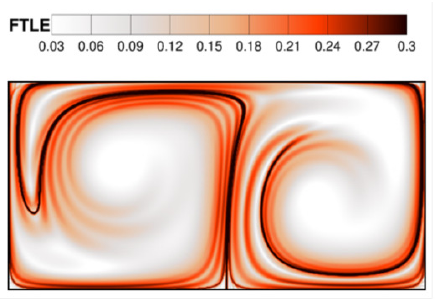}\\(a) Forward FTLE \\\includegraphics[scale=0.9]{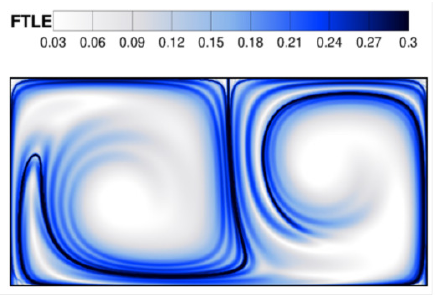}\\(b) Backward FTLE
\end{minipage}
\begin{minipage}{0.34\textwidth}
\centering \includegraphics[scale=0.064]{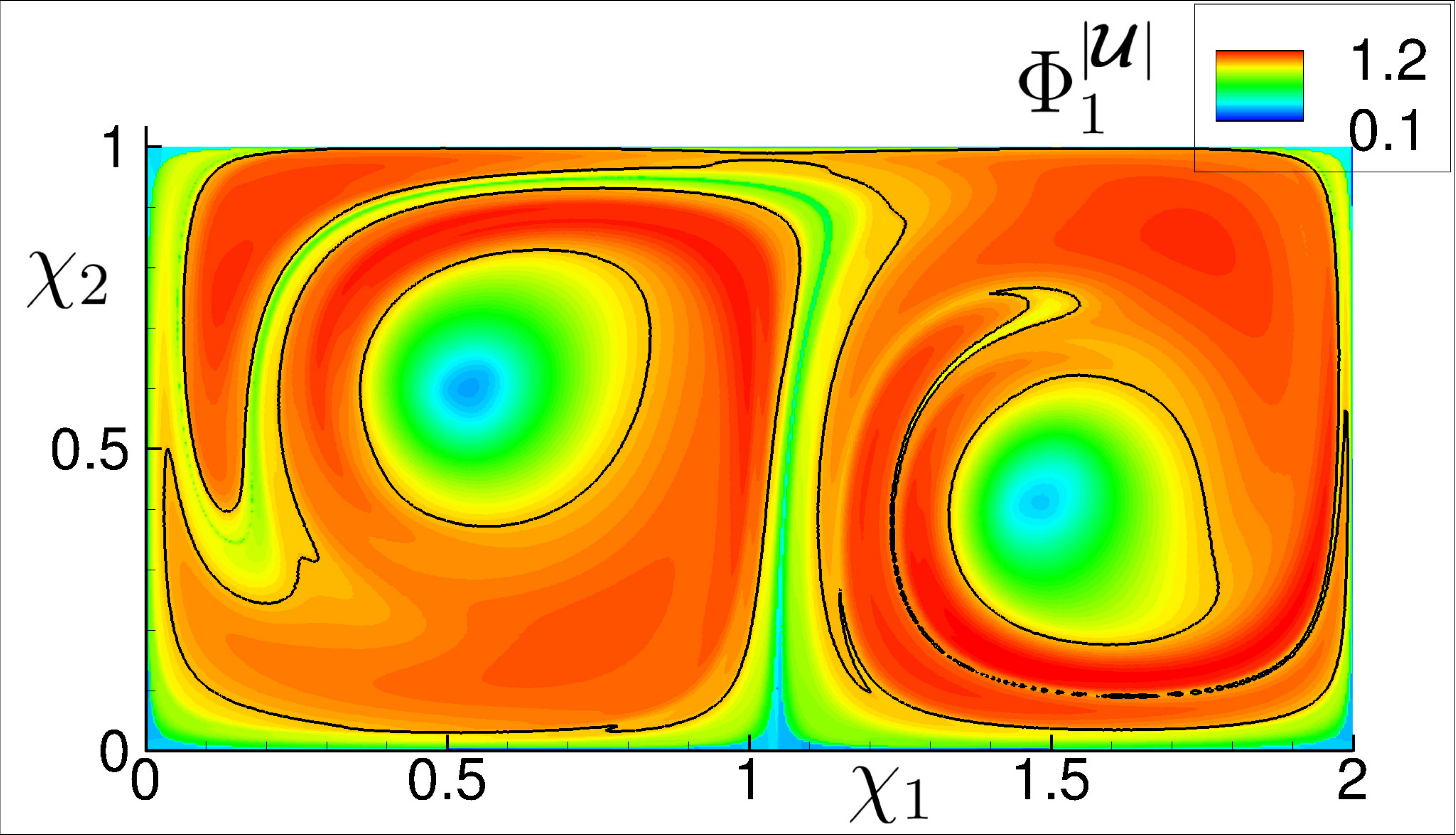}\\(c) LPOD mode\\\includegraphics[scale=0.064]{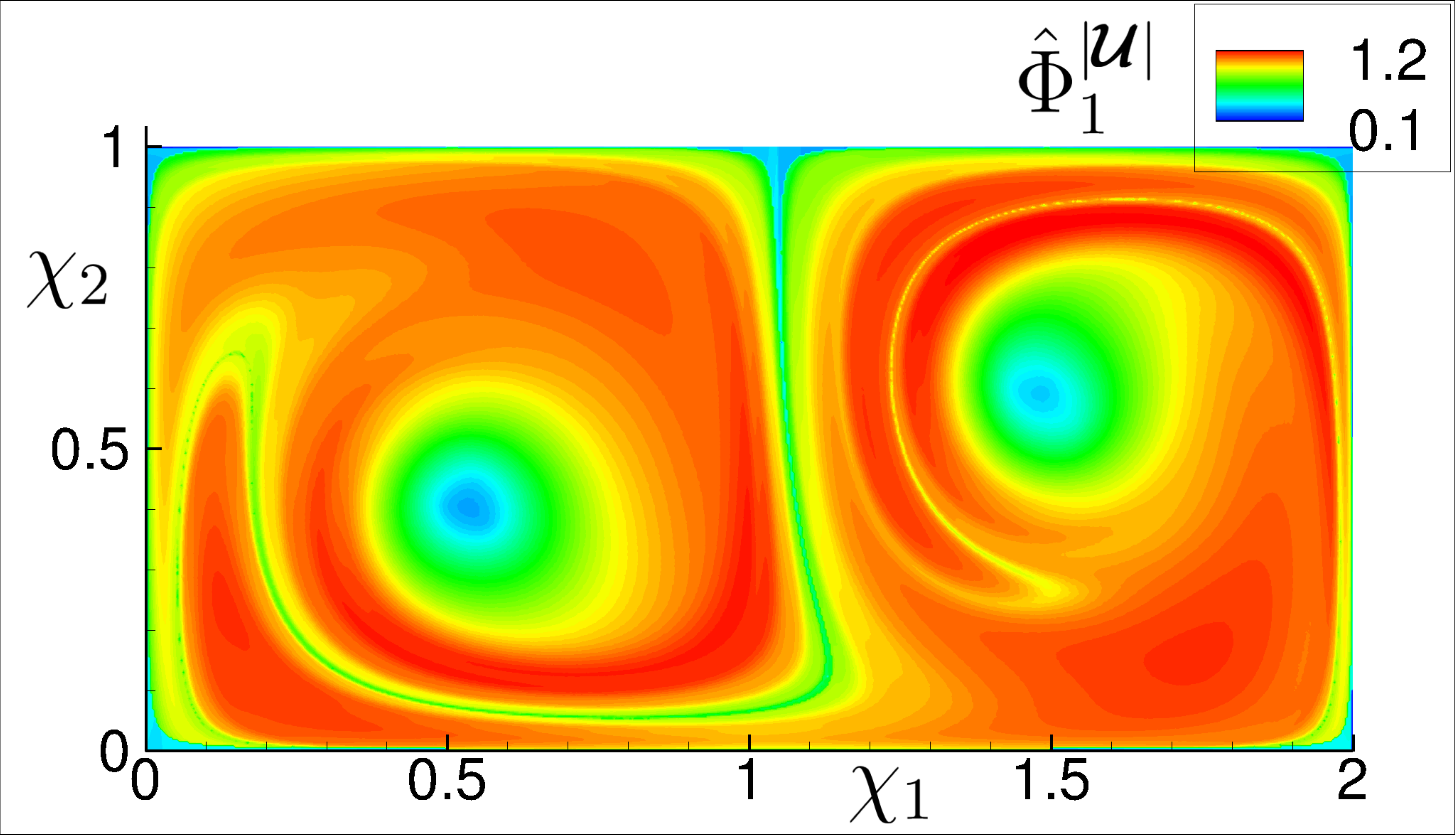}\\(d) Adjoint LPOD mode
\end{minipage}
\begin{minipage}{0.34\textwidth}
\centering \includegraphics[scale=0.064]{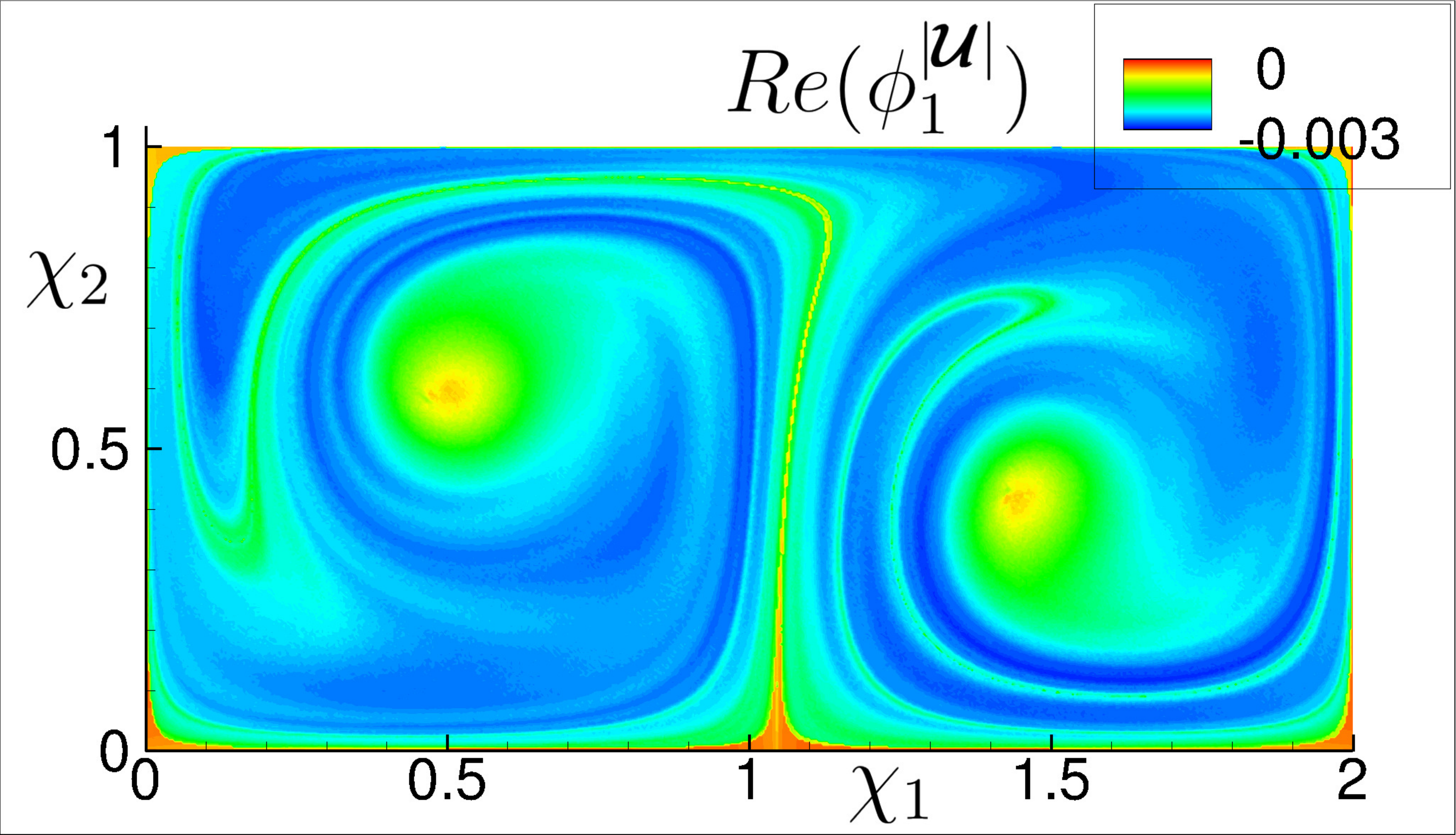}\\(e) LDMD mode\\\includegraphics[scale=0.064]{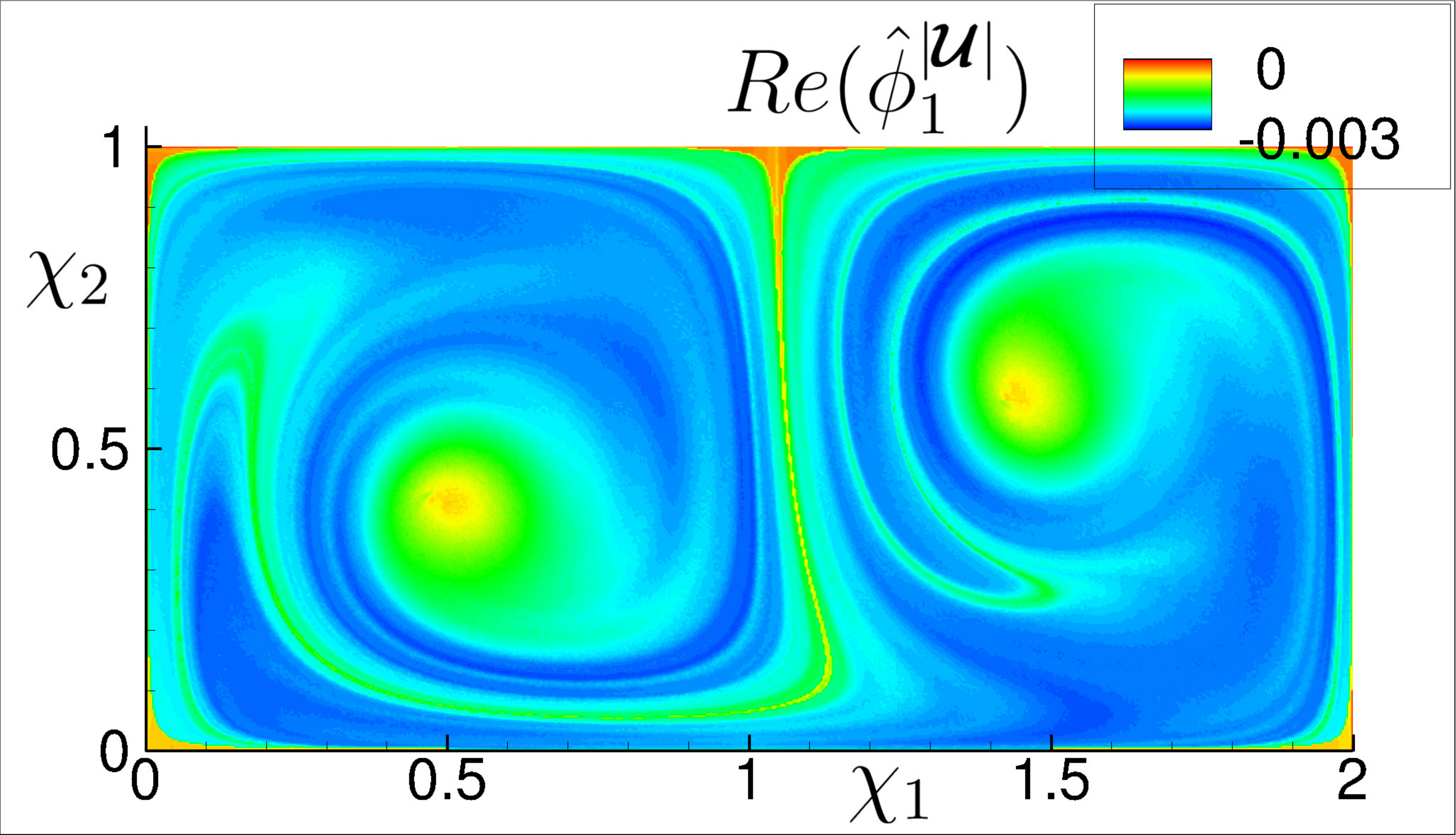}\\(f) Adjoint LDMD mode
\end{minipage}
\caption{Comparison between the FTLE, LPOD and LDMD modes of the double-gyre flow. (a) Forward FTLE and (b) backward FTLE reference data are from~\cite{finn2013integrated}. The first (c) forward (${\Phi}_1^{|\pmb{\mathcal{U}}|}$ with a single contour level of value $1$) and (d) adjoint ($\hat{\Phi}_1^{|\pmb{\mathcal{U}}|}$) LPOD modes as well as the only stationary (c) forward (${\phi}_1^{|\pmb{\mathcal{U}}|}$) and (b) adjoint ($\hat{\phi}_1^{|\pmb{\mathcal{U}}|}$) LDMD modes for the double-gyre flow.}
\label{fig:ftle}
\end{figure}

The FTLE fields for the double-gyre flow in the forward and backward time flow maps are as shown in Fig.~\ref{fig:ftle}(a) and (b), respectively, which are reproduced here from~\cite{finn2013integrated}.
To perform the Lagrangian modal analysis of this flow, the Eulerian snapshots are transformed to a Lagrangian frame of reference.
The LPOD and LDMD modes are then obtained for both forward and backward Lagrangian flow maps.
The first LPOD and adjoint LPOD modes, which correspond to the maximum eigenvalues, are displayed in Fig.~\ref{fig:ftle}(c) and (d), respectively, while the LDMD and adjoint LDMD modes that are stationary (in the finite-time Lagrangian sense; $St=0$) are shown in Fig.~\ref{fig:ftle}(e) and (f), respectively.
The FTLE field, which highlights the maximum eigenvalues of the right Cauchy-Green strain tensor, is expected to be geometrically similar to the first LPOD mode (see Sec.~\ref{sec:LE}), which corresponds to the largest eigenvalue.
The resemblance between the first LPOD modes (Fig.~\ref{fig:ftle}c and d) and the FTLE fields (Fig.~\ref{fig:ftle}a and b) is evident in Fig.~\ref{fig:ftle}, in both the forward and backward time directions.
In the finite-time Lagrangian sense, the first LPOD modes correspond to the mean field of the absolute velocity, while the corresponding LDMD modes display no associated unsteadiness ($St=0$), as shown in Fig.~\ref{fig:ftle}(e) and (f).
The LDMD and adjoint LDMD modes closely match the LPOD and adjoint LPOD modes, exhibiting geometrical similarities with the forward and backward FTLE fields, respectively.

\begin{figure}
\begin{minipage}{0.475\textwidth}
\centering \includegraphics[scale=0.8]{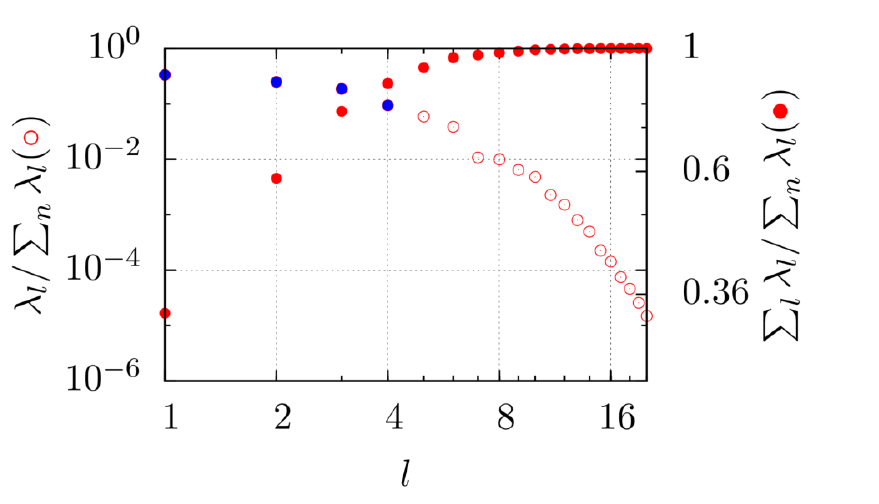}\\(a) LPOD eigenvalues
\end{minipage}
\begin{minipage}{0.475\textwidth}
\centering \includegraphics[scale=0.7]{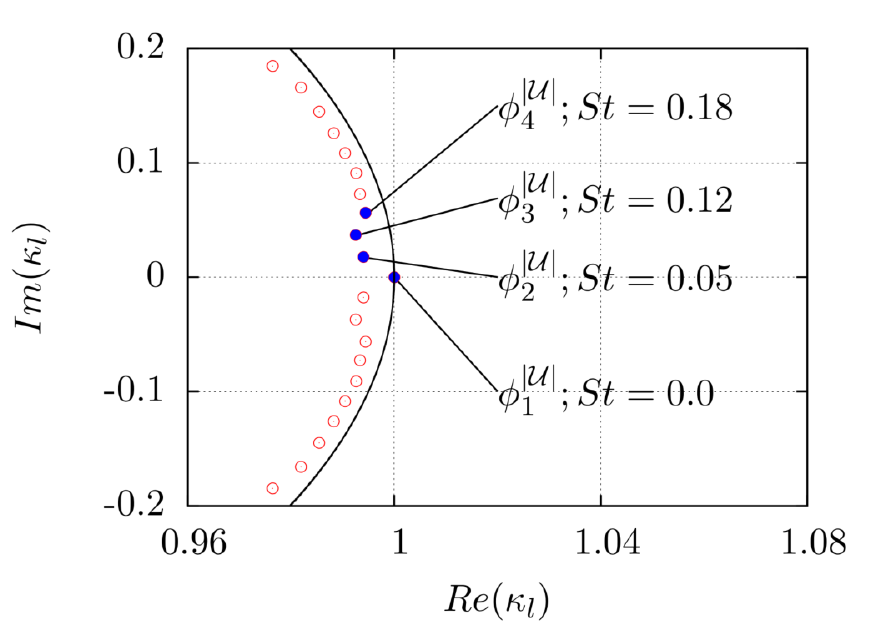}\\(b) LDMD eigenvaues
\end{minipage}
\begin{minipage}{0.325\textwidth}
\centering \includegraphics[scale=0.062]{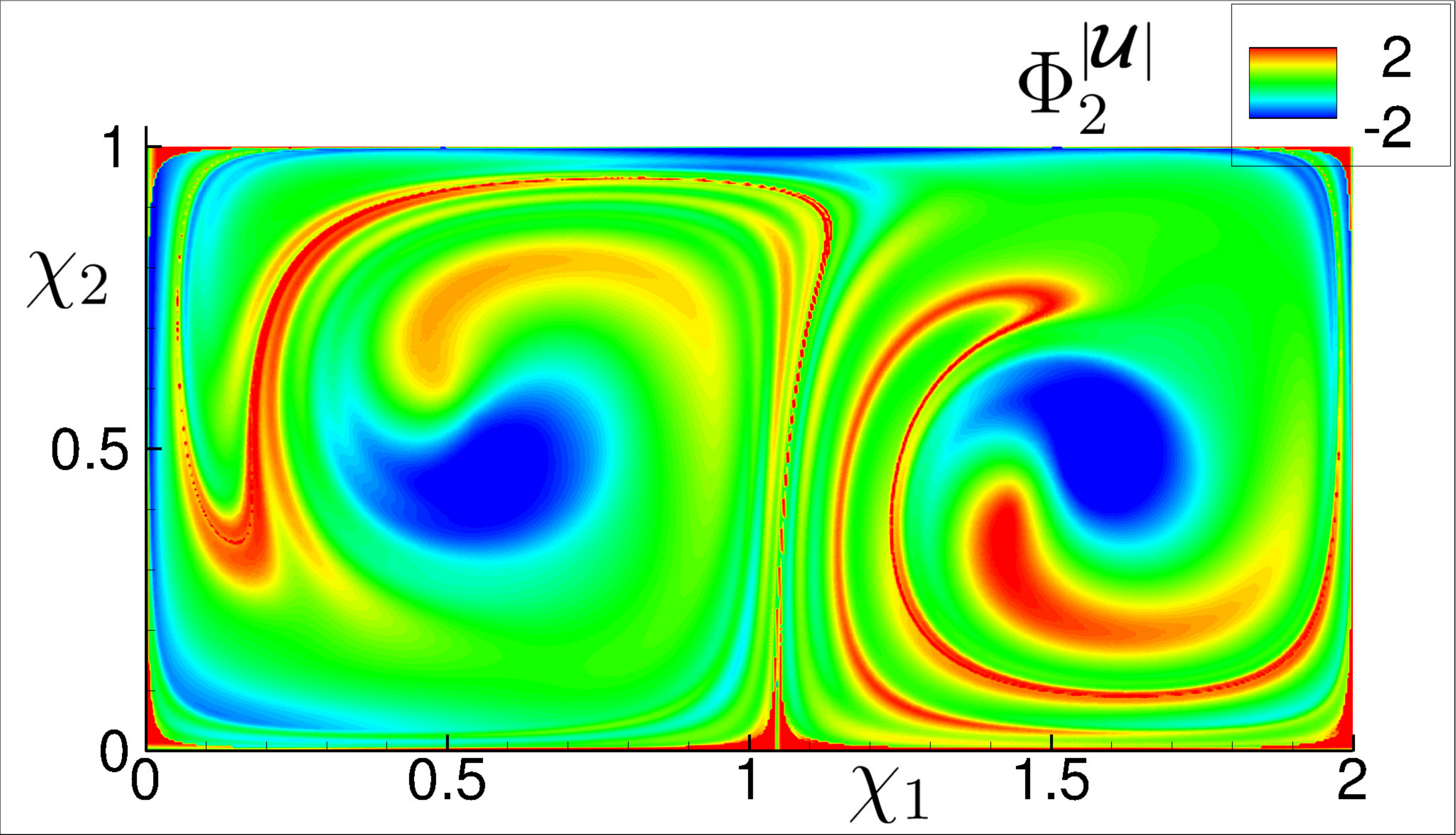}\\(c) LPOD modes\\\includegraphics[scale=0.062]{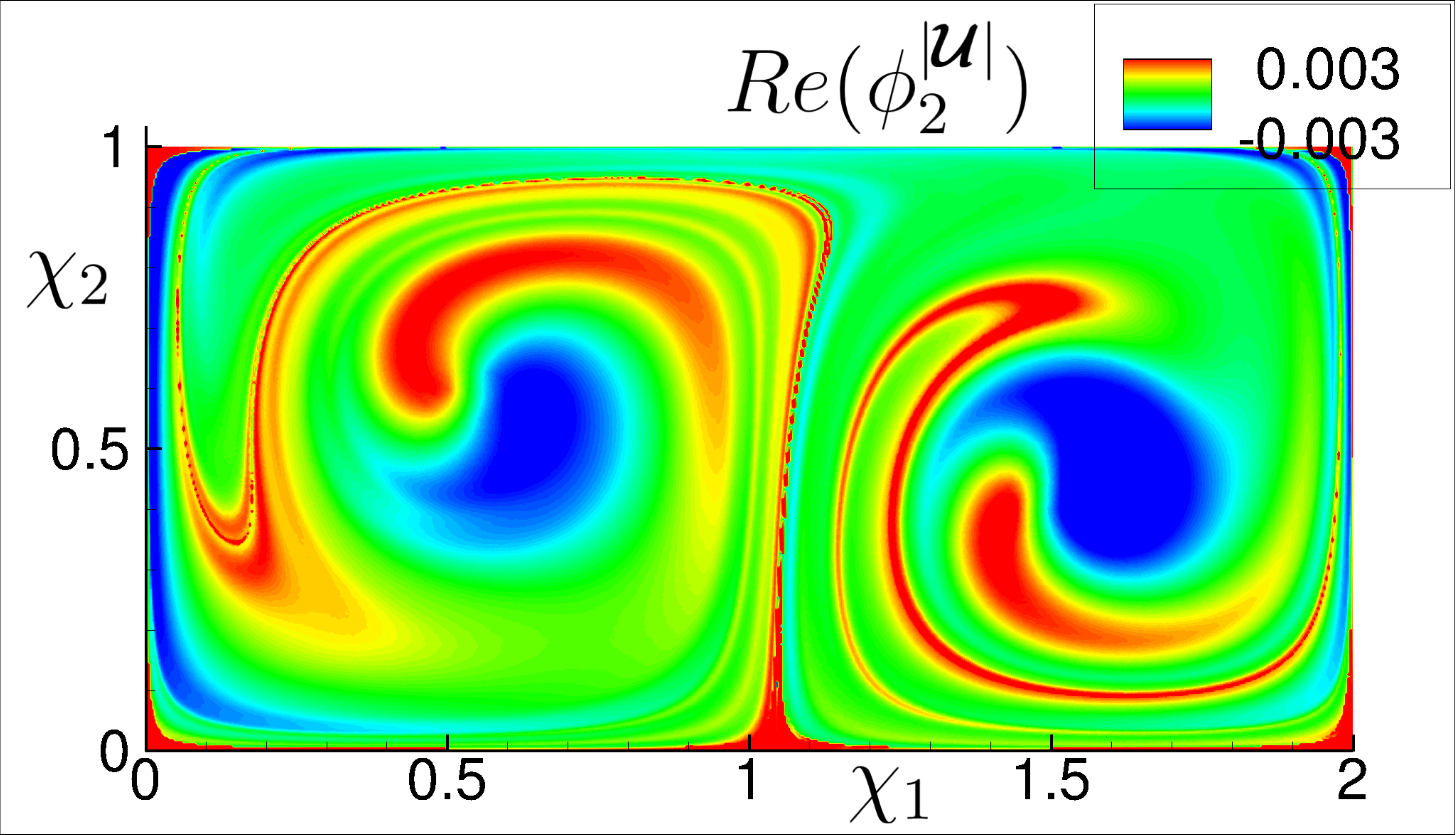}\\(d) LDMD mode
\end{minipage}
\begin{minipage}{0.325\textwidth}
\centering \includegraphics[scale=0.062]{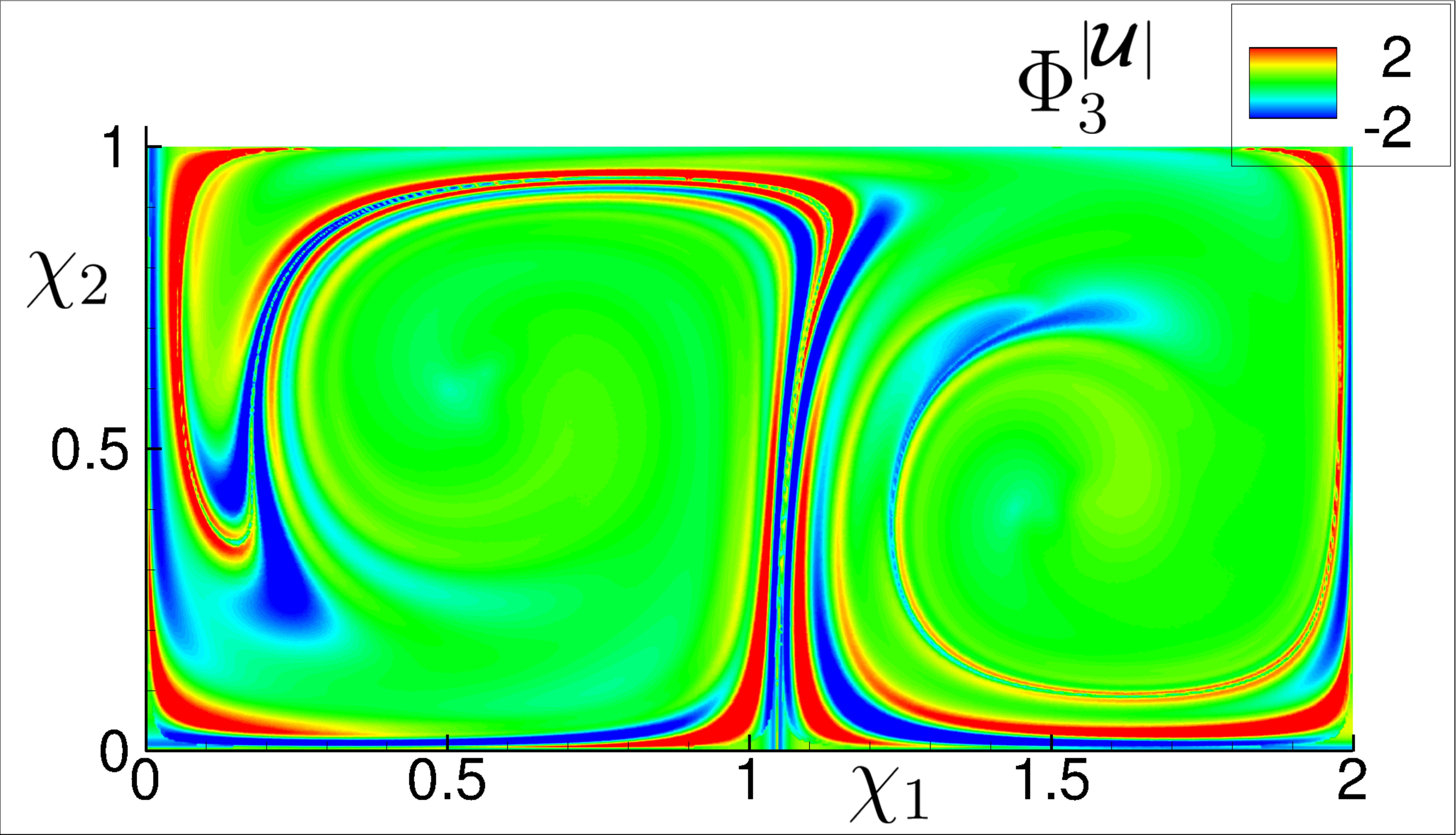}\\(e) LPOD mode\\\includegraphics[scale=0.062]{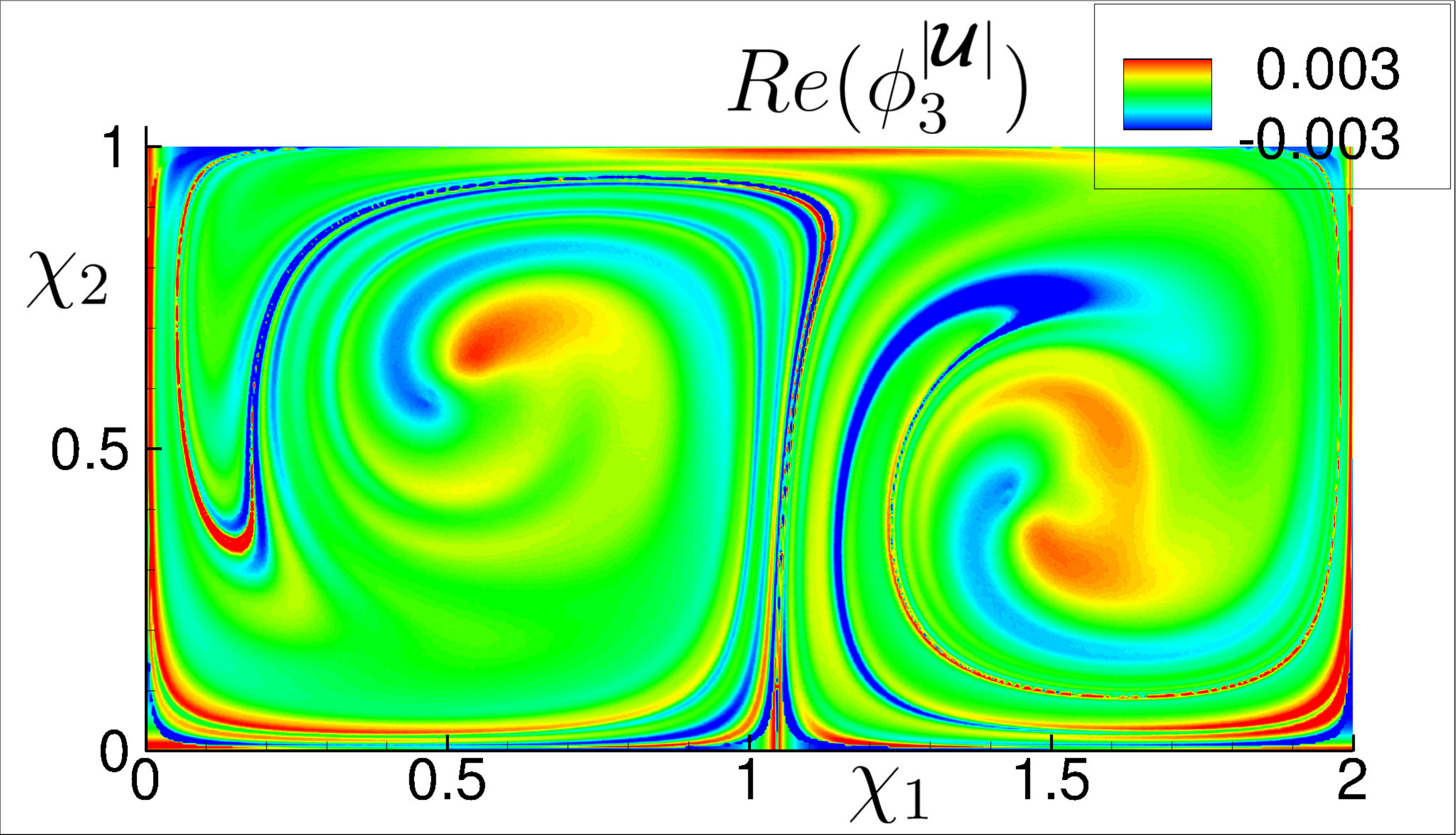}\\(f) LDMD mode
\end{minipage}
\begin{minipage}{0.325\textwidth}
\centering \includegraphics[scale=0.062]{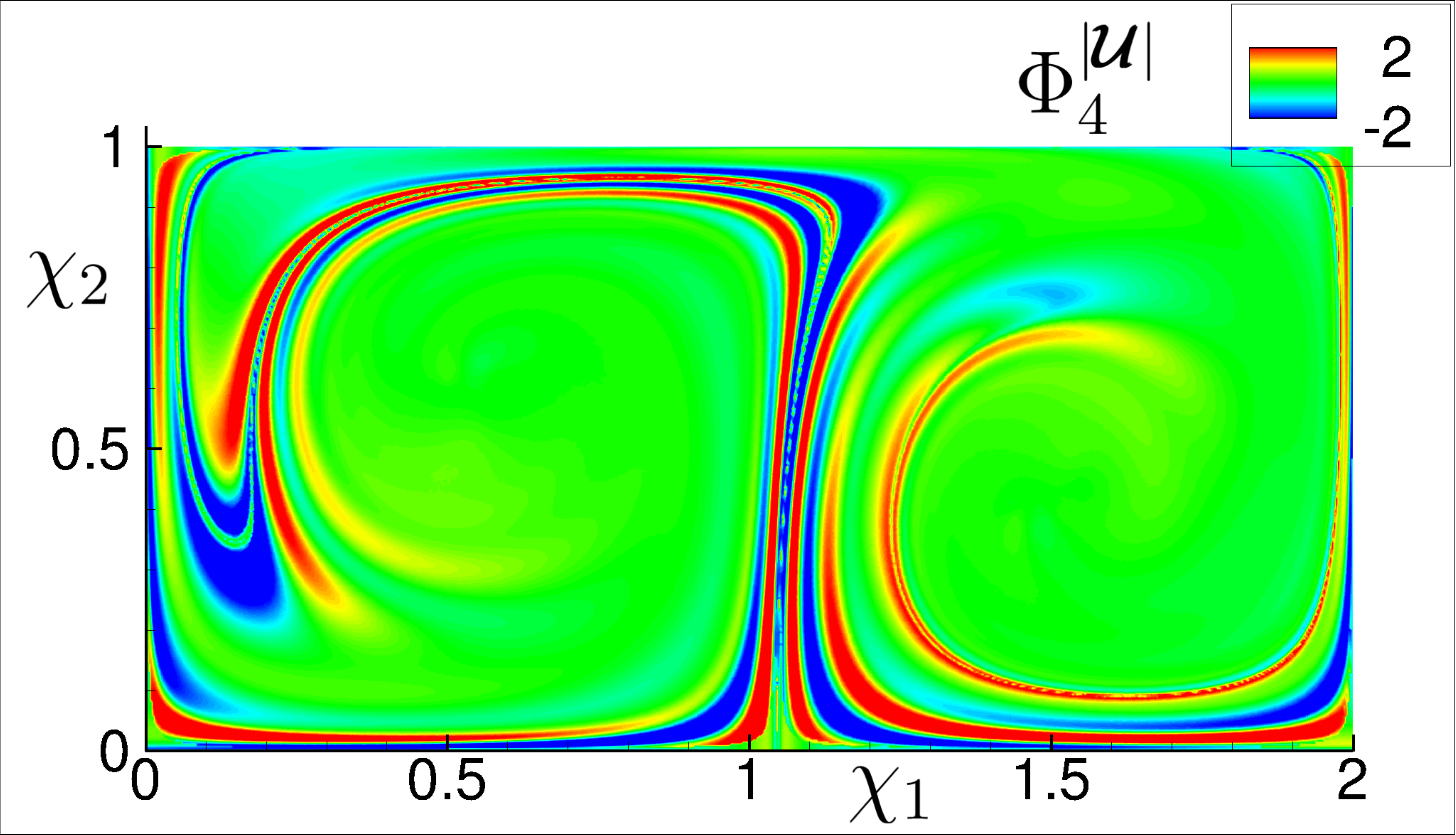}\\(g) LPOD mode\\\includegraphics[scale=0.062]{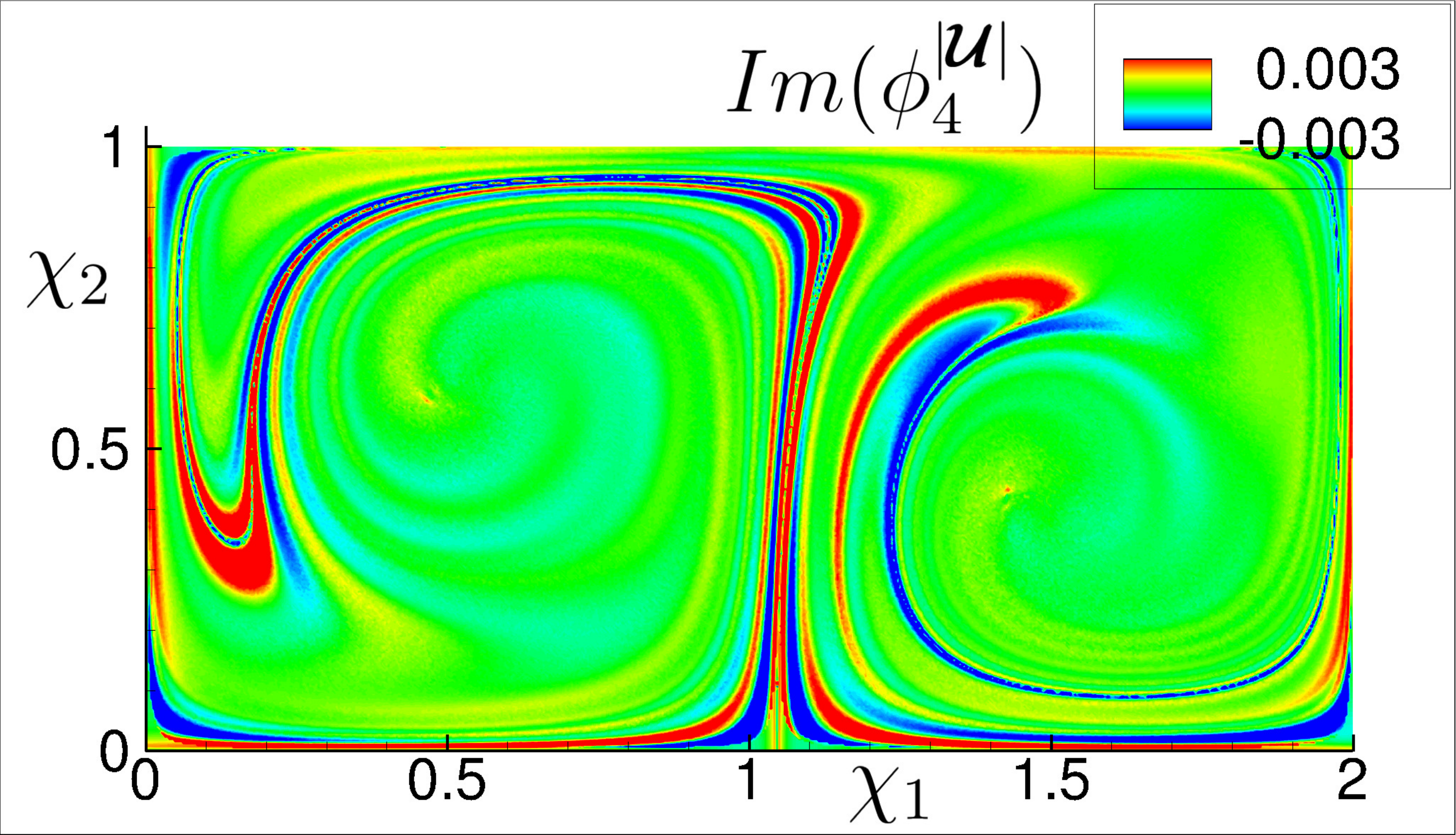}\\(h) LDMD mode
\end{minipage}
\caption{Modal energies/frequencies and higher rank Lagrangian POD/DMD modes of the double-gyre flow. (a) LPOD modal energies (b) LDMD eigenvalues and modal frequencies (c) LPOD mode $\Phi_2^{|\pmb{\mathcal{U}}|}$ (d) LPOD mode $\Phi_3^{|\pmb{\mathcal{U}}|}$ (e) LPOD mode $\Phi_4^{|\pmb{\mathcal{U}}|}$ (f) real part of LDMD mode $\phi_2^{|\pmb{\mathcal{U}}|}$ (g) real part of LDMD mode $\phi_3^{|\pmb{\mathcal{U}}|}$ (h) imaginary part of LDMD mode $\phi_4^{|\pmb{\mathcal{U}}|}$.} 
\label{fig:gyre_modes}
\end{figure}
The simple double-gyre flow leads to a single significant (in terms of the modal energy) LPOD mode, while the higher LPOD modes comprise relatively lower modal energy, as shown in Fig.~\ref{fig:gyre_modes}(a).
For instance, the relative energy of the second LPOD mode is $\lambda_2^{|\pmb{\mathcal{U}}|}/\lambda_1^{|\pmb{\mathcal{U}}|} \approx 0.01$, which decreases further for the higher rank LPOD modes.
The Ritz spectrum of eigenvalues for the LDMD modes is displayed in Fig.~\ref{fig:gyre_modes}(b).
The stationary mode corresponds to $St=0$; in addition, the LDMD modes that have higher modal weights and frequencies of $St=0.05$, $St=0.11$, and $St=0.18$ are marked in the figure (Fig.~\ref{fig:gyre_modes}b).
{The spatial structures of the higher rank LPOD and LDMD modes, displayed in Figs.~\ref{fig:gyre_modes}(c) to (h), exhibit higher modal values (contributions) along the peak values of FTLE field (also FTLE ridges) of Fig.~\ref{fig:ftle}(a), indicating the spatio-temporal dynamics.}

{
The first LPOD mode of Fig.~\ref{fig:ftle}(c) also displays a single contour level at an arbitrary value of $\Phi_1^{|\pmb{\mathcal{U}}|}=1$, delineating the lower velocity magnitude regions of the flow map (material surface) of the double gyre flow.
Clearly, the peak values of the FTLE field are encompassed in $\Phi_1^{|\pmb{\mathcal{U}}|} \lessapprox 1$ region of the first LPOD mode (similarly $Re(\phi_1^{|\pmb{\mathcal{U}}|}) \gtrapprox -0.002$ region of the LDMD mode).
In addition to the high stretching regions of the flow map, the contour level $\Phi_1^{|\pmb{\mathcal{U}}|} = 1$ of the LPOD mode outlines the two core regions of the double gyre flow.
The dynamics of the two gyre cores, in the finite time sense, is apparent by LDMD mode $Re(\phi_2^{|\pmb{\mathcal{U}}|})$ with $St=0.05$ in Fig.~\ref{fig:gyre_modes}(d), which also corresponds to the second most energetic LPOD mode (Fig.~\ref{fig:gyre_modes}a).
The LMA, in general, manifest the spatio-temporal coherence between the different regions of the flow map, where identification of these Lagrangian flow patterns are of paramount significance in the geophysical, chemical, and biological flows, among others.
}

\section{Conclusion} \label{sec:concl}
{
Modal analyses of fluid flows, such as popular proper orthogonal decomposition and dynamic mode decomposition, are typically performed in an Eulerian (fixed) frame of reference, leading to time invariant spatial modes.
These techniques however, face difficulties when the numerical simulations and experiments comprise deforming/moving domains, such as in fluid-structure interaction problems.
To address this issue, we have presented a Lagrangian approach of modal analysis of fluid flows, where the Eulerian flow fields are \textit{a posteriori} transformed to Lagrangian (deforming/moving) flow fields.
For deforming/moving domains, the Lagrangian modal analysis can be alternatively performed in the Lagrangian (material) flow map or moving mesh frame of reference.
Interestingly, in the latter case, the LMA distills out modes that are associated with the domain deformation; in addition, the Lagrangian modes exhibit a large structural similarity with the Eulerian modes on the otherwise static domain.
The LMA (\textit{e.g.}, LPOD/LDMD) procedures elegantly manage the deforming/moving domain, enabling modal analysis with reference to the identity map, which is the initial Lagrangian flow map.

In the material frame of reference, the spatio-temporal material surface/volume is subjected to the LMA. 
The Lagrangian (material) flow map is useful in many circumstances; it is central to  turbulent mixing and transport~\citep{ottino1989kinematics,wiggins2005dynamical}, where features such as fixed points, periodic orbits, stable and unstable manifolds, and chaotic attractors become manifest~\cite{shadden2005definition,haller2015lagrangian}.
In general, most Lagrangian techniques developed to analyze the dynamics of Lagrangian flow map (material surface/volume) naturally relate to or build upon the finite-time Lyapunov exponent, elucidating finite-time dynamics of the material trajectories.
The LMA brings forth a Lagrangian framework for modal decomposition of a spatio-temporal material flow map, providing valuable insights into prominent Lagrangian coherent flow structures and associated dynamics pertinent to the decomposition technique.
For instance, the first LPOD mode is shown to relate closely with the maximum FTLE field, while the higher LPOD modes represent energetically coherent structures of the material flow map; on the other hand, LDMD modes represent dynamically coherent structures of the Lagrangian flow map, exhibiting energy amplification and resonance behavior~\citep{schmid2010dynamic} of not only the primitive variables but also the derived/passive variables, \textit{e.g.}, vorticity, species concentration, temperature.

The Eulerian POD and DMD find restricted usage on steady/base/turbulent mean flows~\citep{schmid2010dynamic,shinde2019transitional}, which are time independent, whereas the Lagrangian POD and DMD can be directly employed for non-uniform steady/base/turbulent mean flows, which are inevitably unsteady in Lagrangian frame of reference.
The LMA on (Eulerian) steady flow yields Lagrangian coherent flow structures that continue to exist in the post-critical (post-bifurcation) flow regime, exhibiting prominent LPOD/LDMD modes that are likely to be engaged in flow transition.
For instance, the LPOD/LDMD modes of the lid-driven cavity at $Re_L=7{,}000$ (pre-bifurcation; Fig.~\ref{fig:LDC_Re7k_lmodes}) and $Re_L=15{,}000$ (post-bifurcation; Fig.~\ref{fig:LDC_Re15k_lmodes}) are strikingly similar, providing insights into the prominent Lagrangian flow features, including those associated with the post-bifurcation unsteadiness.

The notions of finite-time analysis and time-duality pertain to the LMA ansatz, analogous to  hyperbolic trajectories and FTLE.
The forward and adjoint LMA analysis yields Lagrangian (LPOD/LDMD) modes in the upstream and downstream regions of flow, respectively, which potentially leads to  flow sensitivity, receptivity and controllability.
In addition, the LMA remains to be explored for applications that include other features such as barriers in the turbulent mixing and transport, stability characteristics of the steady/base/turbulent mean flows, flow control and optimization in terms of the adjoint formulation, and turbulence and reduced-order modeling.}

\vspace*{0.1in}

\noindent \textbf{Declaration of Interests:} The authors report no conflict of interest.

\section*{Acknowledgment}
We acknowledge support from the Collaborative Center for Aeronautical Sciences sponsored by the Air Force Research Laboratory.

\appendix
\section{Pseudo-algorithm and source code to perform Lagrangian modal analysis} \label{sec:pseudo}

In this section, we present a procedure to perform Lagrangian modal analysis in terms of the Lagrangian proper orthogonal decomposition, considering the simple double gyre flow configuration of Sec.~\ref{sec:gyre}.
The pseudo-algorithm of Table~\ref{tab:algorithm} describes the steps necessary to compute LPOD modes.
Briefly, the input to the algorithm is a set of Eulerian flow fields and associated flow parameters that include the grid, the solution state variables, the number of snapshots ($N_t$), and the time-step ($\delta t$).
A negative value of the time-step is used to perform backward Lagrangian modal analysis, resulting in the Adjoint LPOD modes.
The initial Eulerian flow state $\pmb{u}(\pmb{x}_0,t_0)$ represents the identity map for the Lagrangian flow map, thus $\pmb{\mathcal{U}}(\pmb{\chi}_0,\tau_0)=\pmb{u}(\pmb{x}_0,t_0)$.
An equal number of Lagrangian snapshots ($N_t$) are obtained for the forward ($\delta \tau =\delta t$) or backward ($\delta \tau =-\delta t$) passage of time.
The Lagrangian snapshots can be also extracted from a single snapshot of a steady Eulerian base flow, both in the forward and backward directions.
Lastly, the POD is performed on the Lagrangian flow fields, by computing the correlation matrix via appropriate weight matrix and solving the eigenvalue problem.

\begin{table}
\begin{tabular}{r|l}
\textbf{Inputs:}& Eulerian snapshots, number of snapshots $N_t$, time step $\delta t$ \\
\textbf{Output:} & Lagrangian POD modes \\
\hline
1.& Convert the Eulerian snapshots to the Lagrangian snapshots: \\
  & $\delta \tau = \delta t$ (forward) or $\delta \tau = -\delta t$ (backward) \\
  & $(\pmb{\chi_0},\tau_0)=(\pmb{x}_0,t_0)$\\
  & \textbf{for} $i=1$ to $N_t$ \textbf{do} \\
  & \hspace{5mm} $\pmb{\mathcal{U}}(\pmb{\chi}_{i-1},\tau_{i-1})=\pmb{u}(\pmb{\chi}_{i-1},\tau_{i-1})$ \\
  & \hspace{5mm} $\pmb{\chi}_i = \pmb{\chi}_{i-1} + \pmb{\mathcal{U}}(\pmb{\chi}_{i-1},\tau_{i-1})\times \delta \tau$ \\
  & \hspace{5mm} $\tau_{n}=\tau_{n-1}+\delta \tau$ \\
  & \textbf{end} \\
2.& Lagrangian proper orthogonal decomposition: \\
  & Construct the correlation matrix - \\
  & \textbf{for} $i=1$ to $N_t$ \\
  & \hspace{2mm} \textbf{for} $j=1$ to $N_t$ \\
  & \hspace{5mm} $\pmb{C}(\tau_i,\tau_j)=\pmb{\mathcal{U}}(\pmb{\chi}_i,\tau_i)\pmb{W}_{ij}\pmb{\mathcal{U}}(\pmb{\chi}_j,\tau_j)$ \\
  & \hspace{2mm} \textbf{end} \\
  & \textbf{end} \\
  & Solve the eigenvalue problem - \\
  & $\pmb{C}\pmb{\Psi}_n(\tau)=\lambda_n\pmb{\Psi}_n(\tau)$\\
  & LPOD modes - \\
  & $\pmb{\Phi}_n(\pmb{\chi}_0)=\lambda_n^{-1/2} \sum_{i=1}^{N_t}\pmb{\mathcal{U}}(\pmb{\chi}_i,\tau_i)\Psi_n(\tau_i)$ \\
\end{tabular}
\caption{Pseudo-algorithm for computing Lagrangian proper orthogonal decomposition modes using a set of Eulerian snapshots.}
\label{tab:algorithm}
\end{table}

The source code of the algorithm (Table~\ref{tab:algorithm}) is available in \textit{fortran 90} at \href{https://github.com/Vilas-Shinde/LagrangianModalAnalysis.git}{GitHub link to LagrangianModalAnalysis}.
In addition to the Lagrangian modal analysis sources, the repository includes a subroutine that generates the double gyre flow snapshots in the Eulerian frame of reference.

{
\section{Grid convergence of the lid-driven cavity} \label{sec:grid_ldc}

The geometrical configuration of the lid-driven cavity is displayed in Fig.~\ref{fig:ldc_geo}.
The size of the square cavity is $L_{ref}=L$, the side and bottom surfaces (edges) are prescribed with the no-slip wall and adiabatic boundary conditions.
A uniform velocity $u_1=u_\infty$ in $x_1$ direction is assigned to the top surface, which also uses a Dirichlet condition for all the reference variables.
The specification of regularized velocity profile at the top surface~\citep{shen1991hopf} is not considered, since the effects are not significant~\citep{bergamo2015compressible} for the current purposes.
\begin{figure}
\centering {\includegraphics[scale=0.4]{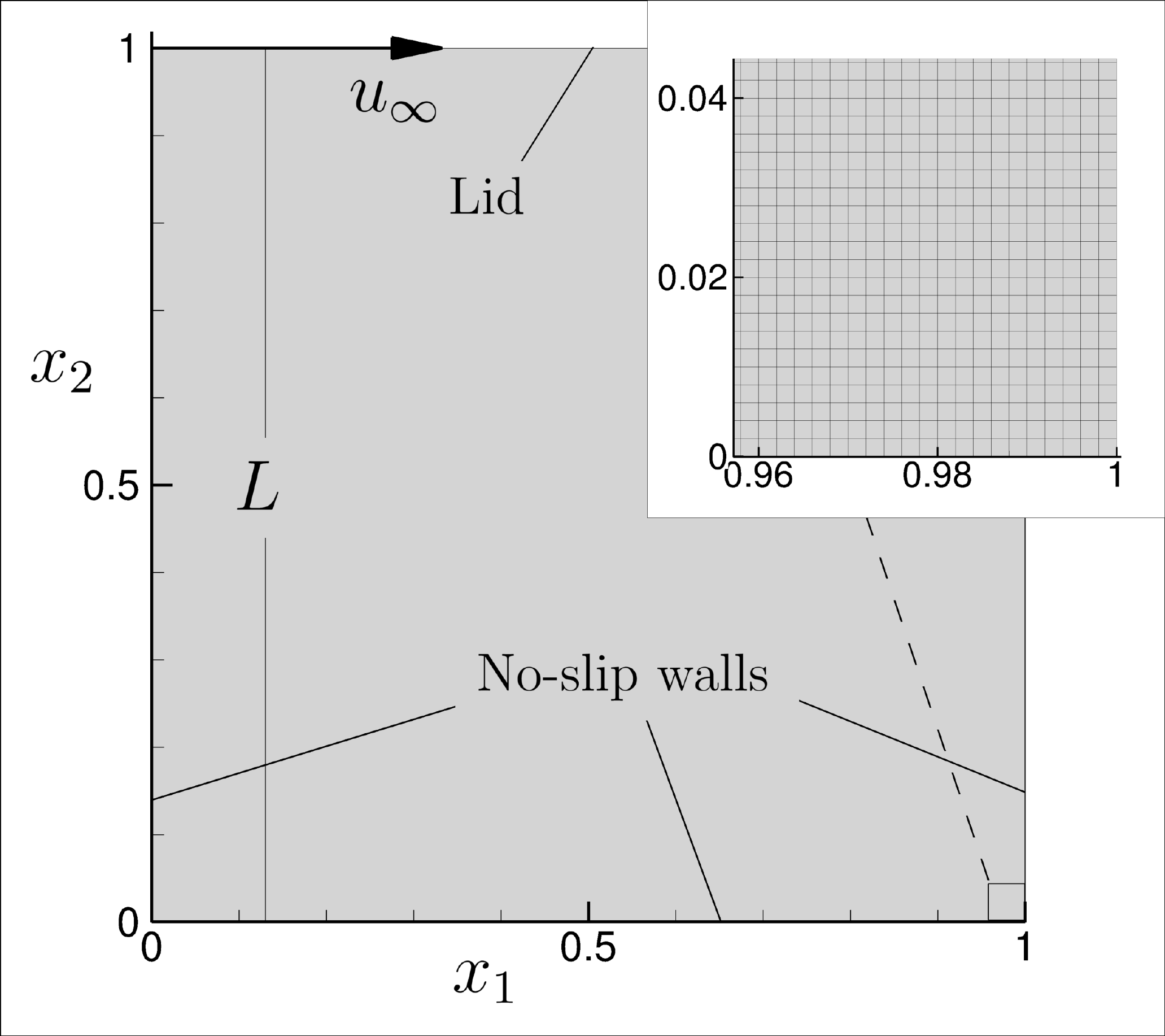}}
\caption{Computational domain of the two-dimensional lid-driven cavity. The inset displays a small region of mesh $\mathtt{M_2}$ with uniform discretization in either direction.}
\label{fig:ldc_geo}
\end{figure}

The computational domain is uniformly discretized in either direction, as shown in the inset of Fig.~\ref{fig:ldc_geo}.
A mesh sensitivity study is performed for $Re_L=15{,}000$ with four meshes, designated $\mathtt{G_1}$, $\mathtt{G_3}$, $\mathtt{G_3}$ and $\mathtt{G_4}$.
The drag and lift coefficients
\[C_d\equiv\frac{F_{x_1}}{\frac{1}{2}\rho_\infty u_\infty^2 L} \text{ and } C_l\equiv\frac{F_{x_2}}{\frac{1}{2}\rho_\infty u_\infty^2 L},\]
where $F_{x_1}$ and $F_{x_2}$ are the forces on the no-slip walls in $x_1$ and $x_2$ directions, respectively, and are noted in Table~\ref{tab:mesh_ldc}.
where the overline ($\overline{\cdot}$) and $rms$ denote the standard time-averaging and root-mean-square of fluctuations operations, respectively.
These parameters display relatively mesh independent results on $\mathtt{G}_3$ and $\mathtt{G}_4$. 
In addition, the table (Table~\ref{tab:mesh_ldc}) displays the Strouhal number, \[ St_L\equiv\frac{fL}{u_\infty}, \] where the flow frequency $f$ and its higher harmonics are estimated using power spectral density of the integrated force $F_{x_1}$ on the cavity.
The Strouhal numbers exhibit converged values for the grid $\mathtt{G}_3$ and $\mathtt{G}_4$, similar to the convergence of the drag and lift coefficients.
\begin{table}
  \begin{center}
\def~{\hphantom{0}}
  \begin{tabular}{lcccccccc}
	Mesh & $n_{x_1} \times n_{x_2}$ 
	& $\overline{C}_d$ & $C_d^{rms}$ & $\overline{C}_l$ & $C_l^{rms}$ & $St_L$ & $^1St_L$ & $^2St_L$ \\ 
	& 
	& $(\times 10^{-2})$ & $(\times 10^{-4})$ & & $(\times 10^{-3})$ & & & \\ [3pt]
	$\mathtt{G}_1$ & $301\times 301$ 
	& $2.70$ &  $2.74$ & $-5.74$ & $4.08$ & $0.14$ & $0.25$ & $0.38$ \\
	$\mathtt{G}_2$ & $501\times 501$ 
	& $2.35$ & $5.94$ & $-5.73$ & $4.07$ & $0.13$ & $0.24$ & $0.37$ \\
	$\mathtt{G}_3$ & $701\times 701$ 
	& $2.06$ & $6.17$ & $-5.72$ & $4.06$ & $0.13$ & $0.24$ & $0.37$ \\
	$\mathtt{G}_4$ & $1{,}001\times 1{,}001$ 
	& $2.09$ & $6.69$ & $-5.72$ & $4.06$ & $0.13$ & $0.24$ & $0.37$ \\
  \end{tabular}
  \caption{Mesh details 
  for the lid-driven cavity at the Reynolds number $Re_L=15{,}000$ and Mach number $M_\infty=0.5$, mesh sensitivity of the time-averaged and root-mean-squared ($rms$) drag and lift coefficients, and the Strouhal number with its higher harmonics.}
  \label{tab:mesh_ldc}
  \end{center}
\end{table}

\begin{figure}
\begin{minipage}{0.5\textwidth}
\centering {\includegraphics[scale=0.85]{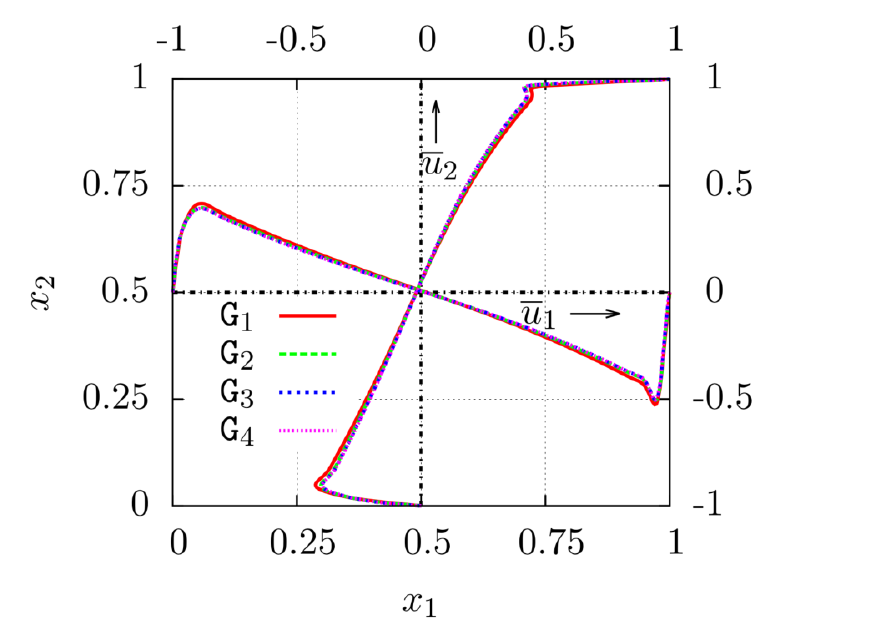}}\\(a)
\end{minipage}
\begin{minipage}{0.5\textwidth}
\centering {\includegraphics[scale=0.8]{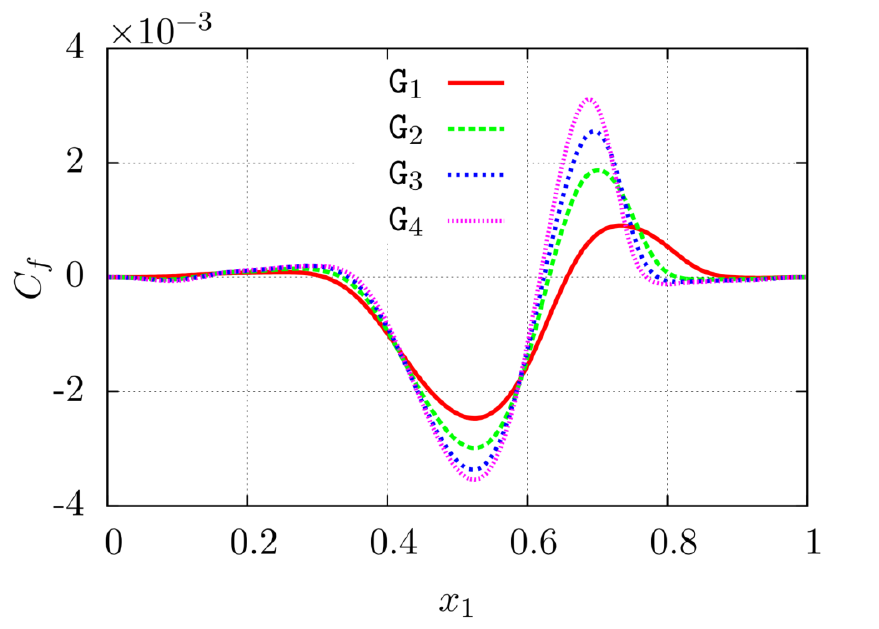}}\\(b)
\end{minipage}
\caption{Grid convergence of the results for the lid-driven cavity at $Re_L=15{,}000$. (a) Time-averaged flow velocity components along the cavity center lines and (b) the skin-friction coefficient on the bottom wall of the cavity for the four grids $\mathtt{G}_1$, $\mathtt{G}_2$, $\mathtt{G}_3$ and $\mathtt{G}_4$.}
\label{fig:ldc_uvmesh}
\end{figure}
Flow velocity components and the skin-friction coefficient, in the time-mean sense, are displayed in Figs.~\ref{fig:ldc_uvmesh}(a) and~\ref{fig:ldc_uvmesh}(b), respectively, for all meshes.
The time-averaged velocity components $\overline{u}_1$ and $\overline{u}_2$ along the vertical ($x_1=0.5$) and horizontal ($x_2=0.5$) center lines, respectively, are nearly identical with respect to meshes $\mathtt{G}_2$, $\mathtt{G}_3$, and $\mathtt{G}_4$.
On the other hand, the skin-friction coefficient on the bottom wall of the cavity, which is defined as 
\begin{equation}\label{eq:skin-friction}
C_f=\frac{\overline{\sigma}_{12}|_{x_2=0}}{\frac{1}{2}\rho_\infty u_\infty^2},
\end{equation}
exhibits some dependence on the mesh refinement from $\mathtt{G}_1$ to $\mathtt{G}_2$, albeit rather locally.
The overall results appear grid converged for meshes $\mathtt{G}_3$ and $\mathtt{G}_4$, thus mesh $\mathtt{G}_3$ is considered for the results and discussion.

\section{Grid convergence of the flow past cylinder} \label{sec:grid_cyl}

The DNS database of flow past cylinder is constructed at a Mach number, $M_\infty=0.5$, for which unsteadiness appears at $Re_D\approx 50$.
Two Reynolds numbers, namely $Re_D=40$ and $Re_D=100$, are chosen to yield pre-bifurcation and post-bifurcation solutions, respectively.
The former condition results in a steady separated flow, while the latter manifests two-dimensional unsteady vortex shedding.
The computational domain, displayed in Fig.~\ref{fig:cyl_m1}, is discretized with a structured cylindrical mesh. 
The cylinder is placed at the origin, with outer boundaries $40D$ away.
The inflow is uniform with a normalized velocity $u_\infty=1.0$, while the outflow is obtained by extrapolation of all variables except pressure, which is maintained constant $p_\infty=1/\gamma M_\infty^2$ in the farfield.
The cylinder surface (perimeter) is prescribed with a no-slip wall and adiabatic boundary conditions.
Periodicity is enforced in the azimuthal direction.
\begin{figure}
  \centering {\includegraphics[scale=0.4]{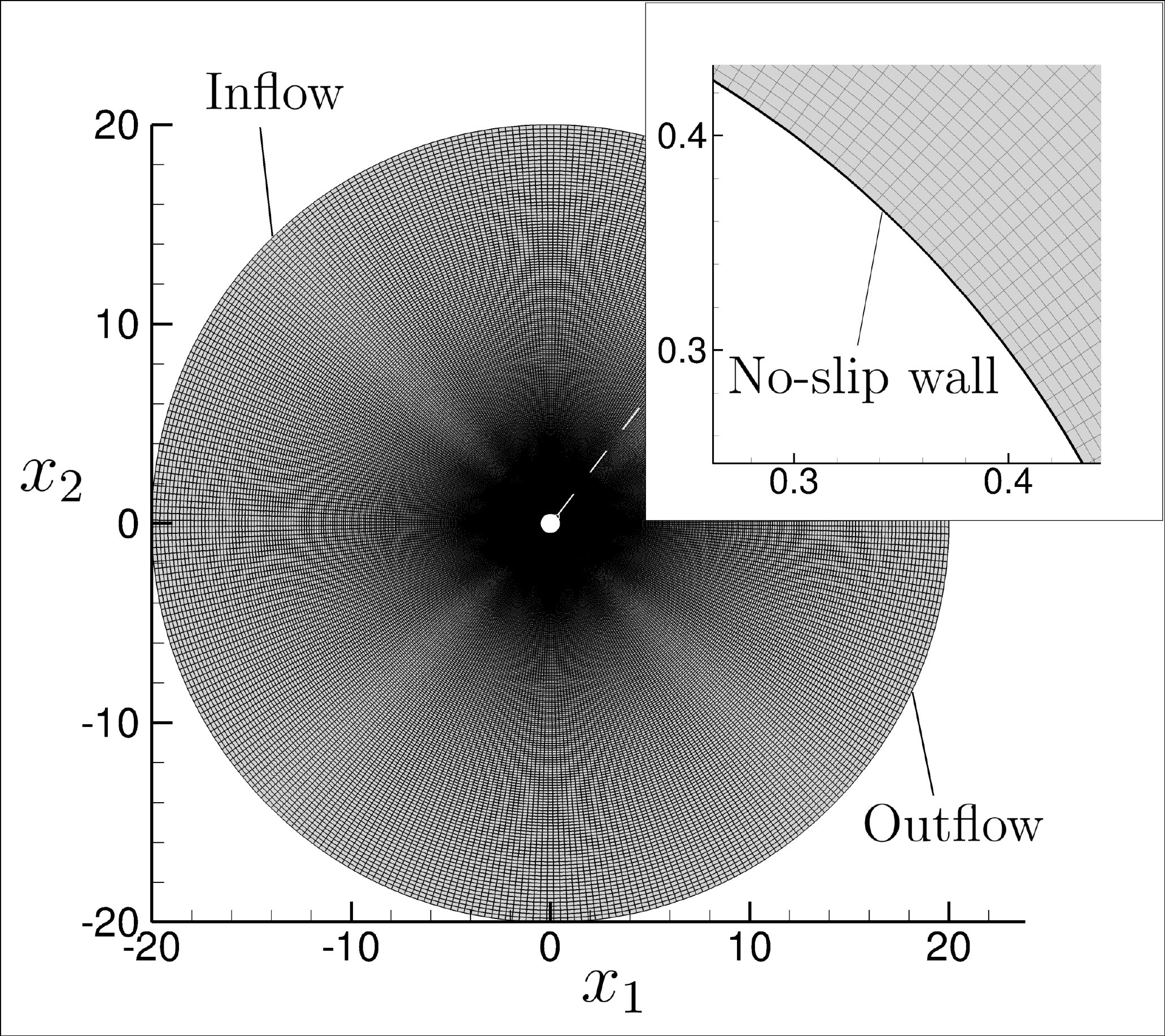}}
  \caption{Computational domain with mesh $\mathtt{M}_1$ to simulate flow past a cylinder of diameter $D$.  The outer boundary is placed at $40D$ from the center. The inset shows the near-wall mesh structure.}
  \label{fig:cyl_m1}
\end{figure}

The sensitivity of the results to spatial resolution is examined  using two meshes,  $\mathtt{M}_1$ and $\mathtt{M}_2$, whose details and effect on the results are provided in Table~\ref{tab:mesh_cyl}.
In mesh $\mathtt{M}_1$, the cylinder surface (perimeter) is discretized using $n_\theta=361$ grid points.
In the radial direction, the first grid point is placed at $\pi D/2(n_\theta-1)$, while the radial extent is divided into $n_r=n_\theta$ points by using a geometric progression with a growth ratio of $r=1.0103$.
For mesh $\mathtt{M}_2$, $n_\theta=541$ is used, resulting in the radial growth ratio of $1.0069$.
\begin{table}
  \begin{center}
\def~{\hphantom{0}}
  \begin{tabular}{lccccccc}
	Mesh & $n_\theta$ & $n_r$ & $r$ & 
	$C_d$ & $C_l$ & $St_D$ \\ [3pt]
	$\mathtt{M}_1$ & $361$ & $361$ & $1.0103$ & 
	$1.5636\pm 0.008$ & $\pm 0.3239$ & $0.1599$ \\
	$\mathtt{M}_2$ & $541$ & $541$ & $1.0069$ & 
	$1.5627\pm 0.008$ & $\pm 0.3267$ & $0.16~~$ \\
  \end{tabular}
  \caption{Details of the computational meshes $\mathtt{M}_1$ and $\mathtt{M}_2$, mesh sensitivity to drag and lift coefficients and Strouhal number for $Re_D=100$ and $M_\infty=0.5$ for the flow past cylinder.}
  \label{tab:mesh_cyl}
  \end{center}
\end{table}

The drag and lift coefficients are defined as:
\[
C_d\equiv \frac{F_d}{\frac{1}{2}\rho_\infty u_\infty^2 D} \text{ and } C_l\equiv \frac{F_l}{\frac{1}{2}\rho_\infty u_\infty^2 D},
\]
where $F_d$ and $F_l$ are the forces in the $x$ and $y$ directions, respectively.
The frequency of unsteady vortex shedding is reported using the Strouhal number, \[ St_D\equiv \frac{fD}{u_\infty}. \]
The drag ($C_d$) and lift ($C_l$) coefficients change little between the two meshes; in addition, the Strouhal number computed using $C_l$ is nearly identical for the two meshes.
The results also match well with the literature~\citep{canuto2015two}.
Thus, we consider the mesh $\mathtt{M}_1$ for further analysis.
}

\bibliographystyle{jfm}
\bibliography{jfm-instructions}

\end{document}